\documentclass[journal,onecolumn]{IEEEtran}
\usepackage{amsfonts}
\usepackage{mathrsfs}
\usepackage{amsmath}
\usepackage{amssymb}
\usepackage{graphicx}
\usepackage{epic}
\usepackage{makecell}
\usepackage{epstopdf}
\usepackage{color}
\usepackage{cite}
\usepackage{bbm}
\usepackage{multirow}
\usepackage{makecell}
\usepackage{mathtools}
\usepackage{enumerate}
\usepackage{hyperref}
\usepackage{lineno}
\usepackage{url}
\usepackage{float}
\usepackage{color}
\usepackage{bm}
\usepackage{comment}
\usepackage{booktabs}
\usepackage{tikz}
\usepackage{indentfirst}
\usepackage{exscale} 
\usepackage{relsize}
\usepackage{pifont}
\usepackage{enumerate}
\newtheorem{claim}{Claim}[section]
\newtheorem{theorem}{Theorem}[section]

\newtheorem{definition}[theorem]{Definition}

\newtheorem{lemma}[theorem]{Lemma}

\newtheorem{remark}[theorem]{Remark}

\newcommand{\C}{\mathcal {C}}

\begin{document}
	
\title{Binary Reconstruction Codes for Correcting One Deletion and One Substitution}
\author{Yuling~Li, Yubo~Sun, Gennian~Ge%
\thanks{This research was supported by the National Key Research and Development Program of China under Grant 2020YFA0712100, the National Natural Science Foundation of China under Grant 12231014, and Beijing Scholars Program.}
\thanks{Y. Li ({\tt 2240501022@cnu.edu.cn}), Y. Sun ({\tt 2200502135@cnu.edu.cn}), and G. Ge ({\tt gnge@zju.edu.cn}) are with the School of Mathematical Sciences, Capital Normal University, Beijing 100048, China.}
}


\maketitle

\begin{abstract}
    In this paper, we investigate binary reconstruction codes capable of correcting one deletion and one substitution. We define the \emph{single-deletion single-substitution ball} function \( \mathcal{B} \) as a mapping from a sequence to the set of sequences that can be derived from it by performing one deletion and one substitution.   
    A binary \emph{$(n,N;\mathcal{B})$-reconstruction code} is defined as a collection of binary sequences of length \( n \) such that the intersection size between the single-deletion single-substitution balls of any two distinct codewords is strictly less than \( N \). This property ensures that each codeword can be uniquely reconstructed from \( N \) distinct elements in its single-deletion single-substitution ball.  
    Our main contribution is to demonstrate that when \( N \) is set to \( 4n - 8 \), \( 3n - 4 \), \( 2n + 9 \), \( n+21 \), \( 31 \),  and \( 7 \), the redundancy of binary \((n,N;\mathcal{B})\)-reconstruction codes can be \( 0 \), \( 1 \), \( 2 \), \( \log\log n + 3 \), \( \log n + 1 \), and \( 3\log n + 4 \), respectively, where the logarithm is on base two.
\end{abstract}



\section{Introduction}\label{I}
\IEEEPARstart{T}{raditional} communication systems involve a sender transmitting a codeword from a codebook through a noisy channel, where the receiver attempts to recover the transmitted codeword from the channel's output. 
This fundamental framework necessitates the use of \emph{error-correcting codes} to ensure reliable communication. Levenshtein's pioneering work \cite{Levenshtein-66-SPD-1D} established optimal codes for correcting single deletion, insertion, or substitution. However, correcting multiple deletion-related errors has presented significant challenges, with low-redundancy code constructions remaining an open problem until recent advances \cite{Brakensiek-18-IT-kD, Cai-21-E, Gabrys-23-IT-DS, Guruswami-21-IT-2D, Gabrys-19-IT-2D, Levenshtein-66-SPD-1D, Li-23-DS, Liu-24-IT, Pi-24-arXiv-2E, Nguyen-24-IT-1D, Song-22-IT, Smagloy-23-IT-DS, Sima-21-IT-kD, Sun-24-IT, Sun-25-IT, Sima-20-ISIT-q, Sima-20-ISIT, Sima-20-ISIT-tD, Song-22-IT-DS, Sima-20-IT-2D, Tenengolts-84-IT-q_D, Ye-24-arXiv, Song-22-ISIT-LD}.

Levenshtein \cite{Levenshtein-01-JCTA-recons} later generalized this framework to multiple identical noisy channels, wherein each channel produces a distinct output. He introduced the \emph{sequence reconstruction problem}, which determines the minimum number of channels required to guarantee the unique reconstruction of the transmitted codeword, given a selected codebook. Subsequent research has resolved this problem for various channel types, including multiple insertions \cite{Sala-17-IT-reconstr-ins, Ye-23-IT,Lan-25}, multiple deletions \cite{Gabrys-18-IT-reconstr-del, Chrisnata-22-IT-reconstr-del, Pham-25-JCTA-reconstr-del, Lan-25}, and combined substitution-insertion channels \cite{Abu-Sini-21-IT-recons}.

Motivated by applications in DNA storage \cite{Church-12-science-DNA, Goldman-13-nature-DNA, Yazdi-15-TMBMC-DNA, Organick-18-nature-DNA} and racetrack memory \cite{Chee-18-IT}, recent work by Cai et al. \cite{Cai-22-IT-recon-edit} and Chrisnata et al. \cite{Chrisnata-22-IT-reconstr-del} has proposed a dual problem of the sequence reconstruction problem, termed \emph{the reconstruction codes problem}. This problem assumes that the number of channels is a fixed parameter and aims to design codes with the smallest possible redundancy such that each codeword can be successfully reconstructed from the outputs of these channels. Significant progress has been made for insertion/deletion channels \cite{Cai-22-IT-recon-edit, Chrisnata-22-IT-reconstr-del, Ye-23-IT, Wu-24-DCC-reconstr, Sun-23-IT-reconstr, Zhang-24-ISIT-reconstr, Sun-24-arXiv} and burst insertion/deletion channels \cite{Sun-23-IT-BDR}. 

In this paper, we address the reconstruction codes problem for channels introducing one deletion and one substitution. Let \( \mathcal{B} \) be the \emph{single-deletion single-substitution ball} that maps a sequence to the set of sequences that can be derived from it by performing one deletion and one substitution. 
We define an \emph{\((n, N; \mathcal{B})\)-reconstruction code} as a set of binary sequences of length \(n\) such that for any distinct codewords \(\bm{x}\) and \(\bm{y}\), it holds that \(
|\mathcal{B}(\bm{x}) \cap \mathcal{B}(\bm{y})| < N\).
In this scenario, each codeword from some specific \emph{$(n,N;\mathcal{B})$-reconstruction code} can be uniquely reconstructed from \( N \) distinct elements belonging to its single-deletion single-substitution ball.

We begin by constructing a redundancy-free reconstruction code, which requires determining the maximum intersection size between distinct single-deletion and single-substitution balls. Let \(d\) (respectively, \(s\)) denote the size of the intersection between the single-deletion (respectively, single-substitution) balls of sequences \(\bm{x}\) and \(\bm{y}\), both of length \(n\). We find that  
\[  
\big| \mathcal{B}(\bm{x}) \cap \mathcal{B}(\bm{y}) \big| \leq (d + s) n + O(1).  
\]  
It is well known from \cite{Levenshtein-01-IT-recons} that \(d \in \{0,1,2\}\) and \(s \in \{0,2\}\). Therefore, the maximum intersection size between distinct single-deletion and single-substitution balls is \(4n + O(1)\), where \(n\) is the length of centers of the error balls. This enables us to derive a redundancy-free \((n, 4n + O(1); \mathcal{B})\)-reconstruction code.  
By imposing suitable constraints to exclude the case where \((d, s) = (2, 2)\), we obtain an \((n, 3n + O(1); \mathcal{B})\)-reconstruction code with only one bit of redundancy.   
By further restricting the number of runs in each sequence belonging to our \((n, 3n + O(1); \mathcal{B})\)-reconstruction code, we derive an \((n, 2n + O(1); \mathcal{B})\)-reconstruction code with two bits of redundancy.   
Moreover, by imposing constraints on the number of runs in each sequence of a single-deletion reconstruction code, we obtain an \((n, n + O(1); \mathcal{B})\)-reconstruction code with $\log\log n+O(1)$ bits of redundancy.  
Finally, we show that a single-edit correcting code, which has redundancy \(\log n + O(1)\), is an \((n, 31; \mathcal{B})\)-reconstruction code, and that the single-deletion single-substitution list-decodable code with list size two developed in \cite{Song-22-ISIT-LD}, with redundancy \(3 \log n + O(1)\), qualifies as an \((n, 7; \mathcal{B})\)-reconstruction code.
We summarize our main contributions in Table~\ref{tab:CON}.   

While this paper was in preparation, we learned that Song et al. \cite{Cai-25-arxiv-sd} also investigated \( (n, N; \mathcal{B}) \)-reconstruction codes, constructing codes with redundancies of \( 1 \) and \( \log n + O(1) \) for \( N = 4n - 8 \) and \( 41 \), respectively.

The remainder of this paper is organized as follows.  
Section~\ref{sec:pre} introduces the relevant notations, definitions, and main tools used throughout the paper.  
Section~\ref{sec:key} presents the key lemmas that characterize the intersection between two distinct single-deletion single-substitution balls under various conditions.  
Section~\ref{sec:constr} describes the construction of our single-deletion single-substitution reconstruction codes.  
Section~\ref{sec:proof} provides the proofs of the lemmas stated in Section~\ref{sec:key}.  
Finally, Section~\ref{sec:concl} concludes the paper and discusses several open problems.  

\begin{table*}[t]
\renewcommand\arraystretch{1.5}
\caption{Redundancies of Our \((n, N; \mathcal{B})\)-Reconstruction Codes for Specific Values of \(N\) } 
\centering
\begin{tabular}{c|c|c} 
\hline
\hline
Parameter $N$ & Code Redundancy & Remark\\
\hline
$N= 4n-8$ & 0 & Theorem \ref{thm:4n}\\
\hline
$N=3n-4$ & 1 & Theorem \ref{thm:3n} \\
\hline
$N=2n+9$ & 2 & Theorem \ref{thm:2n} \\
\hline
$N=n+21$ & $\log\log n+3$ & Theorem \ref{thm:n} \\
\hline
$N=31$ & $\log n+1$ & Theorem \ref{thm:31} \\
\hline
$N=7$ & $3\log n+4$ & Theorem \ref{thm:7} \\
\hline
\hline
\end{tabular}
\label{tab:CON}
\end{table*}

\section{preliminaries}\label{sec:pre}

\subsection{Notations}
Let \([i,j]\) denote the set \(\{i,i+1,\ldots,j\}\) if \(i\leq j\), and the empty set otherwise. Specifically, when \(i=1\), we abbreviate \([1,j]\) as \([j]\). Let \(\Sigma= \{0,1\}\) be the binary alphabet set, and let \(\Sigma^n\) be the set of all sequences of length \(n\) over \(\Sigma\). Furthermore, we define \(\Sigma^{\ast}= \bigcup_{n\geq 0} \Sigma^n\).  

For a sequence \(\bm{x}\in\Sigma^n\), we denote its \(i\)-th entry as \(x_i\) and represent the sequence as \(x_1x_2\cdots x_n\). Moreover, we define its \emph{complement} as \(\overline{\bm{x}}\triangleq \overline{x_1}\overline{x_2}\ldots \overline{x_n}\), where \(\overline{x_i}=1-x_i\) for \(i\in [n]\), and its \emph{reversal} as \(R(\bm{x})\triangleq x_n x_{n-1}\cdots x_1\). Let \(|\bm{x}|\) denote the length of \(\bm{x}\) when \(\bm{x}\) is a sequence, and let \(|\mathcal{C}|\) be the size of \(\mathcal{C}\) when \(\mathcal{C}\) is a set.  

For two sequences \(\bm{x}\in\Sigma^n\) and \(\bm{y}\in\Sigma^m\), we use \(\bm{x}\bm{y}=x_1\cdots x_n y_1\cdots y_m\) to denote their \emph{concatenation}. In particular, we refer to \(\bm{x}^m\) as the sequence obtained by concatenating \(\bm{x}\) \(m\) times. If there exists a set \(\mathcal{I}=\{i_1,i_2,\ldots,i_m\}\) with \(1\le i_1<i_2<\cdots <i_m \leq n\) such that \(\bm{y}=\bm{x}_{\mathcal{I}}\triangleq x_{i_1}x_{i_2}\cdots x_{i_m}\), we say that \(\bm{y}\) is a \emph{subsequence} of \(\bm{x}\). Specifically, when \(\mathcal{I}\) is an interval, \(\bm{y}\) is called a \emph{substring} of \(\bm{x}\).  

For a positive integer \(t\), we say that \(\bm{x}\in \Sigma^n\) has a \emph{period} of \(t\) or is \(t\)-periodic if \(t\) is the smallest integer such that \(x_i=x_{i+t}\) for every \(i\in [n-t]\). Moreover, we say that \(\bm{x}\) is \emph{\(^{\le}t\)-periodic} if it has a period of at most \(t\). A \emph{run} in \(\bm{x}\) is a maximal substring of \(\bm{x}\) of period one, and an \emph{alternating sequence} in \(\bm{x}\) is a maximal substring of \(\bm{x}\) of period two. We use \(r(\bm{x})\) to denote the number of runs in \(\bm{x}\).  
For a positive integer $P$, let $\mathcal{R}(n,P)$ be the set of all binary sequences of length $n$ in which each \(^{\le}2\)-periodic substring is of length at most $P$.

For two sequences \(\bm{x}, \bm{y} \in \Sigma^n\), let \(\mathrm{supp}(\bm{x})\) denote the \emph{support} of \(\bm{x}\), i.e., the set of positions where \(\bm{x}\) is nonzero. Define the \emph{weight} of \(\bm{x}\) as \( \mathrm{wt}(\bm{x}) \triangleq |\mathrm{supp}(\bm{x})|\),  
which counts the number of nonzero entries in \(\bm{x}\). The \emph{Hamming distance} between \(\bm{x}\) and \(\bm{y}\) is given by  \(d_H(\bm{x}, \bm{y}) \triangleq \mathrm{wt}(\bm{x} - \bm{y})\), 
counting the number of coordinates where \(\bm{x}\) and \(\bm{y}\) differ. When \(\bm{x}\) and \(\bm{y}\) are clear from the context, we write \(d_H \triangleq d_H(\bm{x}, \bm{y})\). 
Moreover, define \(\bm{a} \triangleq \bm{a}(\bm{x}, \bm{y}) = \bm{x}_{[1, j_1 - 1]} = \bm{y}_{[1, j_1 - 1]}\) and \(\bm{b} \triangleq \bm{b}(\bm{x}, \bm{y}) = \bm{x}_{[j_{d_H} + 1, n]} = \bm{y}_{[j_{d_H} + 1, n]}\)
as the longest common prefix and suffix of \(\bm{x}\) and \(\bm{y}\), respectively.  

\subsection{Errors and Codes}

Let \(\bm{x}\in \Sigma^n\). A \emph{deletion} transforms it into one of its subsequences of length \(n-1\), and a \emph{substitution} transforms it into a sequence of length \(n\) with a Hamming distance of at most one from \(\bm{x}\). Let \(\mathcal{D}(\bm{x})\) be the \emph{single-deletion ball} of \(\bm{x}\), which contains the set of all sequences obtainable from \(\bm{x}\) after one deletion. Let \(\mathcal{S}(\bm{x})\) be the \emph{single-substitution ball} of \(\bm{x}\), which contains the set of all sequences obtainable from \(\bm{x}\) after one substitution. Finally, let \(\mathcal{B}(\bm{x})\) be the \emph{single-deletion single-substitution ball} of \(\bm{x}\), which contains
the set of all sequences obtainable from \(\bm{x}\) after one deletion and one substitution.  

For notational simplicity, for any sequences \(\bm{x}\) and \(\bm{y}\), we define \(\mathcal{D}(\bm{x},\bm{y})\triangleq \mathcal{D}(\bm{x})\cap \mathcal{D}(\bm{y})\), \(\mathcal{S}(\bm{x},\bm{y})\triangleq \mathcal{S}(\bm{x})\cap \mathcal{S}(\bm{y})\), and \(\mathcal{B}(\bm{x},\bm{y})\triangleq \mathcal{B}(\bm{x})\cap \mathcal{B}(\bm{y})\).  Let $\bm{x}(i,\hat{i})$ be the sequence received from $\bm{x}$ by deleting its $i$-th entry and substituting its $\hat{i}$-th entry.

\begin{definition}  
    A set of sequences \(\mathcal{C}\subseteq \Sigma^n\) is referred to as an \emph{\((n,N;\mathcal{B})\)-reconstruction code} if for any two distinct sequences \(\bm{x},\bm{y}\in \mathcal{C}\), it holds that \(|\mathcal{B}(\bm{x},\bm{y})|<N\).  
\end{definition}  

To evaluate a code \(\mathcal{C}\subseteq \Sigma^n\), we calculate its \emph{redundancy} \(\mathrm{red}
(\mathcal{C})\triangleq n-\log |\mathcal{C}|\), where the logarithm is on base two.  

\begin{remark}  
    It is well known in \cite{Smagloy-23-IT-DS} that the order of deletions and substitutions is commutative in terms of the final result. Therefore, without loss of generality, we adopt the convention that deletions precede substitutions.  
\end{remark}  

The following definitions will be crucial in our code constructions.

\begin{definition}
    For any \( \bm{x}\in\Sigma^n\), for $k\geq 1$, its \emph{$k$-th order VT syndrome} is defined as $\mathrm{VT}^{k}(\bm{x})=\sum_{i=1}^n\sum_{j=1}^i j^{k-1}x_i$.
\end{definition}

\begin{definition}
    The \emph{inversion number} of $\boldsymbol{x}\in \Sigma^n$ is defined as $\mathrm{Inv} (\boldsymbol{x}) \triangleq \left| \{(i,j) : 1 \leq i < j \leq n, x_i > x_j \} \right|$, where the alphabet set $\Sigma$ is ordered such that $0<1$.
\end{definition}

\subsection{Useful Tools}

In this subsection, we review several well-known conclusions that will be critical for our later discussion.

\subsubsection{Error Correcting Codes}
We first review a construction of binary code that can correct one deletion or one substitution.

\begin{lemma}\cite[Theorem 2]{Levenshtein-66-SPD-1D}\label{lem:VT}
    For any $a \in [0,2n-1]$, define $\mathrm{VT}_{a}(n) = \left\{\bm{x} \in \Sigma^n : \mathrm{VT}^{1}(\bm{x}) \equiv a \pmod{2n}\right\}$, it holds that $\mathcal{D}(\bm{x},\bm{y})=\emptyset$ and $\mathcal{S}(\bm{x},\bm{y})=\emptyset$ for any two distinct sequences $\bm{x}, \bm{y}\in \mathrm{VT}_a(n)$.
\end{lemma}

We now review a construction of single-deletion single-substitution list-decodable code with list size two.
\begin{lemma}\cite[Definition 1, Theorem 1, and Lemma 1]{Song-22-ISIT-LD}\label{lem:DS_L}
For any $a_0 \in [0,3]$, $a_1 \in [0,2n-1]$, and $a_2 \in [0,2n^2-1]$, define the code 
\begin{align*}
    \mathcal{C}_L\triangleq \big\{\bm{x}\in \Sigma^n :\mathrm{wt}(\bm{x})\equiv a_0 \pmod{4},~ \mathrm{VT}^k(\bm{x})\equiv a_{k} \pmod{2n^k} \text{ for } k\in \{1,2\} \big\}.
\end{align*}
For any three distinct sequences $\bm{x},\bm{y},\bm{z}\in \mathcal{C}_L$, it holds that $\mathcal{B}(\bm{x})\cap \mathcal{B}(\bm{y})\cap \mathcal{B}(\bm{z})=\emptyset$.
Moreover, if $\boldsymbol{x}(i,\hat{i})= \boldsymbol{y}(j,\hat{j})$, then $\hat{i}, \hat{j}\in [i,j]\cup [j,i]$.
\end{lemma}

\subsubsection{Reconstruction Codes}
Here, we introduce a construction of single-deletion reconstruction code.

\begin{lemma}\cite[Theorem 17]{Cai-22-IT-recon-edit}\label{lem:reconstruction}
    Let $P>0$ be an even integer.
    Let $a_1 \in [0,1]$ and $a_2 \in [0,P/2]$, define the code
    \begin{align*}
        \mathcal{C}_P(a_1,a_2) \triangleq \big\{ \boldsymbol{x}
        \in \mathcal{R}(n,P):\mathrm{wt}(\boldsymbol{x}) \equiv a_1 \pmod{2},~\mathrm{Inv} (\boldsymbol{x}) \equiv a_2 \pmod{1+P/2} \big\},
    \end{align*}
    then for any distinct
    $\bm{x},\bm{y}\in \mathcal{C}_P(a_1,a_2)$, it holds that $|\mathcal{D}(\bm{x},\bm{y})|\leq 1$.
\end{lemma}

\begin{remark}\label{rmk:runlength}
    For any $\bm{x}\in \Sigma^n$, let \(\psi(\bm{x}) \in \Sigma^n\) be defined as
    \(\psi(x)_i = x_i - x_{i-1} \pmod{2}\), where \(x_0 = 0\).  
    It is clear that \(\psi(\cdot)\) is a bijection, and that \(\bm{x} \in \mathcal{R}(n,P)\) if every run in \(\psi(\bm{x})\) has length at most \(P - 1\).  
    Then, by applying an argument similar to that in \cite[Lemma 2]{Schoeny-17-IT-BD}, we can conclude that \(|\mathcal{R}(n,P)| \geq 3 \cdot 2^{n-2}\) when \(P \geq \log n + 3\).  
\end{remark}

\subsubsection{Error Ball}

We first review the sizes of a single-deletion ball and a single-substitution ball.

\begin{lemma}\cite{Levenshtein-66-SPD-1D}\label{lem:size}
    For any $\bm{x}\in \Sigma^n$, it holds that $|\mathcal{D}(\bm{x})|=r(\bm{x})$ and $|\mathcal{S}(\bm{x})|=n+1$, where $r(\bm{x})$ denotes the number of runs in $\bm{x}$.
\end{lemma}

Now we consider the intersection sizes between distinct single-substitution balls and single-deletion balls, respectively.

\begin{lemma}\cite[Corollary 1]{Levenshtein-01-IT-recons}\label{lem:sub}
    Let \(\bm{x}\) and \(\bm{y}\) be distinct sequences in \(\Sigma^{n}\). It holds that \(|S(\bm{x},\bm{y})|\in \{0,2\}\). Specifically,
    \begin{itemize}
        \item If \(d_H(\bm{x},\bm{y})=1\), we can write \(\bm{x}=\bm{a}\alpha \bm{b}\) and \(\bm{y}= \bm{a} \overline{\alpha} \bm{b}\) for some \(\alpha\in \Sigma\). Then, \(|S(\bm{x},\bm{y})|=2\) and \(S(\bm{x},\bm{y})= \{\bm{a}\alpha \bm{b}, \bm{a}\overline{\alpha} \bm{b}\}\);
        \item If \(d_H(\bm{x},\bm{y})=2\), we can write \(\bm{x}=\bm{a}\alpha \bm{c} \beta \bm{b}\) and \(\bm{y}= \bm{a} \overline{\alpha} \bm{c} \overline{\beta} \bm{b}\) for some \(\alpha, \beta\in \Sigma\) and \(\bm{c}\in \Sigma^{\ast}\). Then, \(|S(\bm{x},\bm{y})|=2\) and \(S(\bm{x},\bm{y})= \{\bm{a}\overline{\alpha} \bm{c} \beta \bm{b}, \bm{a}\alpha \bm{c} \overline{\beta} \bm{b}\}\);
        \item If \(d_H(\bm{x},\bm{y})\geq 3\), then \(|S(\bm{x},\bm{y})|=0\).
    \end{itemize}
\end{lemma}

\begin{lemma}\cite[Lemmas 2 and 4]{Chrisnata-22-IT-reconstr-del}\label{lem:del} 
    Let \(\bm{x}\) and \(\bm{y}\) be distinct sequences in \(\Sigma^{n}\). We have \( |\mathcal{D}(\bm{x},\bm{y})|\in \{0,1,2\} \). Specifically, 
    \begin{itemize}
        \item \( |\mathcal{D}(\bm{x},\bm{y})|=2 \) holds if and only if we can write \(\bm{x}=\bm{a}\bm{c}\bm{b}\) and \(\bm{y}=\bm{a}\overline{\bm{c}}\bm{b}\), where \(\bm{c}\) is an alternating sequence of length at least two;
        \item \( |\mathcal{D}(\bm{x},\bm{y})|=1 \) holds if and only if one of the following holds:
        \begin{itemize}
            \item if $d_H(\bm{x},\bm{y})=1$, \(\bm{x}=\bm{a}\alpha\bm{b}\) and \(\bm{y}=\bm{a}\overline{\alpha}\bm{b}\) for some \(\alpha\in \Sigma\);
            \item if $d_H(\bm{x},\bm{y})\geq 2$, $\{\bm{x}, \bm{y}\}= \big\{\bm{a} \overline{\alpha}  \alpha \bm{c}\beta \bm{b}, \bm{a}\alpha \bm{c}\beta  \overline{\beta} \bm{b}\big\}$ with $\overline{\alpha}\alpha  \bm{c}\neq \bm{c}\beta  \overline{\beta}$ and $\alpha  \bm{c}\neq \bm{c}\beta$ for some \(\alpha, \beta\in \Sigma\) and \(\bm{c}\in \Sigma^{\ast}\).
        \end{itemize}
    \end{itemize}
\end{lemma}

As a byproduct of this conclusion, we can characterize the sequence $\bm{z}\in \mathcal{D}(\bm{x},\bm{y})$.

\begin{lemma}\label{lem:del'}
    Let $\bm{x},\bm{y}\in\Sigma^{n}$ be such that $d_H= d_H(\bm{x},\bm{y})\geq 1$. Let $j_1,j_{d_H}$ denote the indices of the first, last bit in which $ \bm{x}$ and $\bm{y}$ differ, respectively. If  $|\mathcal{D}(\bm{x},\bm{y})|=1$, for any $\bm{z}\in \mathcal{D}(\bm{x},\bm{y})$, one of the following holds:
    \begin{itemize}
        \item $\bm{z}=\bm{x}_{[n]\setminus \{j_1\}}= \bm{y}_{[n]\setminus \{j_{d_H}\}}$ and $\bm{x}_{[j_1+1,j_{d_H}]}= \bm{y}_{[j_1,j_{d_H}-1]}$;
        \item $\bm{z}=\bm{x}_{[n]\setminus \{j_{d_H}\}}= \bm{y}_{[n]\setminus \{j_1\}}$ and $\bm{x}_{[j_1,j_{d_H}-1]}= \bm{y}_{[j_1+1,j_{d_H}]}$.
    \end{itemize}
\end{lemma}

The following lemma characterizes the Hamming distance between distinct sequences within a deletion ball.  

\begin{lemma}\label{lem:del_position}\cite[Lemma 5]{Abu-Sini-21-IT-recons}
    Assume \(\bm{x} \in \Sigma^n\) has \(r = r(\bm{x})\) runs. Let \(\bm{x}[i]\) denote the sequence obtained by deleting a symbol from the \(i\)-th run of \(\bm{x}\), then $\mathcal{D}(\bm{x})=\{\bm{x}[1],\bm{x}[2],\cdots,\bm{x}[r]\}$ and $d_{H}(\bm{x}[i],\bm{x}[j])=j-i$ for $1\le i<j\le r$.
\end{lemma}

Based on Lemma \ref{lem:del_position}, we can derive the following conclusion.

\begin{lemma}\label{lem:del_Hamming}
    For any $\bm{u}\in \Sigma^n$ and $\bm{v}\in\Sigma^{n+1}$, let $\mathcal{F}=\{\bm{z}\in \mathcal{D}(\bm{v}):d_H(\bm{z},\bm{u})\leq 1 \}$, it holds that $|\mathcal{F}|\leq 3$.
    Moreover, when $|\mathcal{F}|=3$, we have $\bm{u}\in \mathcal{F}$.
\end{lemma}

\begin{IEEEproof}
    Let $\bm{v}[i]$ be the sequence received by deleting a symbol from the $i$-th run of $\bm{v}$, then $\mathcal{D}(\bm{v})= \big\{\bm{v}[1], \bm{v}[2], \ldots, \bm{v}[r]\big\}$, where $r=r(\bm{v})$.
    Let $i$ be the smallest index such that $d_H(\bm{v}[i],\bm{u})\leq 1$, then for $j\geq i+3$, by Lemma \ref{lem:del_position}, we can compute 
    \begin{align*}
        d_H(\bm{v}[j],\bm{u})\geq d_H(\bm{v}[i],\bm{v}[j])- d_H(\bm{v}[i],\bm{u})\geq j-i-1\geq 2.
    \end{align*}
    This implies that $\mathcal{F}\subseteq \{\bm{v}[i],\bm{v}[i+1],\bm{v}[i+2]\}$.
    When $|\mathcal{F}|=3$, it holds that $\mathcal{F}= \{\bm{v}[i],\bm{v}[i+1],\bm{v}[i+2]\}$.
    By Lemma \ref{lem:del_position}, we have $d_H(\bm{v}[i],\bm{v}[i+2])=2$ and $d_H(\bm{v}[i],\bm{v}[i+1])=d_H(\bm{v}[i+1],\bm{v}[i+2])=1$.
    Since $\bm{v}[i],\bm{v}[i+2]\in \mathcal{F}$, we have $d_H(\bm{v}[i],\bm{u})=d_H(\bm{v}[i+2],\bm{u})=1$. 
    Then, we get $\bm{u}, \bm{v}[i+1]\in \mathcal{S}(\bm{v}[i],\bm{v}[i+2])$.
    Now, by Lemma \ref{lem:sub}, we can conclude that $d_H(\bm{v}[i+1],\bm{u})\in \{0,2\}$.
    It then follows by $\bm{v}[i+1]\in \mathcal{F}$ that $\bm{u}=\bm{v}[i+1]$.
    This completes the proof.
\end{IEEEproof}

\section{Key Lemmas}\label{sec:key}

In this section, we provide several key lemmas, which serve to the construction of our reconstruction codes in the next section.

\begin{lemma}\label{lem:intersection}
    For any two distinct sequences \(\bm{x}, \bm{y} \in \Sigma^n\), let \(d\triangleq |\mathcal{D}(\bm{x}, \bm{y})|\) and \(s\triangleq |\mathcal{S}(\bm{x}, \bm{y})|\), it holds that $|\mathcal{B}(\bm{x},\bm{y})|\leq (d+s)n+O(1)$.
\end{lemma}

The intuition behind this is that the intersection \(\mathcal{B}(\bm{x}, \bm{y})\) can be decomposed into three (possibly overlapping) parts:  
\begin{equation*}
S \triangleq \bigcup_{\bm{z} \in \mathcal{D}(\bm{x}, \bm{y})} \mathcal{S}(\bm{z}), \quad  
D \triangleq \bigcup_{\bm{z} \in \mathcal{S}(\bm{x}, \bm{y})} \mathcal{D}(\bm{z}), \quad  
B \triangleq \mathcal{B}(\bm{x}, \bm{y}) \setminus (D \cup S).  
\end{equation*}
By Lemma \ref{lem:size}, it follows that \(|S| \leq d n\) and \(|D| \leq s n\). The subsequent lemmas show that \(|B| = O(1)\) for all possible values of \(d\) and \(s\). 
In addition, we compute \(|S \cap D|\) to establish a tighter bound on the intersection size \(|\mathcal{B}(\bm{x}, \bm{y})|\).  
By Lemmas \ref{lem:sub} and \ref{lem:del}, it follows that \((d,s) \in \{0,1,2\} \times \{0,2\}\), leading to six cases:  
\[  
(d, s) = (2,2), (1,2), (2,0), (0,2), (1,0), \text{ and } (0,0).  
\]  
These cases are addressed in the following lemmas, respectively, with proofs postponed until Section \ref{sec:proof}.  

\begin{lemma}[The Case of $(d,s)=(2,2)$]\label{lem:(2,2)}
    For \(n \geq 5\), let \(\bm{x}, \bm{y} \in \Sigma^n\) satisfy \(|\mathcal{D}(\bm{x}, \bm{y})| = 2\) and \(|\mathcal{S}(\bm{x}, \bm{y})| = 2\). Then we can express \(\bm{x} = \bm{a} \alpha \overline{\alpha} \bm{b}\) and \(\bm{y} = \bm{a} \overline{\alpha} \alpha \bm{b}\) for some \(\alpha \in \Sigma\).   
    The intersection cardinality satisfies  
    \( |\mathcal{B}(\bm{x}) \cap \mathcal{B}(\bm{y})| \leq 4n - 9 \),
    with equality holding if and only if either \(r(\bm{a}) = 0\) and \(r(\bm{b}) = n - 2\), or \(r(\bm{a}) = n - 2\) and \(r(\bm{b}) = 0\).  
\end{lemma}

\begin{lemma}[The Case of $(d,s)=(1,2)$]\label{lem:(1,2)}
    Let \(\bm{x}, \bm{y} \in \Sigma^n\) satisfy \(|\mathcal{D}(\bm{x}, \bm{y})| = 1\) and \(|\mathcal{S}(\bm{x}, \bm{y})| = 2\). Then, one of the following holds:
    \begin{itemize}  
        \item We can express \(\bm{x} = \bm{a} \alpha \bm{b}\) and \(\bm{y} = \bm{a} \overline{\alpha} \bm{b}\) for some \(\alpha \in \Sigma\). When $n\ge 4$, the intersection cardinality satisfies \( |\mathcal{B}(\bm{x},\bm{y})| \leq 3n - 5 \),
        with equality holding if and only if either \(r(\bm{a}) = 0\) and \(r(\bm{b}) = n - 1\), or \(r(\bm{a}) = n - 1\) and \(r(\bm{b}) = 0\).
        In addition, the intersection cardinality also satisfies 
        \( |\mathcal{B}(\bm{x},\bm{y})| \leq r(\bm{x})+r(\bm{y})+n-1\).    
        
        \item We can express \(\bm{x} = \bm{a} \bm{u} \bm{b}\) and \(\bm{y} = \bm{a} \bm{v} \bm{b}\), where \(\{\bm{u}, \bm{v}\} = \{\alpha^{\ell} \overline{\alpha}, \overline{\alpha} \alpha^{\ell}\}\) for some \(\alpha \in \Sigma\) and \(\ell \geq 2\). When $n\ge 6$, the intersection cardinality satisfies \(|\mathcal{B}(\bm{x},\bm{y})| \leq 3n - 7\) and \( |\mathcal{B}(\bm{x},\bm{y})| \leq r(\bm{x})+r(\bm{y})+ n- 2\).
    \end{itemize}
\end{lemma}

\begin{lemma}[The Case of $(d,s)=(2,0)$]\label{lem:(2,0)}    
    Let \(\bm{x}, \bm{y} \in \Sigma^n\) satisfy \(|\mathcal{D}(\bm{x}, \bm{y})| = 2\) and \(|\mathcal{S}(\bm{x}, \bm{y})| = 0\). Then, we can express \(\bm{x} = \bm{a} \bm{c} \bm{b}\) and \(\bm{y} = \bm{a} \overline{\bm{c}} \bm{b}\), where \(\bm{c}\) is an alternating sequence of length at least three. 
    The intersection cardinality satisfies \(|\mathcal{B}(\bm{x}) \cap \mathcal{B}(\bm{y})| \leq 2n + 8\).  
\end{lemma}

\begin{lemma}[The Case of $(d,s)=(0,2)$]\label{lem:(0,2)}
  Let \(\bm{x}, \bm{y} \in \Sigma^n\) satisfy \(|\mathcal{D}(\bm{x}, \bm{y})| = 0\) and \(|\mathcal{S}(\bm{x}, \bm{y})| = 2\). 
  Then, the intersection cardinality satisfies $|\mathcal{B}(\bm{x},\bm{y})|\leq 2n+4$ and $|\mathcal{B}(\bm{x},\bm{y})|\leq r(\bm{x})+r(\bm{y})+8$.
\end{lemma}

\begin{lemma}[The Case of $(d,s)=(1,0)$]\label{lem:(1,0)}
    Let \(\bm{x}, \bm{y} \in \Sigma^n\) satisfy \(|\mathcal{D}(\bm{x}, \bm{y})| = 1\) and \(|\mathcal{S}(\bm{x}, \bm{y})| = 0\). Then, the intersection cardinality satisfies $|\mathcal{B}(\bm{x},\bm{y})|\le n+20$.
\end{lemma}

\begin{definition}
    A sequence $\bm{z}\in \mathcal{B}(\bm{x},\bm{y})$ is called \emph{good} if there exist indices $i,\hat{i},j,\hat{j}$ such that $\bm{z}=\bm{x}(i,\hat{i})=\bm{y}(j,\hat{j})$ with the property that either $\hat{i}\not\in [i,j]\cup [j,i]$ or $\hat{j}\not\in [i,j]\cup [j,i]$. Otherwise, $\bm{z}$ is termed \emph{bad}.
\end{definition}

\begin{lemma}[The Case of $(d,s)=(0,0)$]\label{lem:(0,0)}
    Let \(\bm{x}, \bm{y} \in \Sigma^n\) satisfy \(|\mathcal{D}(\bm{x}, \bm{y})| = 0\) and \(|\mathcal{S}(\bm{x}, \bm{y})| = 0\).
    Then, the intersection cardinality satisfies $|\mathcal{B}(\bm{x},\bm{y})|\leq 30$.
    Moreover, the number of bad sequences in $\mathcal{B}(\bm{x},\bm{y})$ is at most $6$.
\end{lemma}

Now we are ready to prove Lemma \ref{lem:intersection}.

\begin{IEEEproof}[Proof of Lemma \ref{lem:intersection}]
    Firstly, by Lemmas \ref{lem:sub} and \ref{lem:del}, we have \((d,s) \in \{0,1,2\} \times \{0,2\}\).
    Then by Lemmas \ref{lem:(2,2)}, \ref{lem:(1,2)}, \ref{lem:(2,0)}, \ref{lem:(0,2)}, \ref{lem:(1,0)}, and \ref{lem:(0,0)}, we can conclude that $|\mathcal{B}(\bm{x},\bm{y})|\leq (d+s)n+O(1)$.
    This completes the proof.
\end{IEEEproof}

\section{Reconstruction Codes}\label{sec:constr}

In this section, we present constructions of \((n, N; \mathcal{B})\)-reconstruction codes for specific values of \(N\), namely \(4n-8\), \(3n-4\), \(2n+8\), \(n + 21\), and \(31\).  
Here, we assume \( n > 13 \) so that the inequalities \( 4n - 8 > 3n - 4 > 2n + 8 > n + 21 > 31 \) hold.  

\subsection{The Case of \texorpdfstring{$N=4n+O(1)$}{}}

The first construction is a redundancy-free \((n, 4n-8;\mathcal{B})\)-reconstruction code, as detailed below.  

\begin{theorem}\label{thm:4n}  
    The full space \(\Sigma^n\) serves as an \((n,4n-8;\mathcal{B})\)-reconstruction code.  
\end{theorem}

\begin{IEEEproof}
    For any two distinct sequences \(\bm{x}, \bm{y} \in \Sigma^n\), let \(d=|\mathcal{D}(\bm{x}, \bm{y})|\) and \(s=|\mathcal{S}(\bm{x}, \bm{y})|\).
    By Lemmas \ref{lem:sub} and \ref{lem:del}, we have that \((d,s) \in \{0,1,2\} \times \{0,2\}\).
    For each choice of $(d,s)$, by using the corresponding conclusion in Lemmas \ref{lem:(2,2)}, \ref{lem:(1,2)}, \ref{lem:(2,0)}, \ref{lem:(0,2)}, \ref{lem:(1,0)}, and \ref{lem:(0,0)}, we can conclude that $|\mathcal{B}(\bm{x},\bm{y})|\leq 4n - 9$.
    Then the conclusion follows immediately from the definition of a reconstruction code.  
\end{IEEEproof}

\subsection{The Case of \texorpdfstring{$N=3n+O(1)$}{}}

In this subsection, we consider the construction of $(n, 3n + O(1); \mathcal{B})$-reconstruction codes. By Lemma \ref{lem:intersection}, it suffices to impose a constraint such that for any two distinct sequences $\bm{x}, \bm{y} \in \Sigma^n$, it holds that $|\mathcal{D}(\bm{x},\bm{y})|+|\mathcal{S}(\bm{x},\bm{y})|\neq 4$. 
This can be achieved by requiring that the sequences have the same parity of their inversion number functions, as illustrated below.  

\begin{theorem}\label{thm:3n}
    For any $m\geq 2$ and $a \in [0, m-1]$, the code   
    \[
    \mathcal{C}_{3n-4} = \left\{\bm{x} \in \Sigma^n: \mathrm{Inv} (\boldsymbol{x}) \equiv a \pmod{m}\right\}  
    \]
    forms an $(n, 3n-4; \mathcal{B})$-reconstruction code. Moreover, when $m=2$, by the pigeonhole principle, there exists some choice of $a$ such that $\mathrm{red}(\mathcal{C}_{3n-4}) \leq 1$.  
\end{theorem}  

\begin{IEEEproof}  
    For any two distinct sequences $\bm{x}, \bm{y} \in \mathcal{C}_{3n-4}$, let $d = |\mathcal{D}(\bm{x}, \bm{y})|$ and $s = |\mathcal{S}(\bm{x}, \bm{y})|$. When $(d, s) \neq (2, 2)$, it follows from Lemmas \ref{lem:intersection} and \ref{lem:(1,2)}, \ref{lem:(2,0)}, \ref{lem:(0,2)}, \ref{lem:(1,0)}, and \ref{lem:(0,0)} that $|\mathcal{B}(\bm{x}, \bm{y})| \leq 3n - 5$.  
    Now, we demonstrate that the case where $(d, s) = (2, 2)$ is impossible. Assume $(d, s) = (2, 2)$, by Lemma \ref{lem:(2,2)}, we can write $\bm{x} = \bm{a} \alpha \overline{\alpha} \bm{b}$ and $\bm{y} = \bm{a} \overline{\alpha} \alpha \bm{b}$ for some $\alpha \in \Sigma$. We can then compute   
    \[
    \left| \mathrm{Inv} (\boldsymbol{x})-\mathrm{Inv} (\boldsymbol{y})\right| =  1,
    \]
    which contradicts the constraint imposed on $\bm{x}$ and $\bm{y}$.   
    Thus, $\mathcal{C}_{3n-4}$ is an $(n,3n-4;\mathcal{B})$-reconstruction code.
    Furthermore, when $m=2$, by the pigeonhole principle, there exists some $a$ such that $|\mathcal{C}_{3n-4}|\geq 2^{n-1}$, which implies that $\mathrm{red}(\mathcal{C}_{3n-4})\leq 1$.
    This completes the proof.
\end{IEEEproof}

\begin{remark}\label{rem:3n}
It can be readily verified that for any $m \geq 2$ and $a \in [0, m-1]$, the following codes:
\begin{align*}
   \mathcal{C}_{3n-4}^{'} &= \left\{\bm{x} \in \Sigma^n: \mathrm{VT}(\bm{x}) \equiv a \pmod{m}\right\},\\
   \mathcal{C}_{3n-4}^{''} &= \left\{\bm{x} \in \Sigma^n: \sum_{i=1}^{\lfloor n/2 \rfloor} x_{2i} \equiv a \pmod{m}\right\},
\end{align*}
can also forbid the pair of sequences $\bm{x} = \bm{a} \alpha \overline{\alpha} \bm{b}$ and $\bm{y} = \bm{a} \overline{\alpha} \alpha \bm{b}$. Hence, they are also $(n, 3n\!-\!4; \mathcal{B})$-reconstruction codes, each with a redundancy of at most one bit. Among them, we choose $\mathcal{C}_{3n-4}$ as the single-deletion reconstruction code defined in Lemma~\ref{lem:reconstruction}, which utilizes the inversion number function and serves as a core component in the subsequent construction of $(n, n\!+\!O(1); \mathcal{B})$-reconstruction codes.

\end{remark}

\subsection{The Case of \texorpdfstring{$N=2n+O(1)$}{}}

Similar to the previous subsection, to construct \((n, 2n + O(1); \mathcal{B})\)-reconstruction codes, Lemma \ref{lem:intersection} suggests imposing constraints such that for any two distinct sequences \(\bm{x}, \bm{y} \in \Sigma^n\), it holds that \(|\mathcal{D}(\bm{x},\bm{y})| + |\mathcal{S}(\bm{x},\bm{y})| \leq 2\).  
We first introduce two types of constraints to achieve this.  
\begin{itemize}  
    \item A natural approach is to use a single-deletion correcting code, ensuring that \(|\mathcal{D}(\bm{x},\bm{y})| = 0\) for any two distinct sequences \(\bm{x},\bm{y}\) in the code. Although effective, this method incurs a redundancy of at least \(\log n + O(1)\).  

    \item By analyzing the structures of \(\bm{x}\) and \(\bm{y}\) when \(|\mathcal{D}(\bm{x},\bm{y})| + |\mathcal{S}(\bm{x},\bm{y})| \geq 3\) (as characterized in Lemmas \ref{lem:(2,2)} and \ref{lem:(1,2)}), a more refined strategy involves imposing run-length limited constraints combined with a shifted version of single-deletion correcting code \cite{Schoeny-17-IT-BD} to forbid such structures. This approach reduces the redundancy to \(\log \log n + O(1)\).  
\end{itemize}  

In the following, we present a construction of \((n, 2n + O(1); \mathcal{B})\)-reconstruction code with only two bits of redundancy. This is achieved by restricting the number of runs in sequences belonging to \(\mathcal{C}_{3n-4}\) (defined in Theorem \ref{thm:3n}).  

\begin{lemma}\label{lem:run}
    Let $\mathcal{C}_r\triangleq \big\{ \bm{x}\in \Sigma^n: r(\bm{x})\leq \lceil n/2 \rceil\big\}$, then $|\C_r|\geq 2^{n-1}$.
\end{lemma}

\begin{IEEEproof}
    By \cite[Theorem 2]{Levenshtein-02-ISIT}, we have $|\mathcal{C}_r|= \sum_{i=0}^{\lceil n/2 \rceil-1}2\cdot \binom{n-1}{i}\geq 2^{n-1}$.
\end{IEEEproof}

\begin{theorem}\label{thm:2n}
    For any $m\geq 2$ and $a \in [0, m-1]$, the code   
    \[
    \mathcal{C}_{2n+9} = \left\{\bm{x} \in \mathcal{C}_r: \mathrm{Inv} (\boldsymbol{x}) \equiv a \pmod{m}\right\} 
    \]
    forms an $(n, 2n+9; \mathcal{B})$-reconstruction code. Moreover, when $m=2$, by the pigeonhole principle, there exists some choice of $a$ such that $r(\mathcal{C}_{2n+9}) \leq 2$.  
\end{theorem}  

\begin{IEEEproof}  
    For any two distinct sequences $\bm{x}, \bm{y} \in \mathcal{C}_{2n+9}$, let $d = |\mathcal{D}(\bm{x}, \bm{y})|$ and $s = |\mathcal{S}(\bm{x}, \bm{y})|$. 
    Firstly, we note that $\mathcal{C}_{2n+9}$ is a subcode of $\mathcal{C}_{3n-4}$, implying that $d+s\leq 3$.
    \begin{itemize}
        \item When $d+s\leq 2$, it follows by Lemmas  \ref{lem:(2,0)}, \ref{lem:(0,2)}, \ref{lem:(1,0)}, and \ref{lem:(0,0)} that $|\mathcal{B}(\bm{x}, \bm{y})| \leq 2n +8$.  
        \item When $d+s=3$, or equivalently $(d,s)=(1,2)$, by Lemma \ref{lem:(1,2)}, we have $|\mathcal{B}(\bm{x},\bm{y})|\leq r(\bm{x})+r(\bm{y})+n-1\leq 2n$.  
    \end{itemize}
    Hence, $\mathcal{C}_{2n+9}$ is an $(n,2n+9;\mathcal{B})$-reconstruction code.
    Furthermore, when $m=2$, by Lemma \ref{lem:run} and the pigeonhole principle, there exists some $a$ such that $|\mathcal{C}_{2n+9}|\geq 2^{n-2}$, which implies that $\mathrm{red}(\mathcal{C}_{2n+9})\leq 2$.
    This completes the proof.
\end{IEEEproof}

\subsection{The Case of \texorpdfstring{$N=n+O(1)$}{}}

This subsection presents a construction of \((n, n + O(1); \mathcal{B})\)-reconstruction code with \(\log \log n + O(1)\) bits of redundancy. This is accomplished by restricting the number of runs in sequences from the single-deletion reconstruction code defined in Lemma \ref{lem:reconstruction}, which forms a subcode of the \((n, 2n + 9; \mathcal{B})\)-reconstruction code introduced in Theorem \ref{thm:2n}.

\begin{theorem}\label{thm:n}
    Let $a_1 \in [0,1]$ and $a_2 \in [0,P/2]$, where $P\geq \log n+3$ is an even integer, the code
    \begin{align*}
        \mathcal{C}_{n+21} \triangleq \big\{ \boldsymbol{x}
        \in \mathcal{R}(n,P)\cap \mathcal{C}_r: \mathrm{wt}(\boldsymbol{x}) \equiv a_1 \pmod{2},~\mathrm{Inv} (\boldsymbol{x}) \equiv a_2 \pmod{1+P/2} \big\},
    \end{align*}
    forms an $(n,n+21;\mathcal{B})$-reconstruction code.
    Moreover, by the pigeonhole principle, there exists some choice of $a_1$ and $a_2$ such that $r(\mathcal{C}_{n+21}) \leq \log\log n+3$.  
\end{theorem}  

\begin{IEEEproof}  
    For any two distinct sequences $\bm{x}, \bm{y} \in \mathcal{C}_{n+21}$, let $d = |\mathcal{D}(\bm{x}, \bm{y})|$ and $s = |\mathcal{S}(\bm{x}, \bm{y})|$. 
    First, note that \(\mathcal{C}_{n+21}\) is a subcode of \(\mathcal{C}_{2n+9}\), for which Lemma \ref{lem:intersection} guarantees \(d + s \leq 2\).  
    Also, since \(\mathcal{C}_{n+21} \subseteq \mathcal{C}_P(a_1, a_2)\) (as defined in Lemma \ref{lem:reconstruction}), it follows that \(d \leq 1\).  
    \begin{itemize}
        \item When $d+s\leq 1$, it follows by Lemmas \ref{lem:(1,0)} and \ref{lem:(0,0)} that $|\mathcal{B}(\bm{x}, \bm{y})| \leq n +20$.  
        \item When $d+s=2$ and $d\leq 1$, or equivalently $(d,s)=(0,2)$, by Lemma \ref{lem:(2,0)}, we have $|\mathcal{B}(\bm{x},\bm{y})|\leq r(\bm{x})+r(\bm{y})+8\leq n+8$. 
    \end{itemize}
    Hence, \(\mathcal{C}_{n+21}\) is indeed an \((n, n+21; \mathcal{B})\)-reconstruction code.  
    Furthermore, by Remark \ref{rmk:runlength} and Lemma \ref{lem:run}, we have   
    \(  |\mathcal{R}(n,P) \cap \mathcal{C}_r| \geq 2^{n-2}\).
    Applying the pigeonhole principle, there exist \(a_1\) and \(a_2\) such that  
    \(|\mathcal{C}_{n+21}| \geq \frac{2^{n-2}}{2 \big( \frac{1}{2} \log \log n + 4 \big)}\),
    which implies  \(\mathrm{red}(\mathcal{C}_{n+21}) \leq \log \log n + 3\).
    This completes the proof.
\end{IEEEproof}

\subsection{The Case of \texorpdfstring{$N=O(1)$}{}}

In this subsection, we construct two families of \((n, O(1); \mathcal{B})\)-reconstruction codes.  
The first family is based on a single-edit correcting code introduced in Lemma \ref{lem:VT}.  

\begin{theorem}\label{thm:31}
    For any $a\in [0,2n-1]$, let $\mathrm{VT}_a(n)$ be defined in Lemma \ref{lem:VT}. Then, $\mathcal{C}_{31}\triangleq \mathrm{VT}_a(n)$ is an $(n,31;\mathcal{B})$-reconstruction code.
    Moreover, there exists some choice of $a$ such that $\mathrm{red}(\mathcal{C}_{31})\leq \log n+1$. 
\end{theorem}

\begin{IEEEproof}
    For any two distinct sequences $\bm{x},\bm{y}\in \mathcal{C}_{31}$, by Lemma \ref{lem:VT}, we have $\mathcal{D}(\bm{x},\bm{y})=\emptyset$ and $\mathcal{S}(\bm{x},\bm{y})=\emptyset$.
    Then by Lemma \ref{lem:(0,0)}, we get $|\mathcal{B}(\bm{x},\bm{y})|\leq 30$.
    Hence, $\mathcal{C}_{31}$ is an $(n,31;\mathcal{B})$-reconstruction code.
    Furthermore, by the pigeonhole principle, there exists some $a$ such that $|\mathcal{C}_{31}|\geq \frac{2^{n}}{2n}$, which implies that $\mathrm{red}(\mathcal{C}_{31})\leq \log n+1$.
    This completes the proof.
\end{IEEEproof}

The second family employs a single-deletion single-substitution list-decodable code with list size two, introduced in Lemma \ref{lem:DS_L}.  

\begin{theorem}\label{thm:7}
    For any $a_0\in [0,3]$, $a_1\in [0,2n-1]$, and $a_2\in [0,2n^2-1]$, let $\mathcal{C}_L$ be defined in Lemma \ref{lem:DS_L}. Then $\mathcal{C}_{7}\triangleq \mathcal{C}_L$ is an $(n,7;\mathcal{B})$-reconstruction code.
    Moreover, there exists some choice of $a$ such that $\mathrm{red}(\mathcal{C}_{7})\leq \log 3n+4$. 
\end{theorem}

\begin{IEEEproof}
    We first note that $\mathcal{C}_{7}$ is a subcode of the single-edit correcting code $\mathrm{VT}_a(n)$.
    For any two distinct sequences $\bm{x},\bm{y}\in \mathcal{C}_{7}$, by Lemma \ref{lem:VT}, we have $\mathcal{D}(\bm{x},\bm{y})=\emptyset$ and $\mathcal{S}(\bm{x},\bm{y})=\emptyset$.
    Now, by Lemma \ref{lem:DS_L}, we can conclude that each sequence in the intersection $\mathcal{B}(\bm{x},\bm{y})$ is bad. 
    Then by Lemma \ref{lem:(0,0)}, it holds that $|\mathcal{B}(\bm{x},\bm{y})|\leq 6$.
    Hence, $\mathcal{C}_{7}$ is an $(n,7;\mathcal{B})$-reconstruction code.
    Furthermore, by the pigeonhole principle, there exist some $a_0,a_1,a_2$ such that $|\mathcal{C}_{7}|\geq \frac{2^{n}}{4\cdot 2n \cdot 2n^2}$, which implies that $\mathrm{red}(\mathcal{C}_{7})\leq \log 3n+4$.
    This completes the proof.
\end{IEEEproof}

\section{Missing Proofs}\label{sec:proof}

In this section, we provide detailed proofs for Lemmas \ref{lem:(2,2)}, \ref{lem:(1,2)}, \ref{lem:(2,0)}, \ref{lem:(0,2)}, \ref{lem:(1,0)}, and \ref{lem:(0,0)}.
We first briefly introduce our proof approach.

For any $\bm{x},\bm{y}\in \Sigma^n$, we define the following sets
\begin{equation*}
S \triangleq \bigcup_{\bm{z} \in \mathcal{D}(\bm{x}, \bm{y})} \mathcal{S}(\bm{z}), \quad  
D \triangleq \bigcup_{\bm{z} \in \mathcal{S}(\bm{x}, \bm{y})} \mathcal{D}(\bm{z}), \quad  
B \triangleq \mathcal{B}(\bm{x}, \bm{y}) \setminus (D \cup S).  
\end{equation*}
It is clear that $D, S\subseteq \mathcal{B}(\bm{x},\bm{y})$.
Then by the inclusion-exclusion principle, we can compute
\begin{equation}\label{eq:B}
\begin{aligned}
 |\mathcal{B}(\bm{x},\bm{y})|
 &=|B|+|D\cup S|
 = |B|+|D|+|S|- |D\cap S|. 
\end{aligned}
\end{equation}
Finally, we finish the proof by determining the sizes of $B,D,S$ and $D\cap S$, respectively.

\subsection{Proof of Lemma \ref{lem:(2,2)}}\label{IV-A}

We first characterize the structures of $\bm{x}$ and $\bm{y}$ when $|\mathcal{D}(\bm{x},\bm{y})|=2$ and $|\mathcal{S}(\bm{x},\bm{y})|=2$.
Since $|\mathcal{D}(\bm{x},\bm{y})|=2$, by Lemma \ref{lem:del}, we can express $\bm{x}=\bm{a} \bm{c} \bm{b}$ and $\bm{y}=\bm{a} \overline{\bm{c}} \bm{b}$, where $\bm{c}$ is an alternating sequence of length at least two.
Moreover, since $|\mathcal{S}(\bm{x},\bm{y})|=2$, by Lemma \ref{lem:sub}, we can conclude that the length of $\bm{c}$ is exactly two.
As a result, we can write $\bm{x}=\bm{a}\alpha \overline{\alpha}\bm{b}$ and $\bm{y}=\bm{a} \overline{\alpha} \alpha \bm{b}$ for some $\alpha\in \Sigma$.
In this case, by Lemmas \ref{lem:sub} and \ref{lem:del'}, we can derive 
\begin{gather*}
    \mathcal{S}(\bm{x},\bm{y})= \{\bm{a}\alpha \alpha \bm{b}, \bm{a}\overline{\alpha} \overline{\alpha} \bm{b}\},\quad
    \mathcal{D}(\bm{x},\bm{y})= \{\bm{a}\alpha \bm{b}, \bm{a}\overline{\alpha} \bm{b}\}.
\end{gather*}
This implies that
\begin{equation}\label{eq:(2,2)}
\begin{gathered}
  D= \mathcal{D}(\bm{a}\alpha \alpha \bm{b}) \cup \mathcal{D}(\bm{a}\overline{\alpha} \overline{\alpha} \bm{b}),\quad
  S= \mathcal{S}(\bm{a}\overline{\alpha} \bm{b}) \cup \mathcal{S}(\bm{a}\alpha \bm{b}).
\end{gathered}
\end{equation}
In what follows, we will calculate the sizes of \(S\), \(D\), \(D \cap S\), and \(B \triangleq \mathcal{B}(\bm{x}, \bm{y}) \setminus (D \cup S)\) respectively, which will then allow us to determine the size of \(\mathcal{B}(\bm{x},\bm{y})\) based on Equation (\ref{eq:B}).

\begin{claim}\label{cla:(2,2)_S}
  Let $S$ be defined in Equation (\ref{eq:(2,2)}), we have $|S|= 2n-2$.
\end{claim}

\begin{IEEEproof}
By Lemmas \ref{lem:size} and \ref{lem:sub}, it follows that $|\mathcal{S}(\bm{a}\alpha \bm{b})|=|\mathcal{S}(\bm{a}\overline{\alpha} \bm{b})|=n$ and $\mathcal{S}(\bm{a}\overline{\alpha} \bm{b}) \cap \mathcal{S}(\bm{a}\alpha \bm{b} )=\{\bm{a}\alpha \bm{b},\bm{a}\overline{\alpha} \bm{b}\}$.
Then the conclusion follows by the inclusion-exclusion principle.
\end{IEEEproof}

\begin{claim}\label{cla:(2,2)_D}
 Let $D$ be defined in Equation (\ref{eq:(2,2)}), the following holds:
 \begin{itemize}
   \item if $r(\bm{a})=0$ or $r(\bm{b})=0$, we have $|D|= 2r(\bm{a})+2r(\bm{b})+1 \leq 2n-3$;
   \item if $r(\bm{a})\neq 0$ and $r(\bm{b})\neq 0$, we have $|D|= 2r(\bm{a})+2r(\bm{b})\leq 2n-4$. 
 \end{itemize}
\end{claim}	

\begin{IEEEproof}
By examining the $(|\bm{a}|+1)$-th entries of sequences in $\mathcal{D}(\bm{a} \alpha \alpha \bm{b})$ and $\mathcal{D}(\bm{a} \overline{\alpha} \overline{\alpha} \bm{b})$, we find that $\mathcal{D}(\bm{a} \alpha \alpha \bm{b})\cap \mathcal{D}(\bm{a} \overline{\alpha} \overline{\alpha} \bm{b})=\emptyset$. 
It follows that
\begin{align*}
   |D|
   &=|\mathcal{D}(\bm{a} \alpha \alpha \bm{b})|+|\mathcal{D}(\bm{a} \overline{\alpha} \overline{\alpha} \bm{b})|.  
\end{align*}
By examining the last run of \(\bm{a}\) and the first run of \(\bm{b}\), we can calculate the size of $D$, as detailed in Table \ref{tab:(2,2)_D}. Then the conclusion follows by the fact that $r(\bm{a})+r(\bm{b})\leq |\bm{a}|+ |\bm{b}|= n-2$.
\end{IEEEproof}

\begin{table*}
\caption{The values of $|\mathcal{D}(\bm{a} \alpha \alpha \bm{b})|= r(\bm{a} \alpha \alpha \bm{b})$ and $|\mathcal{D}(\bm{a} \overline{\alpha} \overline{\alpha} \bm{b})|= r(\bm{a} \overline{\alpha} \overline{\alpha} \bm{b})$. Recall that $n\geq 5$, we have $r(\bm{a})+r(\bm{b})\geq 1$. When we refer to the first or last run of a sequence as $\epsilon$, it indicates that its length is zero, implying that it does not contain any run.} 
\centering
\begin{tabular}{c|c|c|c|c} 
\hline 
\hline
the last run of $\bm{a}$ & the first run of $\bm{b}$ & $|\mathcal{D}(\bm{a} \alpha \alpha \bm{b})|$ & $|\mathcal{D}(\bm{a}\overline{\alpha}\overline{\alpha} \bm{b})|$ & $|D|$\\
\hline 
$\epsilon$ & $\alpha$ & $r(\bm{a})+r(\bm{b})$ & $r(\bm{a})+r(\bm{b})+1$ & $2r(\bm{a})+2r(\bm{b})+1$\\
\hline 
$\epsilon$ & $\overline{\alpha}$ & $r(\bm{a})+r(\bm{b})+1$ & $r(\bm{a})+r(\bm{b})$ & $2r(\bm{a})+2r(\bm{b})+1$\\
\hline 
$\alpha$ & $\alpha$ & $r(\bm{a})+r(\bm{b})-1$ & $r(\bm{a})+r(\bm{b})+1$ & $2r(\bm{a})+2r(\bm{b})$\\
\hline
$\alpha$ & $\overline{\alpha}$ & $r(\bm{a})+r(\bm{b})$ & $r(\bm{a})+r(\bm{b})$ & $2r(\bm{a})+2r(\bm{b})$ \\
\hline
$\alpha$ & $\epsilon$ & $r(\bm{a})+r(\bm{b})$ & $r(\bm{a})+r(\bm{b})+1$ & $2r(\bm{a})+2r(\bm{b})+1$ \\
\hline
$\overline{\alpha}$ & $\alpha$ & $r(\bm{a})+r(\bm{b})$ & $r(\bm{a})+r(\bm{b})$ & $2r(\bm{a})+2r(\bm{b})$\\
\hline
$\overline{\alpha}$ & $\overline{\alpha}$ & $r(\bm{a})+r(\bm{b})+1$ & $r(\bm{a})+r(\bm{b})-1$ & $2r(\bm{a})+2r(\bm{b})$\\
\hline 
$\overline{\alpha}$ & $\epsilon$ & $r(\bm{a})+r(\bm{b})+1$ & $r(\bm{a})+r(\bm{b})$ & $2r(\bm{a})+2r(\bm{b})+1$ \\
\hline
\hline
\end{tabular}
\label{tab:(2,2)_D}
\end{table*}

\begin{claim}\label{cla:(2,2)_DS}
Let $D$ and $S$ be defined in Equation (\ref{eq:(2,2)}), we have $3\leq |D\cap S|\leq 6$. Moreover, the following holds:
\begin{itemize}
\item if $r(\bm{a})= 0, r(\bm{b})= 1$ or $r(\bm{a})= 1, r(\bm{b})= 0$, we have $|D\cap S|=3$;
\item if $r(\bm{a})=0, r(\bm{b})\geq 2$ or $r(\bm{a})\geq 2, r(\bm{b})=0$ or $r(\bm{a})=r(\bm{b})=1$, we have $|D\cap S|=4$;
\item if $r(\bm{a})=1, r(\bm{b})\geq 2$ or $r(\bm{a})\geq 2, r(\bm{b})= 1$, we have $|D\cap S|=5$;
\item if $r(\bm{a}), r(\bm{b})\geq 2$, we have $|D\cap S|=6$.
\end{itemize}
\end{claim}
\begin{IEEEproof}
By examining the $(|\bm{a}|+1)$-th entries of sequences in $\mathcal{D}(\bm{a} \alpha \alpha \bm{b})$, $\mathcal{D}(\bm{a} \overline{\alpha} \overline{\alpha} \bm{b})$, $\mathcal{S}(\bm{a}\alpha \bm{b})$, and $\mathcal{S}(\bm{a}\overline{\alpha} \bm{b})$, we have
\begin{gather*}
  \mathcal{D}(\bm{a} \alpha \alpha \bm{b})\cap \mathcal{D}(\bm{a} \overline{\alpha} \overline{\alpha} \bm{b})=\emptyset,\\
  \mathcal{D}(\bm{a} \alpha \alpha \bm{b}) \cap \mathcal{S}(\bm{a}\overline{\alpha} \bm{b})= \{\bm{a} \alpha \bm{b}\} \subseteq \mathcal{D}(\bm{a} \alpha \alpha \bm{b}) \cap \mathcal{S}(\bm{a}\alpha \bm{b}), \\
  \mathcal{D}(\bm{a} \overline{\alpha} \overline{\alpha} \bm{b}) \cap \mathcal{S}(\bm{a}\alpha \bm{b})= \{\bm{a} \overline{\alpha} \bm{b}\} \subseteq\mathcal{D}(\bm{a} \overline{\alpha} \overline{\alpha} \bm{b})\cap \mathcal{S}(\bm{a}\overline{\alpha} \bm{b}).
\end{gather*} 
This implies that 
\begin{align*}
  D\cap S
  &= \big( \mathcal{D}(\bm{a} \alpha \alpha \bm{b}) \cap \mathcal{S}(\bm{a}\alpha \bm{b}) \big) \sqcup \big( \mathcal{D}(\bm{a} \overline{\alpha} \overline{\alpha} \bm{b}) \cap \mathcal{S}(\bm{a} \overline{\alpha} \bm{b}) \big).
\end{align*}
It then suffices to calculate the size of $\mathcal{D}(\bm{a} \alpha \alpha \bm{b}) \cap \mathcal{S}(\bm{a}\alpha \bm{b})$ and $\mathcal{D}(\bm{a} \overline{\alpha} \overline{\alpha} \bm{b}) \cap \mathcal{S}(\bm{a} \overline{\alpha} \bm{b})$, respectively.
 
We first consider the set $\mathcal{D}(\bm{a} \alpha \alpha \bm{b}) \cap \mathcal{S}(\bm{a}\alpha \bm{b})$ and will determine its size by considering the sequence $\bm{a}\alpha\alpha\bm{b}$. For any $\bm{z}\in \mathcal{D}(\bm{a} \alpha \alpha \bm{b}) \cap \mathcal{S}(\bm{a}\alpha \bm{b})$, it holds that $d_H(\bm{z},\bm{a}\alpha \bm{b})\leq 1$.
Then by Lemma \ref{lem:del_position}, there are at most three choices for $\bm{z}$. Specifically, 
\begin{itemize}
 \item $\bm{z}=\bm{a}\alpha\bm{b}$;
 \item $\bm{z}$ is obtained from $\bm{a}\alpha\alpha\bm{b}$ by deleting a symbol from the last $\overline{\alpha}$-run of $\bm{a}$ (if $\bm{a}$ has an $\overline{\alpha}$-run);
 \item $\bm{z}$ is obtained from $\bm{a}\alpha\alpha\bm{b}$ by deleting a symbol from the first $\overline{\alpha}$-run of $\bm{b}$ (if $\bm{b}$ has an $\overline{\alpha}$-run).
\end{itemize}
This implies that $1\leq |\mathcal{D}(\bm{a} \alpha \alpha \bm{b}) \cap \mathcal{S}(\bm{a}\alpha \bm{b})|\leq 3$.
Furthermore, by analyzing the last run of $\bm{a}$ and the first run of $\bm{b}$, we can determine the precise size of the set $\mathcal{D}(\bm{a} \alpha \alpha \bm{b}) \cap \mathcal{S}(\bm{a}\alpha \bm{b})$, as detailed in Table \ref{tab:(2,2)_DS}.

We now consider the set $\mathcal{D}(\bm{a} \overline{\alpha} \overline{\alpha} \bm{b}) \cap \mathcal{S}(\bm{a} \overline{\alpha} \bm{b})$ and will determine its size by considering the sequence $\bm{a}\overline{\alpha}\overline{\alpha}\bm{b}$. For any $\bm{z}\in \mathcal{D}(\bm{a} \overline{\alpha} \overline{\alpha} \bm{b}) \cap \mathcal{S}(\bm{a} \overline{\alpha} \bm{b})$, it holds that $d_H(\bm{z},\bm{a}\overline{\alpha} \bm{b})\leq 1$. 
Then by Lemma \ref{lem:del_position}, there are at most three choices for $\bm{z}$. Specifically, 
 \begin{itemize}
 \item $\bm{z}=\bm{a}\overline{\alpha} \bm{b}$;
 \item $\bm{z}$ is obtained from $\bm{a}\overline{\alpha}\overline{\alpha}\bm{b}$ by deleting a symbol from the last $\alpha$-run of $\bm{a}$ (if $\bm{a}$ has an $\alpha$-run);
 \item $\bm{z}$ is obtained from $\bm{a}\overline{\alpha}\overline{\alpha}\bm{b}$ by deleting a symbol from the first $\alpha$-run of $\bm{b}$ (if $\bm{b}$ has an $\alpha$-run).
 \end{itemize}
This implies that $1\leq |\mathcal{D}(\bm{a} \overline{\alpha} \overline{\alpha} \bm{b}) \cap \mathcal{S}(\bm{a} \overline{\alpha} \bm{b})|\leq 3$.
Furthermore, by analyzing the last run of $\bm{a}$ and the first run of $\bm{b}$, we can determine the precise size of the set $\mathcal{D}(\bm{a} \overline{\alpha} \overline{\alpha} \bm{b}) \cap \mathcal{S}(\bm{a} \overline{\alpha} \bm{b})$, as detailed in Table \ref{tab:(2,2)_DS}.
Then by Table \ref{tab:(2,2)_DS}, the conclusion follows.
\end{IEEEproof}

\begin{table*}
\caption{The values of $|\mathcal{D}(\bm{a} \alpha \alpha \bm{b}) \cap \mathcal{S}(\bm{a}\alpha \bm{b})|$, $|\mathcal{D}(\bm{a}\alpha \bm{b}) \cap \mathcal{S}(\bm{a}\bm{b})|$, $|\mathcal{D}(\bm{a} \overline{\alpha} \overline{\alpha} \bm{b}) \cap \mathcal{S}(\bm{a} \overline{\alpha} \bm{b})|$, and $|\mathcal{D}(\bm{a} \overline{\alpha} \bm{b}) \cap \mathcal{S}(\bm{a} \bm{b})|$.
Let $f\triangleq |\mathcal{D}(\bm{a} \alpha \alpha \bm{b}) \cap \mathcal{S}(\bm{a}\alpha \bm{b})|+ |\mathcal{D}(\bm{a} \overline{\alpha} \overline{\alpha} \bm{b}) \cap \mathcal{S}(\bm{a} \overline{\alpha} \bm{b})|$ and $g\triangleq |\mathcal{D}(\bm{a}\alpha \bm{b}) \cap \mathcal{S}(\bm{a}\bm{b})|+ |\mathcal{D}(\bm{a} \overline{\alpha} \bm{b}) \cap \mathcal{S}(\bm{a} \bm{b})|-1$.
Recall that $n\geq 5$, we have $r(\bm{a})+r(\bm{b})\geq 1$. } 
\centering
\begin{tabular}{c|c|c|c|c|c} 
\hline 
\hline
\multirow{2}{*}{$(r(\bm{a}), r(\bm{b}))$} & \multirow{2}{*}{runs in $\bm{a}$ and $\bm{b}$} & $|\mathcal{D}(\bm{a} \alpha \alpha \bm{b}) \cap \mathcal{S}(\bm{a}\alpha \bm{b})|$ & $|\mathcal{D}(\bm{a} \overline{\alpha} \overline{\alpha} \bm{b}) \cap \mathcal{S}(\bm{a} \overline{\alpha} \bm{b})|$ & \multirow{2}{*}{$f$} & \multirow{2}{*}{$g$}\\
& & or $|\mathcal{D}(\bm{a} \alpha \bm{b}) \cap \mathcal{S}(\bm{a} \bm{b})|$ & or $|\mathcal{D}(\bm{a} \overline{\alpha} \bm{b}) \cap \mathcal{S}(\bm{a} \bm{b})|$ & & \\
\hline 
\multirow{2}{*}{$(0,1)$} & $\bm{b}$ has an $\alpha$-run & $1$ & $2$ & $3$ & $2$\\
\cline{2-6}
& $\bm{b}$ has an $\overline{\alpha}$-run & $2$ & $1$ & $3$ & $2$\\
\hline 
$(0,\geq 2)$ & - & $2$ & $2$ & $4$ & $3$\\
\hline 
\multirow{2}{*}{$(1,0)$} & $\bm{a}$ has an $\alpha$-run & $1$ & $2$ & $3$ & $2$\\
\cline{2-6}
& $\bm{a}$ has an $\overline{\alpha}$-run & $2$ & $1$ & $3$ & $2$\\
\hline
\multirow{4}{*}{$(1,1)$} & $\bm{a}$ has an $\alpha$-run, $\bm{b}$ has an $\alpha$-run & $1$ & $3$ & $4$ & $3$\\
\cline{2-6}
& $\bm{a}$ has an $\alpha$-run, $\bm{b}$ has an $\overline{\alpha}$-run & $2$ & $2$ & $4$ & $3$\\
\cline{2-6}
& $\bm{a}$ has an $\overline{\alpha}$-run, $\bm{b}$ has an $\alpha$-run & $2$ & $2$ & $4$ & $3$\\
\cline{2-6}
& $\bm{a}$ has an $\overline{\alpha}$-run, $\bm{b}$ has an $\overline{\alpha}$-run & $3$ & $1$ & $4$ & $3$\\
\hline
\multirow{2}{*}{$(1,\geq 2)$} & $\bm{a}$ has an $\alpha$-run & $2$ & $3$ & $5$ & $4$\\
\cline{2-6}
& $\bm{a}$ has an $\overline{\alpha}$-run & $3$ & $2$ & $5$ & $4$\\
\hline
$(\geq 2,0)$ & - & $2$ & $2$ & $4$ & $3$\\
\hline
\multirow{2}{*}{$(\geq 2,1)$} & $\bm{b}$ has an $\alpha$-run & $2$ & $3$ & $5$ & $4$\\
\cline{2-6}
& $\bm{b}$ has an $\overline{\alpha}$-run & $3$ & $2$ & $5$ & $4$\\
\hline 
$(\geq 2,\geq 2)$ & - & $3$ & $3$ & $6$ & $5$\\
\hline
\hline
\end{tabular}
\label{tab:(2,2)_DS}
\end{table*}

\begin{claim}\label{cla:(2,2)_B}
Let $D,S$ be defined in Equation (\ref{eq:(2,2)}), we have $|B|= |\mathcal{B}(\bm{x}, \bm{y}) \setminus (D \cup S)|\leq 2$. Moreover, the following holds:
\begin{itemize}
  \item if $r(\bm{a})=0$, or $r(\bm{b})=0$, or $r(\bm{a})=r(\bm{b})=1$ with distinct runs in \(\bm{a}\) and \(\bm{b}\), then $|B|=0$;
  \item if $r(\bm{a})=1, r(\bm{b})\geq 2$, or $r(\bm{a})\geq 2, r(\bm{b})= 1$, or $r(\bm{a})=r(\bm{b})= 1$ with the same run in $\bm{a}$ and $\bm{b}$, then $|B|=1$;
  \item if $r(\bm{a})\geq 2, r(\bm{b})\geq 2$, then $|B|=2$.
\end{itemize}
\end{claim}

\begin{IEEEproof}
Recall that $\bm{x}= \bm{a} \alpha \overline{\alpha} \bm{b}$ and $\bm{y}= \bm{a} \overline{\alpha} \alpha \bm{b}$.
Let $\bm{x}_i\triangleq \bm{x}_{[n]\setminus \{i\}}$ and $\bm{y}_i\triangleq \bm{y}_{[n]\setminus \{i\}}$ for $i\in [n]$, we have
\begin{align*}
    \mathcal{B}(\bm{x},\bm{y})
    &= \bigcup_{i=1}^{n} \bigcup_{j=1}^{n} \mathcal{S}(\bm{x}_i, \bm{y}_j).
\end{align*}
We first determine the intersection $\mathcal{S}(\bm{x}_i, \bm{y}_j)$  and distinguish between the following six cases according to the values of $i$ and $j$.
Let $m\triangleq |\bm{a}|$.
\begin{itemize}
  \item If $i=m+1$ or $j=m+2$, we have 
      \begin{align*}
        \mathcal{S}(\bm{x}_i, \bm{y}_j)\subseteq \mathcal{S}(\bm{a}\overline{\alpha} \bm{b}) \subset S.
      \end{align*}
  \item If $i=m+2$ or $j=m+1$, we have 
      \begin{align*}
        \mathcal{S}(\bm{x}_i, \bm{y}_j)\subseteq \mathcal{S}(\bm{a}\alpha \bm{b}) \subset S.
      \end{align*}
  \item If $i,j \leq m$, we have $\bm{x}_i= \bm{a}_i \alpha \overline{\alpha} \bm{b}$ and $\bm{y}_j= \bm{a}_j \overline{\alpha} \alpha \bm{b}$. By examining the $m$-th entries of sequences in the intersection $\mathcal{S}(\bm{x}_i, \bm{y}_j)$, we can conclude that
        \begin{align*}
        \mathcal{S}(\bm{x}_i, \bm{y}_j)
        \subseteq \{\bm{a}_i \alpha \alpha \bm{b}, \bm{a}_j \overline{\alpha} \overline{\alpha} \bm{b}\}
        \subseteq \mathcal{D}(\bm{a}\alpha \alpha \bm{b}) \cup \mathcal{D}(\bm{a}\overline{\alpha} \overline{\alpha} \bm{b})= D.
      \end{align*}
      
  \item If $i,j \geq m+3$, we have $\bm{x}_i= \bm{a} \alpha \overline{\alpha} \bm{b}_{i-m-2}$ and $\bm{y}_j= \bm{a} \overline{\alpha} \alpha \bm{b}_{j-m-2}$. By examining the $(m+1)$-th entries of sequences in the intersection $\mathcal{S}(\bm{x}_i, \bm{y}_j)$, we can conclude that
      \begin{align*}
        \mathcal{S}(\bm{x}_i, \bm{y}_j)
        \subseteq \{\bm{a} \alpha \alpha \bm{b}_{i-m-2}, \bm{a} \overline{\alpha} \overline{\alpha} \bm{b}_{j-m-2}\}
        \subseteq \mathcal{D}(\bm{a}\alpha \alpha \bm{b}) \cup \mathcal{D}(\bm{a}\overline{\alpha} \overline{\alpha} \bm{b})= D.
      \end{align*}
  \item If $i\leq m$ and $j\geq m+3$, we can assume that $\bm{x}_i\neq \bm{x}_{m+1}$ and $\bm{y}_j\neq \bm{y}_{m+2}$, i.e., $\bm{a}_i \alpha \neq \bm{a}$ and $\bm{b}\neq \alpha \bm{b}_{j-m-2}$, since the other scenarios have been considered in the first case.
      As a result, we assume that both $\bm{a}$ and $\bm{b}$ have an $\overline{\alpha}$-run.
      Observe that $\bm{x}_i= \bm{a}_i \alpha \overline{\alpha} \bm{b}$ and $\bm{y}_j= \bm{a} \overline{\alpha} \alpha \bm{b}_{j-m-2}$, we can compute $d_H(\bm{x}_i,\bm{y}_j)= d_H(\bm{a}_i \alpha , \bm{a})+ d_H(\bm{b}, \alpha \bm{b}_{j-m-2})\geq 2$.
      It then suffices to consider the scenario where $d_H(\bm{x}_i,\bm{y}_j)=2$, or equivalently $d_H(\bm{a}_i \alpha, \bm{a})= d_H(\bm{b}, \alpha \bm{b}_{j-m-2})=1$, as in the other cases, the intersection $\mathcal{S}(\bm{x}_i, \bm{y}_j)$ is empty.
      In this case, by Lemma \ref{lem:sub} we have
      \begin{align*}
        \mathcal{S}(\bm{x}_i, \bm{y}_j)
        =\{\bm{a}_i \alpha \overline{\alpha} \alpha \bm{b}_{j-m-2}, \bm{a} \overline{\alpha} \bm{b} \}.
      \end{align*}
      We now consider the sequences $\bm{a} \alpha$ and $\alpha \bm{b}$.
      Since $d_H(\bm{a}_i \alpha, \bm{a})= d_H(\bm{b}, \alpha \bm{b}_{j-m-2})=1$, it follows by Lemma \ref{lem:del_position} that there is exactly one choice for $\bm{a}_i\alpha$ and $\alpha \bm{b}_{j-m-2}$, namely, $\bm{a}'\triangleq \bm{a}_i \alpha\neq \bm{a}$ is obtained from $\bm{a}\alpha$ by deleting a symbol of its last $\overline{\alpha}$-run and $\bm{b}'\triangleq \alpha\bm{b}_{j-m-2}\neq \bm{b}$ is obtained from $\alpha\bm{b}$ by deleting a symbol of its first $\overline{\alpha}$-run.
      Since $\bm{a}' \neq \bm{a}$ and $\bm{b}\neq \bm{b}'$, it can be easily checked that $\bm{a}' \overline{\alpha} \bm{b}'\notin \mathcal{S}(\bm{a}\overline{\alpha} \bm{b}) \cup \mathcal{S}(\bm{a}\alpha \bm{b}) \cup \mathcal{D}(\bm{a}\alpha \alpha \bm{b}) \cup \mathcal{D}(\bm{a}\overline{\alpha} \overline{\alpha} \bm{b})= D \cup S$.
      Consequently, we obtain
    \begin{align*}
        \left( \bigcup_{i=1}^{m}\bigcup_{j=m+3}^n \mathcal{S}(\bm{x}_i, \bm{y}_j) \right) \setminus (D\cup S)
        =\{ \bm{a}' \overline{\alpha} \bm{b}' \}.
    \end{align*}
  \item If $i\geq m+3$ and $j\leq m$, 
      by considering the reversal of $\bm{x}$ and $\bm{y}$ and using the conclusion of the previous case, we can conclude that 
      \begin{align*}
        \left( \bigcup_{i=1}^{m}\bigcup_{j=m+3}^n \mathcal{S}(\bm{x}_i, \bm{y}_j) \right) \setminus (D\cup S)
        =\{ \bm{a}'' \alpha \bm{b}''\},
    \end{align*}
    where $\bm{a}''\neq \bm{a}$ is obtained from $\bm{a}\overline{\alpha}$ by deleting a symbol of its last $\alpha$-run and $\bm{b}''\neq \bm{b}$ is obtained from $\overline{\alpha}\bm{b}$ by deleting a symbol of its first $\alpha$-run.
\end{itemize}
Finally, we prove the conclusion by analyzing the runs in $\bm{a}$ and the runs in $\bm{b}$, as discussed as follows:
\begin{itemize}
  \item if $r(\bm{a})=0$, or $r(\bm{b})=0$, or $r(\bm{a})=r(\bm{b})=1$ with distinct runs in \(\bm{a}\) and \(\bm{b}\), then both $\bm{a}' \overline{\alpha} \bm{b}'$ and $\bm{a}'' \alpha \bm{b}''$ do not exist, implying that $|B|=0$;
  \item if $r(\bm{a})=1, r(\bm{b})\geq 2$, or $r(\bm{a})\geq 2, r(\bm{b})= 1$, or $r(\bm{a})=r(\bm{b})= 1$  with the same runs in $\bm{a}$ and $\bm{b}$, then exactly one of $\bm{a}' \overline{\alpha} \bm{b}'$ and $\bm{a}'' \alpha \bm{b}''$ exists, implying that $|B|=1$;
  \item if $r(\bm{a})\geq 2, r(\bm{b})\geq 2$, then both $\bm{a}' \overline{\alpha} \bm{b}'$ and $\bm{a}'' \alpha \bm{b}''$ exist, implying that $|B|=2$.
\end{itemize}
\end{IEEEproof}

Combining Equation (\ref{eq:B}) with Claims \ref{cla:(2,2)_S}, \ref{cla:(2,2)_D}, \ref{cla:(2,2)_DS}, and \ref{cla:(2,2)_B}, we obtain the following: 
\begin{itemize}
  \item if $r(\bm{a})\le 1, r(\bm{b})\le 1$, then $|S|=2n-2$, $|D|\le  2r(\bm{a})+2r(\bm{b})+1\le 5$, $|D\cap S|\ge 3$, and $|B|\le 1$, implying that $|\mathcal{B}(\bm{x},\bm{y})|= |B|+|D|+|S|- |D\cap S|\le  2n+1$;
  \item if $r(\bm{a})=0, r(\bm{b})\geq 2$, or $r(\bm{a})\geq 2, r(\bm{b})=0$, then $|S|=2n-2$, $|D|= 2r(\bm{a})+2r(\bm{b})+1\leq 2n-3$, $|D\cap S|=4$, and $|B|=0$, implying that $|\mathcal{B}(\bm{x},\bm{y})|= |B|+|D|+|S|- |D\cap S|\leq 4n-9$ and $|\mathcal{B}(\bm{x},\bm{y})|= 4n-9$ holds if and only if $r(\bm{a})+r(\bm{b})=n-2$; 
  \item if $r(\bm{a})=1, r(\bm{b})\geq 2$, or $r(\bm{a})\geq 2, r(\bm{b})= 1$, then $|S|=2n-2$, $|D|= 2r(\bm{a})+2r(\bm{b})\leq 2n-4$, $|D\cap S|=5$, and $|B|\leq 1$, implying that $|\mathcal{B}(\bm{x},\bm{y})|= |B|+|D|+|S|- |D\cap S| \leq 4n-10$;
  \item if $r(\bm{a})\geq 2, r(\bm{b})\geq 2$, then $|S|=2n-2$, $|D|= 2r(\bm{a})+2r(\bm{b})\leq 2n-4$, $|D\cap S|=6$, and $|B|=2$, implying that $|\mathcal{B}(\bm{x},\bm{y})|= |B|+|D|+|S|- |D\cap S| \leq 4n-10$.
\end{itemize}
In summary, when $n\ge 5$, we have $|\mathcal{B}(\bm{x},\bm{y})|\leq 4n-9$ with equality holding if and only if either $r(\bm{a})=0, r(\bm{b})=n-2$ or $r(\bm{a})=n-2, r(\bm{b})=0$.
This completes the proof of Lemma \ref{lem:(2,2)}.

\subsection{Proof of Lemma \ref{lem:(1,2)}}\label{subsec:(1,2)}

The proof follows a strategy similar to that of Lemma \ref{lem:(2,2)}, we include
it here for completeness.
We first characterize the structures of $\bm{x}$ and $\bm{y}$ when $|\mathcal{D}(\bm{x},\bm{y})|=1$ and $|\mathcal{S}(\bm{x},\bm{y})|=2$.
Since $|\mathcal{S}(\bm{x},\bm{y})|=2$, by Lemma \ref{lem:sub}, we have $d_H(\bm{x},\bm{y})\leq 2$. 
\begin{itemize}
    \item If $d_H(\bm{x},\bm{y})=1$, by Lemma \ref{lem:del}, we can express $\bm{x}=\bm{a}\alpha \bm{b}$ and $\bm{y}=\bm{a}\overline{\alpha}\bm{b}$ for some $\alpha \in \{0,1\}$.
    \item If $d_H(\bm{x},\bm{y})=2$, by Lemma \ref{lem:del}, we can express $\{\bm{x}, \bm{y}\}= \big\{\bm{a}\overline{\alpha}\alpha  \bm{c}\beta \bm{b}, \bm{a}\alpha \bm{c}\beta  \overline{\beta} \bm{b}\big\}$ with $\overline{\alpha}\alpha  \bm{c}\neq \bm{c}\beta  \overline{\beta}$ for some $\alpha, \beta\in \Sigma$ and $\bm{c}\in \Sigma^{\ast}$.
    Since $d_H(\bm{x},\bm{y})=2$, we have $\alpha\bm{c}= \bm{c}\beta$.
    Consider the sequence $\alpha \bm{c} \beta$, by Lemma \ref{lem:del_position}, we can conclude that $r(\alpha \bm{c} \beta)=1$.
    This implies that $\{\bm{x}, \bm{y}\}= \big\{\bm{a}\alpha^{\ell} \overline{\alpha} \bm{b}, \bm{a}\overline{\alpha} \alpha^{\ell} \bm{b}\big\}$ for some $\alpha\in \Sigma$ and $\ell\geq 2$.
\end{itemize}

We discuss these two cases separately, and obtain the following conclusions.

\begin{lemma}\label{lem:(1,2)'}
    Let $\bm{x}=\bm{a}\alpha\bm{b}, \bm{y}=\bm{a}\overline{\alpha}\bm{b}\in\Sigma^n$, where $\alpha\in \Sigma$, for $n\ge 4$, it holds that $|\mathcal{B}(\bm{x}) \cap \mathcal{B}(\bm{y})|\leq 3n-5$, with equality holding if and only if $r(\bm{a})=0, r(\bm{b})=n-1$, or $r(\bm{a})=n-1, r(\bm{b})=0$, and that \( |\mathcal{B}(\bm{x},\bm{y})| \leq r(\bm{x})+r(\bm{y})+n-1\).
\end{lemma}

\begin{lemma}\label{lem:(1,2)''}
    Let $\bm{x}= \bm{a}\alpha^{\ell} \overline{\alpha} \bm{b}, \bm{y}= \bm{a}\overline{\alpha}\alpha^{\ell}  \bm{b}\in \Sigma^n$, where $\alpha\in \Sigma$ and $\ell\geq 2$, for $n\ge 6$, it holds that $|\mathcal{B}(\bm{x}) \cap \mathcal{B}(\bm{y})|\leq 3n-7$ and \( |\mathcal{B}(\bm{x},\bm{y})| \leq r(\bm{x})+r(\bm{y})+n-2\).
\end{lemma}

By Lemmas \ref{lem:(1,2)'} and \ref{lem:(1,2)''}, the conclusion of Lemma \ref{lem:(1,2)} follows.
In the following, we prove Lemmas \ref{lem:(1,2)'} and \ref{lem:(1,2)''}, respectively.

\smallskip

\subsubsection{Proof of Lemma \ref{lem:(1,2)'}}

Firstly, by Lemmas \ref{lem:sub} and \ref{lem:del'}, we can derive 
\begin{gather*}
    \mathcal{S}(\bm{x},\bm{y})= \{\bm{a} \alpha \bm{b}, \bm{a} \overline{\alpha} \bm{b}\},\quad
    \mathcal{D}(\bm{x},\bm{y})= \{\bm{a} \bm{b}\}.
\end{gather*}
This implies that
\begin{equation}\label{eq:(1,2)}
\begin{gathered}
  D= \mathcal{D}(\bm{a}\alpha  \bm{b}) \cup \mathcal{D}(\bm{a} \overline{\alpha} \bm{b}),\quad
  S= \mathcal{S}(\bm{a}\bm{b}).
\end{gathered}
\end{equation}
In what follows, we will calculate the sizes of \(S\), \(D\), \(D \cap S\), and \(B \triangleq \mathcal{B}(\bm{x}, \bm{y}) \setminus (D \cup S)\) respectively, which will then allow us to determine the size of \(\mathcal{B}(\bm{x},\bm{y})\) based on Equation (\ref{eq:B}).

\begin{claim}\label{cla:(1,2)_S}
 Let $S$ be defined in Equation (\ref{eq:(1,2)}), then $|S|=n$.
\end{claim}

\begin{IEEEproof}
    The conclusion follows by Lemma \ref{lem:size} directly.
\end{IEEEproof}

\begin{claim}\label{cla:(1,2)_D}
 Let $D$ be defined in Equation (\ref{eq:(1,2)}), we have $|D|=r(\bm{x})+r(\bm{y})-1$. Moreover, the following holds:
 \begin{itemize}
   \item if $r(\bm{a})=0$ or $r(\bm{b})=0$, we have $|D|= 2r(\bm{a})+2r(\bm{b}) \leq 2n-2$;
   \item if $r(\bm{a})\neq 0$ and $r(\bm{b})\neq 0$, we have $|D|= 2r(\bm{a})+2r(\bm{b})-1\leq 2n-3$. 
 \end{itemize}   
\end{claim}

\begin{IEEEproof}
  By Lemma \ref{lem:del}, we have $\mathcal{D}(\bm{a}\alpha  \bm{b}) \cap \mathcal{D}(\bm{a} \overline{\alpha} \bm{b})=\{\bm{a}\bm{b}\}$. It follows that
\begin{align*}
   |D|
   &=|\mathcal{D}(\bm{a} \alpha \bm{b})|+|\mathcal{D}(\bm{a} \overline{\alpha} \bm{b})|-1
   =r(\bm{a} \alpha \bm{b})+r(\bm{a} \overline{\alpha} \bm{b})-1
   =r(\bm{x})+r(\bm{y})-1.
\end{align*}
Observe that $r(\bm{a} \alpha \bm{b})= r(\bm{a} \alpha \alpha \bm{b})$, $r(\bm{a}\overline{\alpha} \bm{b})= r(\bm{a} \overline{\alpha}\overline{\alpha} \bm{b})$, and $r(\bm{a})+r(\bm{b})\leq |\bm{a}|+ |\bm{b}|= n-1$.
Then by Table \ref{tab:(2,2)_D}, the conclusion follows. 
\end{IEEEproof}

\begin{claim}\label{cla:(1,2)_DS}
Let $D$ and $S$ be defined in Equation (\ref{eq:(1,2)}), we have $2\leq |D\cap S|\leq 5$. Moreover, the following holds:
\begin{itemize}
\item if $r(\bm{a})=0, r(\bm{b})=1$ or $r(\bm{b})=0, r(\bm{a})=1$, then $|D\cap S|=2$;
\item if $r(\bm{a})=0, r(\bm{b})\ge 2$ or $r(\bm{b})=0, r(\bm{a})\ge 2$ or $r(\bm{a})=r(\bm{b})=1$, then $|D\cap S|=3$;
\item if $r(\bm{a})=1, r(\bm{b})\ge 2$ or $r(\bm{b})=1, r(\bm{a})\ge 2$, then $|D\cap S|=4$;
\item if $r(\bm{a}), r(\bm{b})\ge 2$, $|D\cap S|=5$.
\end{itemize}   
\end{claim}

\begin{IEEEproof}
By Lemma \ref{lem:del}, we have $\mathcal{D}(\bm{a}\alpha  \bm{b}) \cap \mathcal{D}(\bm{a} \overline{\alpha} \bm{b})=\{\bm{a}\bm{b}\}$. Observe that $\bm{a}\bm{b}\in \mathcal{S}(\bm{a}  \bm{b})$.
Then by the inclusion-exclusion principle, we can compute
\begin{align*}
|D\cap S|=\big|\mathcal{D}(\bm{a} \alpha \bm{b})\cap \mathcal{S}(\bm{a}  \bm{b})\big|+ \big|\mathcal{D}(\bm{a}\overline{\alpha} \bm{b})\cap \mathcal{S}(\bm{a}  \bm{b}) \big|-1.  
\end{align*}

We first consider the set $\mathcal{D}(\bm{a} \alpha \bm{b}) \cap \mathcal{S}(\bm{a} \bm{b})$ and will determine its size by considering the sequence $\bm{a}\alpha\bm{b}$. For any $\bm{z}\in \mathcal{D}(\bm{a} \alpha  \bm{b}) \cap \mathcal{S}(\bm{a} \bm{b})$, it holds that $d_H(\bm{z},\bm{ab})\leq 1$. 
Then by Lemma \ref{lem:del_position}, there are at most three choices for $\bm{z}$. Specifically, 
\begin{itemize}
 \item $\bm{z}=\bm{a}\bm{b}$;
 \item $\bm{z}$ is obtained from $\bm{a}\alpha\bm{b}$ by deleting a symbol from the last $\overline{\alpha}$-run of $\bm{a}$ (if $\bm{a}$ has an $\overline{\alpha}$-run);
 \item $\bm{z}$ is obtained from $\bm{a}\alpha\bm{b}$ by deleting a symbol from the first $\overline{\alpha}$-run of $\bm{b}$ (if $\bm{b}$ has an $\overline{\alpha}$-run).
\end{itemize}
This implies that $1\leq |\mathcal{D}(\bm{a} \alpha \bm{b}) \cap \mathcal{S}(\bm{a}\bm{b})|\leq 3$.
Furthermore, by analyzing the last run of $\bm{a}$ and the first run of $\bm{b}$, we can determine the precise size of the set $\mathcal{D}(\bm{a} \alpha  \bm{b}) \cap \mathcal{S}(\bm{a} \bm{b})$, as detailed in Table \ref{tab:(2,2)_DS}.

We now consider the set $\mathcal{D}(\bm{a}  \overline{\alpha} \bm{b}) \cap \mathcal{S}(\bm{a} \bm{b})$ and will determine its size by considering the sequence $\bm{a}\overline{\alpha}\bm{b}$. For any $\bm{z}\in \mathcal{D}(\bm{a}  \overline{\alpha} \bm{b}) \cap \mathcal{S}(\bm{a}  \bm{b})$, it holds that $d_H(\bm{z},\bm{ab})\leq 1$. 
Then by Lemma \ref{lem:del_position}, there are at most three choices for $\bm{z}$. Specifically, 
 \begin{itemize}
 \item $\bm{z}=\bm{a}\bm{b}$;
 \item $\bm{z}$ is obtained from $\bm{a}\overline{\alpha}\overline{\alpha}\bm{b}$ by deleting a symbol from the last $\alpha$-run of $\bm{a}$ (if $\bm{a}$ has an $\alpha$-run);
 \item $\bm{z}$ is obtained from $\bm{a}\overline{\alpha}\bm{b}$ by deleting a symbol from the first $\alpha$-run of $\bm{b}$ (if $\bm{b}$ has an $\alpha$-run).
 \end{itemize}
This implies that $1\leq |\mathcal{D}(\bm{a}  \overline{\alpha} \bm{b}) \cap \mathcal{S}(\bm{a}  \bm{b})|\leq 3$.
Furthermore, by analyzing the runs in $\bm{a}$ and the runs in $\bm{b}$, we can determine the precise size of the set $\mathcal{D}(\bm{a} \overline{\alpha} \bm{b}) \cap \mathcal{S}(\bm{a}  \bm{b})$, as detailed in Table \ref{tab:(2,2)_DS}. 
Then by Table \ref{tab:(2,2)_DS}, the conclusion follows. 
\end{IEEEproof}

\begin{claim}\label{cla:(1,2)_B}
Let $D,S$ be defined in Equation (\ref{eq:(1,2)}), we have $|B|=|\mathcal{B}(\bm{x}, \bm{y}) \setminus (D \cup S)|\leq 2$. Moreover, the following holds:
\begin{itemize}
  \item if $r(\bm{a})=0$, or $r(\bm{b})=0$, or $r(\bm{a})=r(\bm{b})=1$ with the same run in \(\bm{a}\) and \(\bm{b}\), then $|B|=0$;
  \item if $r(\bm{a})=1, r(\bm{b})\geq 2$, or $r(\bm{a})\geq 2, r(\bm{b})= 1$, or $r(\bm{a})=r(\bm{b})= 1$ with distinct runs in $\bm{a}$ and $\bm{b}$, then $|B|=1$;
  \item if $r(\bm{a})\geq 2, r(\bm{b})\geq 2$, then $|B|=2$.
\end{itemize}
\end{claim}

\begin{IEEEproof}
    Recall that $\bm{x}= \bm{a}\alpha \bm{b}$ and $\bm{y}= \bm{a}\overline{\alpha} \bm{b}$.
    Let $\bm{x}_i\triangleq \bm{x}_{[n]\setminus \{i\}}$ and $\bm{y}_i\triangleq \bm{y}_{[n]\setminus \{i\}}$ for $i\in [n]$, we have
    \begin{align*}
    \mathcal{B}(\bm{x},\bm{y})
    &= \bigcup_{i=1}^{n} \bigcup_{j=1}^{n} \mathcal{S}(\bm{x}_i, \bm{y}_j).
    \end{align*}
    We first determine the intersection $\mathcal{S}(\bm{x}_i, \bm{y}_j)$. Let $m\triangleq |\bm{a}|$.
    \begin{itemize}
	\item If $i=m+1$ or $j=m+1$, we have 
	\begin{align*}
		\mathcal{S}(\bm{x}_i, \bm{y}_j)\subseteq \mathcal{S}(\bm{a} \bm{b})=S.
	\end{align*}
    
	\item If $i,j\leq m$, by examining the $m$-th entries of sequences in the intersection $\mathcal{S}(\bm{x}_i, \bm{y}_j)$, we can conclude that
	\begin{align*}
		\mathcal{S}(\bm{x}_i, \bm{y}_j)
		\subseteq \{\bm{a}_j \alpha \bm{b}, \bm{a}_i \overline{\alpha}  \bm{b}\}
		\subseteq \mathcal{D}(\bm{a}\alpha  \bm{b}) \cup \mathcal{D}(\bm{a} \overline{\alpha} \bm{b})=D.
	\end{align*}
	
	\item If $ i,j\ge m+2$, by examining the $(m+1)$-th entries of sequences in the intersection $\mathcal{S}(\bm{x}_i, \bm{y}_j)$, we can conclude that
	\begin{align*}
		\mathcal{S}(\bm{x}_i, \bm{y}_j)\subseteq \{\bm{a} \alpha  \bm{b}_{j-m-1}, \bm{a} \overline{\alpha}  \bm{b}_{i-m-1}\}
		\subseteq \mathcal{D}(\bm{a}\alpha  \bm{b}) \cup \mathcal{D}(\bm{a} \overline{\alpha} \bm{b})=D.
	\end{align*}
	
	\item If $ i\leq m$ and $ j\ge m+2$, we can assume that $\bm{x}_i\neq \bm{x}_{m+1}$ and $\bm{y}_j\neq \bm{y}_{m+2}$, i.e., $\bm{a}_i \alpha \neq \bm{a}$ and $\bm{b}\neq \overline{\alpha} \bm{b}_{j-m-2}$, since the other scenarios are analogous to the first case.
      As a result, we assume that $\bm{a}$ has an $\overline{\alpha}$-run and $\bm{b}$ has an $\alpha$-run.
      Observe that $\bm{x}_i= \bm{a}_i \alpha \bm{b}$ and $\bm{y}_j= \bm{a} \overline{\alpha} \bm{b}_{j-m-2}$, we can compute $d_H(\bm{x}_i,\bm{y}_j)= d_H(\bm{a}_i \alpha , \bm{a})+ d_H(\bm{b}, \overline{\alpha} \bm{b}_{j-m-2})\geq 2$.
      It then suffices to consider the scenario where $d_H(\bm{x}_i,\bm{y}_j)=2$, or equivalently $d_H(\bm{a}_i \alpha, \bm{a})= d_H(\bm{b}, \overline{\alpha} \bm{b}_{j-m-2})=1$, as in the other cases, the intersection $\mathcal{S}(\bm{x}_i, \bm{y}_j)$ is empty.
      In this case, by Lemma \ref{lem:sub}, we have
      \begin{align*}
        \mathcal{S}(\bm{x}_i, \bm{y}_j)
        =\{\bm{a}_i \alpha \overline{\alpha} \bm{b}_{j-m-2}, \bm{a} \bm{b} \}.
      \end{align*}
      We now consider the sequences $\bm{a} \alpha$ and $\overline{\alpha} \bm{b}$.
      Since $d_H(\bm{a}_i \alpha, \bm{a})= d_H(\bm{b}, \overline{\alpha} \bm{b}_{j-m-2})=1$, it follows by Lemma \ref{lem:del_position} that there is exactly one choice for $\bm{a}_i\alpha$ and $\overline{\alpha} \bm{b}_{j-m-2}$, namely, $\bm{a}'\triangleq \bm{a}_i \alpha\neq \bm{a}$ is obtained from $\bm{a}\alpha$ by deleting a symbol of its last $\overline{\alpha}$-run and $\bm{b}'\triangleq \overline{\alpha}\bm{b}_{j-m-2}\neq \bm{b}$ is obtained from $\overline{\alpha} \bm{b}$ by deleting a symbol of its first $\alpha$-run.
      Since $\bm{a}' \neq \bm{a}$ and $\bm{b}\neq \bm{b}'$, it can be easily checked that $\bm{a}'\bm{b}'\notin \mathcal{S}(\bm{a}\bm{b}) \cup \mathcal{D}(\bm{a}\alpha \bm{b}) \cup \mathcal{D}(\bm{a}\overline{\alpha} \bm{b})= D \cup S$.
      Consequently, we obtain
    \begin{align*}
        \left( \bigcup_{i=1}^{m}\bigcup_{j=m+2}^n \mathcal{S}(\bm{x}_i, \bm{y}_j) \right) \setminus (D\cup S)
        =\{ \bm{a}'  \bm{b}'\}.
    \end{align*}

	\item If $ i\geq m+2$ and $ j\leq m$, by considering the reversal of $\bm{x}$ and $\bm{y}$ and using the conclusion of the previous case, we can conclude that 
      \begin{align*}
        \left( \bigcup_{i=1}^{m}\bigcup_{j=m+2}^n \mathcal{S}(\bm{x}_i, \bm{y}_j) \right) \setminus (D\cup S)
        =\{ \bm{a}'' \bm{b}''\},
    \end{align*}
    where $\bm{a}''\neq \bm{a}$ is obtained from $\bm{a}\overline{\alpha}$ by deleting a symbol of its last $\alpha$-run and $\bm{b}''\neq \bm{b}$ is obtained from $\alpha\bm{b}$ by deleting a symbol of its first $\overline{\alpha}$-run.
\end{itemize}
Finally, we prove the lemma by analyzing the runs in $\bm{a}$ and the runs in $\bm{b}$, as discussed as follows:
\begin{itemize}
\item if $r(\bm{a})=0$, or $r(\bm{b})=0$, or $r(\bm{a})=r(\bm{b})=1$ with the same run in \(\bm{a}\) and \(\bm{b}\), then both $\bm{a}' \bm{b}'$ and $\bm{a}'' \bm{b}''$ do not exist, implying that $|B|=0$;
\item if $r(\bm{a})=1, r(\bm{b})\geq 2$, or $r(\bm{a})\geq 2, r(\bm{b})= 1$, or $r(\bm{a})=r(\bm{b})= 1$  with distinct runs in $\bm{a}$ and $\bm{b}$, then exactly one of $\bm{a}' \bm{b}'$ and $\bm{a}'' \bm{b}''$ exists, implying that $|B|=1$;
\item if $r(\bm{a})\geq 2, r(\bm{b})\geq 2$, then both $\bm{a}' \bm{b}'$ and $\bm{a}'' \bm{b}''$ exist, implying that $|B|=2$.
\end{itemize}
\end{IEEEproof}

Now, combining Equation (\ref{eq:B}) with Claims \ref{cla:(1,2)_D}, \ref{cla:(1,2)_S}, \ref{cla:(1,2)_DS}, and \ref{cla:(1,2)_B}, we have $|\mathcal{B}(\bm{x},\bm{y})|= |B|+|D|+|S|- |D\cap S|\leq 2+r(\bm{x})+r(\bm{y})-1+n-2= r(\bm{x})+r(\bm{y})+n-1$.
Moreover, the following holds:
\begin{itemize}
\item if $r(\bm{a})\leq 1, r(\bm{b})\leq 1$, then $|S|=n$, $|D|\leq 2r(\bm{a})+2r(\bm{b})= 4$, $|D\cap S|\geq 2$, and $|B|\leq 1$, implying that $|\mathcal{B}(\bm{x},\bm{y})|= |B|+|D|+|S|- |D\cap S|= n+3$;
\item if $r(\bm{a})=0, r(\bm{b})\geq 2$, or $r(\bm{a})\geq 2, r(\bm{b})=0$, then $|S|=n$, $|D|= 2r(\bm{a})+2r(\bm{b})\leq 2n-2$, $|D\cap S|=3$, and $|B|=0$, implying that $|\mathcal{B}(\bm{x},\bm{y})|= |B|+|D|+|S|- |D\cap S|\leq 3n-5$, with equality holding if and only if $r(\bm{a})+r(\bm{b})=n-1$; 
\item if $r(\bm{a})=1, r(\bm{b})\geq 2$, or $r(\bm{a})\geq 2, r(\bm{b})= 1$, then $|S|=n$, $|D|= 2r(\bm{a})+2r(\bm{b})-1\leq 2n-3$, $|D\cap S|=4$, and $|B|\leq 1$, implying that $|\mathcal{B}(\bm{x},\bm{y})|= |B|+|D|+|S|- |D\cap S| \leq 3n-6$;
\item if $r(\bm{a})\geq 2, r(\bm{b})\geq 2$, then $|S|=n$, $|D|= 2r(\bm{a})+2r(\bm{b})-1\leq 2n-3$, $|D\cap S|=5$, and $|B|=2$, implying that $|\mathcal{B}(\bm{x},\bm{y})|= |B|+|D|+|S|- |D\cap S| \leq 3n-6$.
\end{itemize}
In summary, when $n\ge 4$, we have $|\mathcal{B}(\bm{x},\bm{y})|\leq 3n-5$ with equality holding if and only if either $r(\bm{a})=0, r(\bm{b})=n-1$ or $r(\bm{a})=n-1, r(\bm{b})=0$.
This completes the proof of Lemma \ref{lem:(1,2)'}.

\smallskip

\subsubsection{Proof of Lemma \ref{lem:(1,2)''}}

Firstly, by Lemmas \ref{lem:sub} and \ref{lem:del}, we can derive 
\begin{gather*}
    \mathcal{S}(\bm{x},\bm{y})= \{\bm{a} \alpha^{\ell+1} \bm{b}, \bm{a}\overline{\alpha} \alpha^{\ell-1} \overline{\alpha} \bm{b}\},\quad
    \mathcal{D}(\bm{x},\bm{y})= \{\bm{a}\alpha^{\ell} \bm{b}\}.
\end{gather*}
This implies 
\begin{equation}\label{eq:(1,2)'}
\begin{gathered}
  D= \mathcal{D}(\bm{a} \alpha^{\ell+1} \bm{b}) \cup \mathcal{D}( \bm{a}\overline{\alpha} \alpha^{\ell-1} \overline{\alpha} \bm{b}), \quad
  S=\mathcal{S}(\bm{a}\alpha^{\ell} \bm{b}).
\end{gathered}
\end{equation}
In what follows, we will calculate the sizes of \(S\), \(D\), \(D \cap S\), and \(B \triangleq \mathcal{B}(\bm{x}, \bm{y}) \setminus (D \cup S)\) respectively, which will then allow us to determine the size of \(\mathcal{B}(\bm{x},\bm{y})\) based on Equation (\ref{eq:B}).

\begin{claim}\label{cla:(1,2)_S'}
 Let $S$ be defined in Equation (\ref{eq:(1,2)'}), then $|S|=n$.
\end{claim}

\begin{IEEEproof}
    The conclusion follows by Lemma \ref{lem:size} directly.
\end{IEEEproof}

\begin{claim}\label{cla:(1,2)_D'}
  Let $D$ be defined in Equation (\ref{eq:(1,2)'}), the following holds:
 \begin{itemize}
 \item if $r(\bm{a})=0, r(\bm{b})=0$, we have $|D|=4$;
   \item if $r(\bm{a})=0,r(\bm{b})\ge 1$ or $r(\bm{b})=0,r(\bm{a})\ge 1$, we have $|D|= 2r(\bm{a})+2r(\bm{b})+3 \leq 2n-2\ell+1\leq 2n-3$;
   \item if $r(\bm{a})\neq 0, r(\bm{b})\neq 0$, we have $|D|= 2r(\bm{a})+2r(\bm{b})+2\leq 2n-2\ell\leq 2n-4$. 
 \end{itemize}
\end{claim}

\begin{IEEEproof}
By examining the $(|\bm{a}|+1)$-th and the $(|\bm{a}|+\ell)$-th entries of sequences in $\mathcal{D}(\bm{a} \alpha^{\ell+1} \bm{b})$ and $\mathcal{D}(\bm{a} \overline{\alpha} \alpha^{\ell-1}\overline{\alpha} \bm{b})$, we find that 
\begin{equation}\label{eq:empty}
 \mathcal{D}(\bm{a} \alpha^{\ell+1} \bm{b})\cap \mathcal{D}(\bm{a} \overline{\alpha} \alpha^{\ell-1}\overline{\alpha} \bm{b})=\emptyset.
\end{equation}
It follows that
\begin{align*}
|D|
&=|\mathcal{D}(\bm{a} \alpha^{\ell+1} \bm{b})|+|\mathcal{D}(\bm{a} \overline{\alpha} \alpha^{\ell-1}\overline{\alpha} \bm{b})|.
\end{align*}
By examining the last run of \(\bm{a}\) and the first run of \(\bm{b}\), we can calculate the size of $D$, as detailed in Table \ref{tab:D2}. Then the conclusion follows by the fact that $r(\bm{a})+r(\bm{b})\leq |\bm{a}|+ |\bm{b}|= n-\ell-1$.   
\end{IEEEproof}

\begin{remark}\label{rmk:run}
  In Claim \ref{cla:(1,2)_D'}, we can establish that $|D|= r(\bm{x})+r(\bm{y})$. 
  The reason is as follows:
  \begin{itemize}
    \item If $\bm{b}$ is empty, we have $r(\bm{a} \alpha^{\ell+1} \bm{b})= r(\bm{a} \alpha^{\ell} \overline{\alpha} \bm{b})-1$ and $r(\bm{a} \overline{\alpha} \alpha^{\ell-1}\overline{\alpha} \bm{b})= r(\bm{a} \overline{\alpha} \alpha^{\ell}\bm{b})+1 $, implying that $r(\bm{a} \alpha^{\ell+1} \bm{b})+ r(\bm{a} \overline{\alpha} \alpha^{\ell-1}\overline{\alpha} \bm{b})= r(\bm{a} \alpha^{\ell} \overline{\alpha} \bm{b})+ r(\bm{a} \overline{\alpha} \alpha^{\ell} \bm{b})= r(\bm{x})+r(\bm{y})$.
        
    \item If $\bm{b}$ starts with an $\alpha$-run, we have $r(\bm{a} \alpha^{\ell+1} \bm{b})= r(\bm{a} \alpha^{\ell} \overline{\alpha} \bm{b})-2$ and $r(\bm{a} \overline{\alpha} \alpha^{\ell-1}\overline{\alpha} \bm{b})= r(\bm{a} \overline{\alpha} \alpha^{\ell} \bm{b})+2$, implying that $r(\bm{a} \alpha^{\ell+1} \bm{b})+ r(\bm{a} \overline{\alpha} \alpha^{\ell-1}\overline{\alpha} \bm{b})= r(\bm{a} \alpha^{\ell} \overline{\alpha} \bm{b})+ r(\bm{a} \overline{\alpha} \alpha^{\ell} \bm{b})= r(\bm{x})+r(\bm{y})$.   
    
    \item If $\bm{b}$ starts with an $\overline{\alpha}$-run, we have $r(\bm{a} \alpha^{\ell+1} \bm{b})= r(\bm{a} \alpha^{\ell} \overline{\alpha} \bm{b})$ and $r(\bm{a} \overline{\alpha} \alpha^{\ell-1}\overline{\alpha} \bm{b})= r(\bm{a} \overline{\alpha} \alpha^{\ell} \bm{b})$, implying that $r(\bm{a} \alpha^{\ell+1} \bm{b})+ r(\bm{a} \overline{\alpha} \alpha^{\ell-1}\overline{\alpha} \bm{b})= r(\bm{a} \alpha^{\ell} \overline{\alpha} \bm{b})+ r(\bm{a} \overline{\alpha} \alpha^{\ell} \bm{b})= r(\bm{x})+r(\bm{y})$.
  \end{itemize}
\end{remark}

\begin{table*}
\caption{The values of $|\mathcal{D}(\bm{a} \alpha^{\ell+1}  \bm{b})|= r(\bm{a} \alpha^{\ell+1} \bm{b})$ and $|\mathcal{D}(\bm{a} \overline{\alpha}\alpha^{\ell-1}\overline{\alpha} \bm{b})|= r(\bm{a} \overline{\alpha}\alpha^{\ell-1}\overline{\alpha}\bm{b})$. When we refer to the first or last run of a sequence as $\epsilon$, it indicates that its length is zero, implying that it does not contain any run.} 
\centering
\begin{tabular}{c|c|c|c|c} 
\hline 
\hline
the last run of $\bm{a}$ & the first run of $\bm{b}$ & $|\mathcal{D}(\bm{a} \alpha^{\ell+1} \bm{b})|$ & $|\mathcal{D}(\bm{a}\overline{\alpha}\alpha^{\ell-1}\overline{\alpha}\bm{b})|$ & $|D|$\\
\hline 
$\epsilon$ & $\epsilon$ & $1$ & $3$ & $4$ \\
\hline 
$\epsilon$ & $\alpha$ & $r(\bm{a})+r(\bm{b})$ & $r(\bm{a})+r(\bm{b})+3$ & $2r(\bm{a})+2r(\bm{b})+3$\\
\hline 
$\epsilon$ & $\overline{\alpha}$ & $r(\bm{a})+r(\bm{b})+1$ & $r(\bm{a})+r(\bm{b})+2$ & $2r(\bm{a})+2r(\bm{b})+3$\\
\hline 
$\alpha$ & $\alpha$ & $r(\bm{a})+r(\bm{b})-1$ & $r(\bm{a})+r(\bm{b})+3$ & $2r(\bm{a})+2r(\bm{b})+2$\\
\hline
$\alpha$ & $\overline{\alpha}$ & $r(\bm{a})+r(\bm{b})$ & $r(\bm{a})+r(\bm{b})+2$ & $2r(\bm{a})+2r(\bm{b})+2$ \\
\hline
$\alpha$ & $\epsilon$ & $r(\bm{a})+r(\bm{b})$ & $r(\bm{a})+r(\bm{b})+3$ & $2r(\bm{a})+2r(\bm{b})+3$ \\
\hline
$\overline{\alpha}$ & $\alpha$ & $r(\bm{a})+r(\bm{b})$ & $r(\bm{a})+r(\bm{b})+2$ & $2r(\bm{a})+2r(\bm{b})+2$\\
\hline
$\overline{\alpha}$ & $\overline{\alpha}$ & $r(\bm{a})+r(\bm{b})+1$ & $r(\bm{a})+r(\bm{b})+1$ & $2r(\bm{a})+2r(\bm{b})+2$\\
\hline 
$\overline{\alpha}$ & $\epsilon$ & $r(\bm{a})+r(\bm{b})+1$ & $r(\bm{a})+r(\bm{b})+2$ & $2r(\bm{a})+2r(\bm{b})+3$ \\
\hline
\hline
\end{tabular}
\label{tab:D2}
\end{table*}

\begin{claim}\label{cla:(1,2)_DS'}
 Let $D$ and $S$ be defined in Equation (\ref{eq:(1,2)'}), we have $3\leq |D\cap S|\leq 5$. Moreover, if $r(\bm{a})\geq 2$ or $r(\bm{b})\geq 2$, then $|D\cup S|\geq 4$.
\end{claim}

\begin{IEEEproof}
Since by Equation (\ref{eq:empty}) that $\mathcal{D}(\bm{a} \alpha^{\ell+1} \bm{b})\cap \mathcal{D}(\bm{a} \overline{\alpha} \alpha^{\ell-1}\overline{\alpha} \bm{b})=\emptyset$, we can derive 
\begin{align*}
|D\cap S|=|\mathcal{D}(\bm{a} \alpha^{\ell+1} \bm{b})\cap \mathcal{S}(\bm{a}\alpha^{\ell} \bm{b}) |+ |\mathcal{D}(\bm{a}\overline{\alpha}\alpha^{\ell-1}\overline{\alpha} \bm{b})\cap \mathcal{S}(\bm{a}\alpha^{\ell} \bm{b})|.
\end{align*}
It remains to consider the set $\mathcal{D}(\bm{a} \alpha^{\ell+1} \bm{b})\cap \mathcal{S}(\bm{a}\alpha^{\ell} \bm{b})$ and $\mathcal{D}(\bm{a}\overline{\alpha}\alpha^{\ell-1}\overline{\alpha} \bm{b})\cap \mathcal{S}(\bm{a}\alpha^{\ell} \bm{b})$, respectively.

We first consider the set $\mathcal{D}(\bm{a} \alpha^{\ell+1} \bm{b})\cap \mathcal{S}(\bm{a}\alpha^{\ell} \bm{b})$ and will determine its size by considering the sequence $\bm{a}\alpha^{\ell+1}\bm{b}$.
For any $\bm{z}\in \mathcal{D}(\bm{a} \alpha^{\ell+1} \bm{b})\cap \mathcal{S}(\bm{a}\alpha^{\ell} \bm{b})$, it holds that $d_H(\bm{z},\bm{a}\alpha^{\ell} \bm{b})\leq 1$.
Then by Lemma \ref{lem:del_position}, we can conclude that there are at most three choices for $\bm{z}$. Specifically,
\begin{itemize}
    \item $\bm{z}=\bm{a}\alpha^{\ell}\bm{b}$;

    \item $\bm{z}$ is obtained from $\bm{a}\alpha^{\ell+1}\bm{b}$ by deleting a symbol from the last $\overline{\alpha}$-run of $\bm{a}$ (if $\bm{a}$ has an $\overline{\alpha}$-run);

    \item $\bm{z}$ is obtained from $\bm{a}\alpha^{\ell+1}\bm{b}$ by deleting a symbol from the first $\overline{\alpha}$-run of $\bm{b}$ (if $\bm{b}$ has an $\overline{\alpha}$-run).
\end{itemize}
This implies that $1\leq |\mathcal{D}(\bm{a} \alpha^{\ell+1} \bm{b})\cap \mathcal{S}(\bm{a}\alpha^{\ell} \bm{b})|\leq 3$.

We now consider the set $\mathcal{D}(\bm{a}\overline{\alpha}\alpha^{\ell-1}\overline{\alpha} \bm{b})\cap \mathcal{S}(\bm{a}\alpha^{\ell} \bm{b})$ and will determine its size by examining the $|\bm{a}|+1$-th and the $(|\bm{a}|+\ell)$-th entries of sequences in this set.
For any $\bm{z}\in \mathcal{D}(\bm{a}\overline{\alpha}\alpha^{\ell-1}\overline{\alpha} \bm{b})\cap \mathcal{S}(\bm{a}\alpha^{\ell} \bm{b})$. \begin{itemize}
    \item If $\bm{z}_{|\bm{a}|+1}= \overline{\alpha}$, we can conclude that $\bm{z}= \bm{a} \overline{\alpha} \alpha^{\ell-1} \bm{b}$;

    \item If $\bm{z}_{|\bm{a}|+1}= \alpha$, we can conclude that $\bm{z}_{|\bm{a}|+\ell}= \overline{\alpha}$. Then we get $\bm{z}= \bm{a} \alpha^{\ell-1} \overline{\alpha} \bm{b}$.
\end{itemize}
This implies that $|\mathcal{D}(\bm{a}\overline{\alpha}\alpha^{\ell-1}\overline{\alpha} \bm{b})\cap \mathcal{S}(\bm{a}\alpha^{\ell} \bm{b})|=2$.
Then the conclusion follows.
\end{IEEEproof}

\begin{claim}\label{cla:(1,2)_B'}
Let $D,S$ be defined in Equation (\ref{eq:(1,2)'}), we have $|B|= |\mathcal{B}(\bm{x}, \bm{y}) \setminus (D \cup S)|\leq 1$. Moreover, $|B|=1$ if and only if both $\bm{a}$ and $\bm{b}$ have an $\alpha$-run.
\end{claim}

\begin{IEEEproof}
Recall that $\bm{x}= \bm{a}\alpha^{\ell} \overline{\alpha} \bm{b}, \bm{y}= \bm{a}\overline{\alpha} \alpha^{\ell} \bm{b}\in \Sigma^n$, where $\alpha\in \Sigma$ and $\ell\geq 2$. 
Let $\bm{x}_i\triangleq \bm{x}_{[n]\setminus \{i\}}$ and $\bm{y}_i\triangleq \bm{y}_{[n]\setminus \{i\}}$ for $i\in [n]$, we have
\begin{align*}
    \mathcal{B}(\bm{x},\bm{y})
    &= \bigcup_{i=1}^{n} \bigcup_{j=1}^{n} \mathcal{S}(\bm{x}_i, \bm{y}_j).
\end{align*}
We first determine the intersection $\mathcal{S}(\bm{x}_i, \bm{y}_j)$. Let $m\triangleq |\bm{a}|$.
\begin{itemize}
    \item If $i=m+\ell+1$ or $j=m+1$, we have 
    \begin{align*}
        \mathcal{S}(\bm{x}_i, \bm{y}_j)\subseteq \mathcal{S}(\bm{a}\alpha^{\ell}\bm{b})=S.
    \end{align*}
    
\item If $m+1\leq i\leq m+\ell$, we have $\bm{x}_i= \bm{a} \alpha^{\ell-1} \overline{\alpha} \bm{b}$.
We further consider the choice for $j$.
Since we have discussed $j=m+1$ in the first case, we assume that $\bm{y}_j \neq \bm{y}_{m+1}$.
\begin{itemize}
    \item If $j\leq m$, since $\bm{y}_j \neq \bm{y}_{m+1}$, we can assume that $\bm{a}_j \overline{\alpha} \neq \bm{a}$. Now, we can compute $d_H(\bm{x}_i,\bm{y}_j)= d_H(\bm{a},\bm{a}_j \overline{\alpha})+ d_H(\alpha^{\ell-1}\overline{\alpha} \bm{b},  \alpha^{\ell}\bm{b})\geq 2$.
    It then suffices to consider the  scenario where $d_H(\bm{x}_i,\bm{y}_j)= 2$, or equivalently $d_H(\bm{a},\bm{a}_j \overline{\alpha})= d_H(\alpha^{\ell-1}\overline{\alpha} \bm{b},  \alpha^{\ell}\bm{b})=1$, as in the other cases, the intersection $\mathcal{S}(\bm{x}_i, \bm{y}_j)$ is empty. In this case, by Lemma \ref{lem:sub}, we have 
    \begin{align*}
        \mathcal{S}(\bm{x}_i, \bm{y}_j) \subseteq \{\bm{a}\alpha^{\ell}\bm{b}, \bm{a}_j \overline{\alpha}\alpha^{\ell-1}\overline{\alpha} \bm{b}\} \subseteq D\cup S.
    \end{align*}
    
    \item If $m+2\leq j\leq m+\ell+1$, we have $\bm{y}_j= \bm{a} \overline{\alpha} \alpha^{\ell-1} \bm{b}$. Now, we can compute $d_H(\bm{x}_i,\bm{y}_j)= d_H(\alpha^{\ell-1}\overline{\alpha},  \overline{\alpha}\alpha^{\ell-1})= 2$. Then by Lemma \ref{lem:sub}, we have 
    \begin{align*}
        \mathcal{S}(\bm{x}_i, \bm{y}_j)=\{\bm{a}\alpha^{\ell}\bm{b}, \bm{a} \overline{\alpha} \alpha^{\ell-2}\overline{\alpha} \bm{b}\} \subseteq D\cup S.
     \end{align*}
     
    \item If $j\geq m+\ell+2$, we have $\bm{y}_j= \bm{a} \overline{\alpha} \alpha^{\ell} \bm{b}_{j-m-\ell-1}$. We can assume that $\bm{y}_j\neq \bm{y}_{m+\ell+1}$, i.e., $\alpha \bm{b}_{j-m-\ell-1}\neq \bm{b}$, since the other scenarios have discussed previously. Now, we can compute $d_H(\bm{x}_i,\bm{y}_j)= d_H(\bm{a}\alpha^{\ell-1}\overline{\alpha}, \bm{a} \overline{\alpha} \alpha^{\ell-1})+ d_H(\bm{b},  \alpha\bm{b}_{j-m-\ell-1})\ge 3$.
    Then we get 
    \begin{align*}
        \mathcal{S}(\bm{x}_i, \bm{y}_j)=\emptyset.
    \end{align*}
\end{itemize}
Consequently, we can conclude that 
\begin{align*}
  \bigcup_{i=m+1}^{m+\ell}\bigcup_{j=1}^n \mathcal{S}(\bm{x}_i, \bm{y}_j)\subseteq D\cup S.
\end{align*}

\item If $m+2\leq j\leq m+\ell+1$, by considering the reversal of $\bm{x}$ and $\bm{y}$ and using the conclusion of the previous case, we can conclude that
\begin{align*}
\bigcup_{j=m+2}^{m+\ell+1}\bigcup_{i=1}^{n} \mathcal{S}(\bm{x}_i, \bm{y}_j) \subseteq D\cup S.
\end{align*}

\item If $ i,j\leq m$, we have $d_H(\bm{x}_i,\bm{y}_j)=d_H(\bm{a}_i,\bm{a}_j)+d_H(\alpha^{\ell} \overline{\alpha}, \overline{\alpha}\alpha^{\ell})= d_H(\bm{a}_i,\bm{a}_j)+2$. 
It then suffices to consider the  scenario where $d_H(\bm{x}_i,\bm{y}_j)= 2$, or equivalently $\bm{a}_i=\bm{a}_j$, as in the other cases, the intersection $\mathcal{S}(\bm{x}_i, \bm{y}_j)$ is empty.
In this case, by Lemma \ref{lem:sub}, we have
\begin{align*}
    \mathcal{S}(\bm{x}_i, \bm{y}_j)
    \subseteq \{\bm{a}_i \alpha^{\ell+1}  \bm{b}, \bm{a}_i \overline{\alpha}\alpha^{\ell-1} \overline{\alpha} \bm{b}\}
    \subseteq D.
\end{align*}

\item If $ i,j\ge m+\ell+2$, by considering the reversal of $\bm{x}$ and $\bm{y}$ and using the conclusion of the previous case, we can conclude that
\begin{align*}
    \mathcal{S}(\bm{x}_i, \bm{y}_j)\subseteq \{\bm{a} \alpha^{\ell+1}  \bm{b}_{j-m-\ell-1}, \bm{a} \overline{\alpha}\alpha^{\ell-1} \overline{\alpha} \bm{b}_{j-m-\ell-1}\}
    \subseteq D.
\end{align*}
        
\item If $ i\leq m$ and $ j\ge m+\ell+2$, we have $\bm{x}_i= \bm{a}_i \alpha^{\ell} \overline{\alpha} \bm{b}$ and $\bm{y}_j= \bm{a} \overline{\alpha} \alpha^{\ell} \bm{b}_{j-m-\ell-1}$.
We assume that $\bm{x}_i\neq \bm{x}_{m+1}$ and $\bm{y}_j\neq \bm{y}_{m+\ell+1}$, i.e., $\bm{a}_i \alpha \neq \bm{a}$ and $\alpha \bm{b}_{j-m-\ell-1}\neq \bm{b}$, since the other scenarios have discussed previously. 
Now we can compute $d_H(\bm{x}_i,\bm{y}_j)=d_H(\bm{a}_i\alpha,\bm{a})+d_H(\alpha^{\ell-1}\overline{\alpha },\overline{\alpha}\alpha^{\ell-1})+d_H( \bm{b},\alpha \bm{b}_{j-l-m-1})\geq 4$.
This implies that 
\begin{align*}
    \mathcal{S}(\bm{x}_i, \bm{y}_j)=\emptyset.
\end{align*}
        
\item If $ i\ge m+\ell+2$ and $ j\leq m$, we have $\bm{x}_i= \bm{a} \alpha^{\ell} \overline{\alpha} \bm{b}_{i-m-\ell-1}$ and $\bm{y}_j= \bm{a}_j \overline{\alpha} \alpha^{\ell} \bm{b}$. 
We assume that $\bm{x}_i\neq \bm{x}_{m+\ell+1}$ and $\bm{y}_j\neq \bm{y}_{m+1}$, i.e., $\overline{\alpha} \bm{b}_{i-m-\ell-1} \neq \bm{b}$ and $\bm{a}_{j} \overline{\alpha}\neq \bm{a}$, since these scenarios have discussed previously. 
As a result, we assume that both $\bm{a}$ and $\bm{b}$ have an $\alpha$-run.
We can compute $d_H(\bm{x}_i,\bm{y}_j)= d_H(\bm{a}, \bm{a}_j \overline{\alpha})+ d_H(\overline{\alpha} \bm{b}_{i-l-m-1},\bm{b})\geq 2$.
It then suffices to consider the scenario where $d_H(\bm{x}_i,\bm{y}_j)=2$, or equivalently $d_H(\bm{a}, \bm{a}_j \overline{\alpha})= d_H(\overline{\alpha} \bm{b}_{i-m-\ell-1},\bm{b})=1$, as in the other cases, the intersection $\mathcal{S}(\bm{x}_i, \bm{y}_j)$ is empty.
In this case, by Lemma \ref{lem:sub}, we have
\begin{align*}
    \mathcal{S}(\bm{x}_i, \bm{y}_j)
    =\{\bm{a}_j \overline{\alpha} \alpha^{\ell} \overline{\alpha} \bm{b}_{i-l-m-1}, \bm{a} \alpha^{\ell} \bm{b}\}.
\end{align*}
We now consider the sequences $\bm{a} \overline{\alpha}$ and $\overline{\alpha} \bm{b}$. 
Since $d_H(\bm{a}, \bm{a}_j \overline{\alpha})= d_H(\overline{\alpha} \bm{b}_{i-m-\ell-1},\bm{b})=1$, it follows by Lemma \ref{lem:del_position} that there is exactly one choice for $\bm{a}_j \overline{\alpha}$ and $\overline{\alpha} \bm{b}_{i-m-\ell-1}$, namely, $\bm{a}'\triangleq \bm{a}_j \overline{\alpha}\neq \bm{a}$ is obtained from $\bm{a} \overline{\alpha}$ by deleting a symbol of its last $\alpha$-run and $\bm{b}'\triangleq \overline{\alpha} \bm{b}_{i-m-\ell-1}\neq \bm{b}$ is obtained from $\overline{\alpha} \bm{b}$ by deleting a symbol of its first $\alpha$-run.
Since $\bm{a}' \neq \bm{a}$ and $\bm{b}\neq \bm{b}'$, it can be easily checked that $\bm{a}' \alpha^{\ell} \bm{b}'\notin \mathcal{D}(\bm{a}\alpha^{\ell+1} \bm{b}) \cup \mathcal{D}(\bm{a}\overline{\alpha}\alpha^{\ell-1}\overline{\alpha} \bm{b})\cup \mathcal{S}(\bm{a} \alpha^{\ell} \bm{b})=D \cup S$.
Consequently, we obtain
\begin{align*}
    \left( \bigcup_{i=m+\ell+2}^{n}\bigcup_{j=1}^{m} \mathcal{S}(\bm{x}_i, \bm{y}_j) \right) \setminus (D\cup S)
    \subseteq\{\bm{a}' \alpha^{\ell} \bm{b}'\}.
\end{align*}
\end{itemize}
Finally, we prove the lemma by analyzing the runs in $\bm{a}$ and $\bm{b}$, as discussed below:
\begin{itemize}
 \item If one of $\bm{a}$ and $\bm{b}$ does not have an $\alpha$-run, then $\bm{a}'\alpha^{\ell} \bm{b}'$ does not exist, implying that $|B|=0$.
  \item If both $\bm{a}$ and $\bm{b}$ have an $\alpha$-run, then $\bm{a}'\alpha^{\ell} \bm{b}'$ exists, implying that $|B|=1$.
\end{itemize}
\end{IEEEproof}

Now, combining Equation (\ref{eq:B}) with Remark \ref{rmk:run} and Claims \ref{cla:(1,2)_S'}, \ref{cla:(1,2)_D'}, \ref{cla:(1,2)_DS'}, and \ref{cla:(1,2)_B'}, we have $|\mathcal{B}(\bm{x},\bm{y})|= |B|+|D|+|S|- |D\cap S|\leq 1+r(\bm{x})+r(\bm{y})+ n- 3= r(\bm{x})+r(\bm{y})+ n- 2$.
Moreover, the following holds:
\begin{itemize}
 \item If $r(\bm{a})\leq 1, r(\bm{b})\leq 1$, then $|S|=n$, $|D|\leq 7$, $|D\cap S|\geq 3$, and $|B|\leq 1$, implying that $|\mathcal{B}(\bm{x},\bm{y})|= |B|+|D|+|S|- |D\cap S|= n+5$;
 \item If $r(\bm{a})=0, r(\bm{b})\geq 2$ or $r(\bm{a})\ge 2, r(\bm{b})=0$, then $|S|=n$, $|D|\leq 2n-3$, $|D\cap S|=4$, and $|B|=0$, implying that $|\mathcal{B}(\bm{x},\bm{y})|= |B|+|D|+|S|- |D\cap S|\leq 3n-7$;
 \item If $r(\bm{a})\geq 1, r(\bm{b})\geq 2$ or $r(\bm{a})\ge 2, r(\bm{b})\geq 1$, then $|S|=n$, $|D|\leq 2n-4$, $|D\cap S|\geq 4$, and $|B|\leq 1$, implying that $|\mathcal{B}(\bm{x},\bm{y})|= |B|+|D|+|S|- |D\cap S|\leq 3n-7$.
\end{itemize}
In summary, when $n\ge 6$, we have $|\mathcal{B}(\bm{x},\bm{y})|\leq 3n-7$. This completes the proof of Lemma \ref{lem:(1,2)''}.

\subsection{Proof of Lemma \ref{lem:(2,0)}}\label{subsec:(2,0)}

The proof follows a strategy similar to that of Lemma \ref{lem:(2,2)}, we include it here for completeness. 
We first characterize the structures of $\bm{x}$ and $\bm{y}$ when $|\mathcal{D}(\bm{x},\bm{y})|=2$ and $|\mathcal{S}(\bm{x},\bm{y})|=0$.
Since $|\mathcal{D}(\bm{x},\bm{y})|=2$, by Lemma \ref{lem:del}, we can express $\bm{x}=\bm{a} \bm{c} \bm{b}$ and $\bm{y}=\bm{a} \overline{\bm{c}} \bm{b}$, where $\bm{c}$ is an alternating sequence of length at least two.
Moreover, since $|\mathcal{S}(\bm{x},\bm{y})|=2$, by Lemma \ref{lem:sub}, we can conclude that the length of $\bm{c}$ is at least three.
In this case, by Lemmas \ref{lem:sub} and \ref{lem:del'}, we can derive 
\begin{gather*}
    \mathcal{S}(\bm{x},\bm{y})= \emptyset,\quad
    \mathcal{D}(\bm{x},\bm{y})= \{\bm{a}\bm{c}_1 \bm{b}, \bm{a}\overline{\bm{c}}_1 \bm{b}\},
\end{gather*}
where $\bm{c}_1\triangleq \bm{c}_{[2,|\bm{c}|]}$ is an alternating sequence of length at least two.
This implies that
\begin{equation}\label{eq:(2,0)}
\begin{gathered}
  D= \emptyset,\quad
  S= \mathcal{S}(\bm{a}\bm{c}_1\bm{b}) \cup \mathcal{S}(\bm{a} \overline{\bm{c}}_1 \bm{b}).
\end{gathered}
\end{equation}
Then by Equation (\ref{eq:B}), we can calculate 
\begin{equation}\label{eq:(2,0)_B}
    |\mathcal{B}(\bm{x},\bm{y})|= |B|+|S|,
\end{equation}
where $B \triangleq \mathcal{B}(\bm{x}, \bm{y}) \setminus S$.
In what follows, we will calculate the sizes of \(S\) and \(B\) respectively, which will then allow us to determine the size of \(\mathcal{B}(\bm{x},\bm{y})\) based on Equation (\ref{eq:(2,0)_B}).

\begin{claim}\label{cla:(2,0)_S}
    Let $S$ be defined in (\ref{eq:(2,0)}), we have $|S|=2n$ if $|\bm{c}|\geq 4$ and $|S|=2n-2$ if $|\bm{c}|=3$.
\end{claim}

\begin{IEEEproof}
    By Lemma \ref{lem:size}, we have $|\mathcal{S}(\bm{a}\bm{c}_1\bm{b})|= |\mathcal{S}(\bm{a} \overline{\bm{c}}_1 \bm{b})|=n$.
    Moreover, by Lemma \ref{lem:sub}, we can conclude that $|\mathcal{S}(\bm{a}\bm{c}_1\bm{b}) \cap \mathcal{S}(\bm{a} \overline{\bm{c}}_1 \bm{b})|=2$ if $|\bm{c}|=3$ and $|\mathcal{S}(\bm{a}\bm{c}_1\bm{b}) \cap \mathcal{S}(\bm{a} \overline{\bm{c}}_1 \bm{b})|=0$ if $|\bm{c}|\geq 4$.
    Then the conclusion follows by the inclusion-exclusion principle.
\end{IEEEproof}

\begin{claim}\label{cla:(2,0)_B}
    Let $S$ be defined in Equation (\ref{eq:(2,0)}), we have $|B|= |\mathcal{B}(\bm{x}, \bm{y}) \setminus S|\leq 8$.
\end{claim}

\begin{IEEEproof}
Recall that $\bm{x}=\bm{a} \bm{c} \bm{b}$ and $\bm{y}=\bm{a} \overline{\bm{c}} \bm{b}$, where $\bm{c}$ is an alternating sequence of length at least three.
Let $\bm{x}_i\triangleq \bm{x}_{[n]\setminus \{i\}}$ and $\bm{y}_i\triangleq \bm{y}_{[n]\setminus \{i\}}$ for $i\in [n]$, we have
\begin{align*}
    \mathcal{B}(\bm{x},\bm{y})
    &= \bigcup_{i=1}^{n} \bigcup_{j=1}^{n} \mathcal{S}(\bm{x}_i, \bm{y}_j).
\end{align*}
We first consider the intersection $\mathcal{S}(\bm{x}_i, \bm{y}_j)$.
Let $m\triangleq |\bm{a}|$ and $\ell\triangleq |\bm{c}|\geq 3$.
\begin{itemize}
    \item If $i=m+1$ or $j=m+\ell$, we have 
    \[\mathcal{S}(\bm{x}_i, \bm{y}_j)\subseteq \mathcal{S}(\bm{a}\bm{c}_{1} \bm{b})\subseteq S.
    \]
    \item If $i=m+\ell$ or $j=m+1$, we have 
    \[\mathcal{S}(\bm{x}_i, \bm{y}_j)\subseteq \mathcal{S}(\bm{a}\overline{\bm{c}}_{1} \bm{b})\subseteq S.
    \]
    
    \item If $i\le m$, since we have discussed $i=m+1$ in the first case, we assume that $\bm{x}_i= \bm{a}_i \bm{c} \bm{b}\neq \bm{x}_{m+1}$, i.e., $\bm{a}_i c_1 \neq \bm{a}$.
    We further consider the choice for $j$.
    \begin{itemize}
        \item If $j\leq m$, we have $\bm{x}= \bm{a}_i \bm{c} \bm{b}$ and $\bm{y}=\bm{a}_j \overline{\bm{c}} \bm{b}$. We can compute  $d_H(\bm{x}_i,\bm{y}_j)\geq d_H(\bm{c},\overline{\bm{c}})=\ell>2$, which implies that
        \begin{align*}
            \mathcal{S}(\bm{x}_i, \bm{y}_j)=\emptyset.
        \end{align*}
    
        \item If $m+2\le j\le m+\ell-1$, we have $\bm{y}= \bm{a} \overline{\bm{c}}_{[1,j-m-1]} \overline{\bm{c}}_{[j-m+1,\ell]}\bm{b}= \bm{a} \bm{c}_{[2,j-m]} \overline{\bm{c}}_{[j-m+1,\ell]}\bm{b}$.
        Recall that $\bm{x}= \bm{a}_i \bm{c} \bm{b}= \bm{a}_i c_1 \bm{c}_{[2,j-m]} \bm{c}_{[j-m+1,\ell]}\bm{b}$, where $\bm{a}_{i}c_1\neq \bm{a}$, we can calculate $d_H(\bm{x}_i,\bm{y}_j)= d_H(\bm{a}_ic_1,\bm{a})+ \ell+m-j\geq 2$. 
        It then suffices to consider the scenario where $d_H(\bm{x}_i,\bm{y}_j)=2$, or equivalently $d_H(\bm{a}_ic_1,\bm{a})=1$ and $j=m+\ell-1$, as in the other cases, the intersection $\mathcal{S}(\bm{x}_i,\bm{y}_j)$ is empty.
        In this case, by Lemma \ref{lem:sub}, we have
        \[\mathcal{S}(\bm{x}_i, \bm{y}_j)=\{\bm{a}_ic_1 \bm{c}_{[2,\ell-1]}\overline{c}_{\ell}\bm{b},\bm{a}\bm{c}_1\bm{b}\}.
        \]
        We now consider the sequence $\bm{a}c_1$.
        Since $d_H(\bm{a},\bm{a}_ic_1)=1$, it follows by Lemma \ref{lem:del_position} that there is exactly one choice for $\bm{a}_ic_1$, namely, $\bm{a}'\triangleq \bm{a}_{i}c_1\neq \bm{a}$ is obtained from $\bm{a}c_1$ by deleting a symbol of its last $\overline{c}_1$-run. 
        Since $\bm{a'}\neq \bm{a}$, we have $d_H(\bm{a'}\bm{c}_{[2,\ell-1]}\overline{c}_{\ell}\bm{b},\bm{a}\bm{c}_{1} \bm{b})\ge 2$ and $d_H(\bm{a'}\bm{c}_{[2,\ell-1]}\overline{c}_{\ell}\bm{b},\bm{a}\overline{\bm{c}}_{1} \bm{b})\ge 2$.
        This implies that $\bm{a'}\bm{c}_{[2,{\ell}-1]}\overline{c}_{\ell}\bm{b}\not\in \mathcal{S}(\bm{a}\bm{c}_{1} \bm{b}) \cup \mathcal{S}(\bm{a}\overline{\bm{c}}_{1} \bm{b})=S$.
        Consequently, we obtain
        \begin{align*}
           \left( \bigcup_{i=1}^{m}\bigcup_{j= m+2}^{m+\ell-1} \mathcal{S}(\bm{x}_i, \bm{y}_j) \right)\setminus S=\{\bm{a'}\bm{c}_{[2,\ell-1]}\overline{c}_{\ell}\bm{b}\}.
            \end{align*}
    \item If $j\ge m+\ell+1$, since we have discussed $j=m+\ell$ in the first case, we assume that $\bm{y}_j= \bm{a}\overline{\bm{c}}_{[1,\ell-1]}\overline{c}_{\ell}\bm{b}_{j-m-\ell}\neq \bm{y}_{m+\ell}$, i.e., $\overline{c}_{\ell} \bm{b}_{j-m-\ell}\neq \bm{b}$.
    Recall that $\bm{x}_i= \bm{a}_i c_1 \bm{c}_{[2,\ell]} \bm{b}= \bm{a}_i c_1 \overline{\bm{c}}_{[1,\ell-1]} \bm{b}$, we can compute $d_H(\bm{x}_i,\bm{y}_j)= d_H(\bm{a}_{i}c_1, \bm{a})+ d_H(\bm{b},\overline{c}_{\ell}\bm{b}_{j-m-\ell})\geq 2$.
    It then suffices to consider the scenario where $d_H(\bm{x}_i,\bm{y}_j)=2$, or equivalently $d_H(\bm{a}_{i}c_1, \bm{a})= d_H(\bm{b},\overline{c}_{\ell}\bm{b}_{j-m-\ell})=1$, as in the other cases, the intersection $\mathcal{S}(\bm{x}_i, \bm{y}_j)$ is empty.
    In this case, by Lemma \ref{lem:sub}, we have
    \begin{align*}
        \mathcal{S}(\bm{x}_i, \bm{y}_j)
        =\{\bm{a}\bm{c}_1\bm{b},\bm{a'}\bm{c}_{1}\bm{b'}\}.
    \end{align*}
    We now consider the sequences $\bm{a} c_1$ and $\overline{c}_{\ell} \bm{b}$.
    Since $d_H(\bm{a}_{i}c_1, \bm{a})= d_H(\overline{c}_{\ell}\bm{b}_{j-m-\ell}, \bm{b})=1$, it follows by Lemma \ref{lem:del_position} that there is exactly one choice for $\bm{a}_i c_1$ and $\overline{c}_{\ell}\bm{b}_{j-m-\ell}$, namely, $\bm{a}'\triangleq \bm{a}_i c_1\neq \bm{a}$ is obtained from $\bm{a}c_1$ by deleting a symbol from its last $\overline{c}_1$-run and $\bm{b}'= \overline{c}_{\ell}\bm{b}_{j-m-\ell}\neq \bm{b}$ is obtained from $\overline{c}_{\ell}\bm{b}$ by deleting a symbol from its first $c_{\ell}$-run.
    Since $\bm{a}'\neq \bm{a}$ and $\bm{b}'\neq \bm{b}$, it can be easily checked that $\bm{a'}\bm{c}_{1}\bm{b'} \not\in \mathcal{S}(\bm{a}\bm{c}_{1} \bm{b}) \cup \mathcal{S}(\bm{a}\overline{\bm{c}}_{1} \bm{b})=S$.
    Consequently, we obtain 
        \begin{align*}
          \left( \bigcup_{i=1}^{m}\bigcup_{j= m+\ell+1}^{n} \mathcal{S}(\bm{x}_i, \bm{y}_j) \right)\setminus S=\{\bm{a'}\bm{c}_{1}\bm{b'}\}.
        \end{align*}
    \end{itemize}
In summary, when $i\le m$, we have 
\begin{align*}
  \left| \left( \bigcup_{i=1}^{m}\bigcup_{j= 1}^{n} \mathcal{S}(\bm{x}_i, \bm{y}_j)\right)\setminus S\right|\le 2.  
\end{align*}
    
\item If $i\ge m+\ell+1$, by considering the reversal of $\bm{a}$ and $\bm{b}$ and using the conclusion of the previous case, we can obtain that
\begin{align*}
\left| \left( \bigcup_{i=m+\ell+1}^{n}\bigcup_{j=1}^{n} \mathcal{S}(\bm{x}_i, \bm{y}_j) \right) \setminus S\right|\le 2.  
\end{align*}

\item If $m+2\le i\le m+\ell-1$ and $j\le m$, by swapping $\bm{x}$ and $\bm{y}$ and using the conclusion of the case where $i\le m$ and $m+2\le j\le m+\ell-1$, we can conclude that
\begin{align*}
\left| \left( \bigcup_{i=m+2}^{m+\ell-1}\bigcup_{j= 1}^{m} \mathcal{S}(\bm{x}_i, \bm{y}_j)\right) \setminus S\right|\le 1. 
\end{align*}  

\item If $m+2\le i\le m+\ell-1$ and $j\ge m+\ell+1$, by considering the reversal of $\bm{x}$ and $\bm{y}$ and using the conclusion of the previous case, we can obtain that 
 \begin{align*}
\left|\left( \bigcup_{i=m+2}^{m+\ell-1}\bigcup_{j= m+\ell+1}^{n} \mathcal{S}(\bm{x}_i, \bm{y}_j) \right)\setminus S\right|\le 1. 
\end{align*}

\item If $m+2\le i,j\le m+\ell-1$, let $\bm{c}'=\bm{c}_{[2,\ell-1]}$ be an alternating sequence of length $\ell-2\geq 1$, we have $\bm{x}= \bm{a}c_1 \bm{c}_{i-m-1}' c_{\ell} \bm{b}$ and $\bm{y}= \bm{a}\overline{c}_1\overline{\bm{c}}_{j-m-1}' \overline{c}_{\ell} \bm{b}$.
We can compute $d_H(\bm{x}_i,\bm{y}_j)= d_H( \bm{c}_{i-m-1}', \overline{\bm{c}}_{j-m-1}')+2$. 
It then suffices to consider the scenario where $d_H(\bm{x}_i,\bm{y}_j)=2$, or equivalently $\bm{c}_{i-m-1}'=\overline{\bm{c}}_{j-m-1}'$, as in the other cases, the intersection $\mathcal{S}(\bm{x}_i,\bm{y}_j)$ is empty.
In this case, by Lemma \ref{lem:del}, we get $\bm{c}_{i-m-1}'\in \{\bm{c}_{[3,\ell-1]}, \bm{c}_{[2,\ell-2]}\}$.
Further by Lemma \ref{lem:sub}, we obtain
\begin{align*}
    \bigcup_{i=m+2}^{m+\ell-1}\bigcup_{j=m+2}^{m+\ell-1} \mathcal{S}(\bm{x}_i,\bm{y}_j)
    &=\big\{\bm{a}c_1 \bm{c}_{[3,\ell-1]} \overline{c}_{\ell} \bm{b},~\bm{a} \overline{c}_1 \bm{c}_{[3,\ell-1]} c_{\ell} \bm{b},~\bm{a}c_1 \bm{c}_{[2,\ell-2]} \overline{c}_{\ell} \bm{b},~\bm{a} \overline{c}_1 \bm{c}_{[2,\ell-2]} c_{\ell}\bm{b} \big\}\\
    &=\big\{\bm{a}c_1 \bm{c}_{[3,\ell-1]} \overline{c}_{\ell} \bm{b},~\bm{a} \bm{c}_1 \bm{b},~\bm{a} \bm{c}_{\ell} \bm{b},~\bm{a} \overline{c}_1 \bm{c}_{[2,\ell-2]} c_{\ell}\bm{b} \big\}.
\end{align*}
Since
$\bm{a} \bm{c}_1 \bm{b} \in \mathcal{S}(\bm{a} \bm{c}_1 \bm{b})\subseteq S$ and $\bm{a} \bm{c}_{\ell} \bm{b}\in \mathcal{S}(\bm{a} \bm{c}_{\ell} \bm{b})\subseteq S$, we derive
\begin{align*}
    \left|\left(\bigcup_{i=m+2}^{m+\ell-1}\bigcup_{j=m+2}^{m+\ell-1} \mathcal{S}(\bm{x}_i,\bm{y}_j) \right)\setminus S\right|\leq 2.
\end{align*}
\end{itemize}
Therefore, we can conclude that \(|\mathcal{B}(\bm{x},\bm{y}) \setminus S|\le 8\), which completes the proof.
\end{IEEEproof}

Now, combining Equation \ref{eq:(2,0)_B} with Claims \ref{cla:(2,0)_S} and \ref{cla:(2,0)_B}, we have $|\mathcal{B}(\bm{x},\bm{y})|=|S|+|B|\leq 2n+8$.
This completes the proof.
    
\subsection{Proof of Lemma \ref{lem:(0,2)}}\label{subsec:(0,2)}

The proof follows a strategy similar to that of Lemma \ref{lem:(2,2)}, we include it here for completeness. 
We first characterize the structures of $\bm{x}$ and $\bm{y}$ when $|\mathcal{D}(\bm{x},\bm{y})|=0$ and $|\mathcal{S}(\bm{x},\bm{y})|=2$.
Since $|\mathcal{D}(\bm{x},\bm{y})|=0$, by Lemma \ref{lem:del}, we have $d_H(\bm{x},\bm{y})\geq 2$.
Moreover, since $|\mathcal{S}(\bm{x},\bm{y})|=2$, we can conclude that $d_H(\bm{x},\bm{y})= 2$.
As a result, we can express $\bm{x}=\bm{a}\alpha \bm{v} \beta \bm{b}$ and $\bm{x}=\bm{a}\overline{\alpha} \bm{v} \overline{\beta} \bm{b}$, where $\alpha,\beta \in \Sigma$ and $\bm{v}\in \Sigma^{\ast}$, such that $\bm{v}\beta \neq \overline{\alpha} \bm{v}$ and $\alpha \bm{v}\neq \bm{v}\overline{\beta}$.
In this case, by Lemmas \ref{lem:sub} and \ref{lem:del'}, we can derive 
\begin{gather*}
    \mathcal{S}(\bm{x},\bm{y})= \{\bm{a}\overline{\alpha} \bm{v}\beta \bm{b}, \bm{a}\alpha \bm{v}\overline{\beta} \bm{b}\},\quad
    \mathcal{D}(\bm{x},\bm{y})= \emptyset.
\end{gather*}
This implies that
\begin{equation}\label{eq:(0,2)}
\begin{gathered}
  D= \mathcal{D}(\bm{a}\overline{\alpha} \bm{v}\beta \bm{b}) \cup \mathcal{D}(\bm{a}\alpha \bm{v}\overline{\beta} \bm{b}), \quad S= \emptyset.
\end{gathered}
\end{equation}
Then by Equation (\ref{eq:B}), we can calculate 
\begin{equation}\label{eq:(0,2)_B}
    |\mathcal{B}(\bm{x},\bm{y})|= |B|+|D|,
\end{equation}
where $B \triangleq \mathcal{B}(\bm{x}, \bm{y}) \setminus D$.
In what follows, we will calculate the sizes of \(D\) and \(B\) respectively, which will then allow us to determine the size of \(\mathcal{B}(\bm{x},\bm{y})\) based on Equation (\ref{eq:(0,2)_B}).

\begin{claim}\label{cla:(0,2)_D}
    Let $D$ be defined in Equation (\ref{eq:(0,2)}), we have $|D|\leq 2n$ and $|D|\leq r(\bm{x})+r(\bm{y})+4$.
\end{claim}

\begin{IEEEproof}
By Lemma \ref{lem:size}, we have $|\mathcal{D}(\bm{a} \overline{\alpha} \bm{v}\beta \bm{b})|= r(\bm{a} \overline{\alpha} \bm{v}\beta \bm{b})\leq n$ and $|\mathcal{D}(\bm{a} \alpha\bm{v}\overline{\beta} \bm{b})|=r(\bm{a} \alpha\bm{v}\overline{\beta} \bm{b})\leq n$.
Then we get $|D|\leq 2n$.
Moreover, observe that $r(\bm{a} \overline{\alpha}\bm{v}\beta \bm{b})\leq r(\bm{a} \alpha \bm{v}\beta \bm{b}) +2= r(\bm{x})+2$ and $r(\bm{a} \alpha\bm{v}\overline{\beta} \bm{b})\leq r(\bm{a} \overline{\alpha} \bm{v}\overline{\beta} \bm{b})+2= r(\bm{y})+2$, we get $|D|\leq r(\bm{x})+r(\bm{y})+4$.
\end{IEEEproof}


\begin{claim}\label{cla:(0,2)_B}
    Let $D$ be defined in Equation (\ref{eq:(0,2)}), we have $|B|= |\mathcal{B}(\bm{x}, \bm{y}) \setminus D|\leq 4$.
\end{claim}

\begin{IEEEproof}
Recall that $\bm{x}=\bm{a}\alpha \bm{v} \beta \bm{b}$ and $\bm{x}=\bm{a}\overline{\alpha} \bm{v} \overline{\beta} \bm{b}$, where $\alpha,\beta \in \Sigma$, $\bm{v}\in \Sigma^{\ast}$, $\bm{v}\beta \neq \overline{\alpha} \bm{v}$, and $\alpha \bm{v}\neq \bm{v}\overline{\beta}$.
Let $\bm{x}_i\triangleq \bm{x}_{[n]\setminus \{i\}}$ and $\bm{y}_i\triangleq \bm{y}_{[n]\setminus \{i\}}$ for $i\in [n]$, we have 
\begin{align*}
    \mathcal{B}(\bm{x},\bm{y})
    &= \bigcup_{i=1}^{n} \bigcup_{j=1}^{n} \mathcal{S}(\bm{x}_i, \bm{y}_j).
\end{align*}
We divide our proof into the following four cases. Let $m\triangleq |\bm{a}|$ and $\ell \triangleq |\bm{v}|$.
\begin{itemize}
\item If $i,j\leq m+\ell+1$, we have $\bm{x}= (\bm{a}\alpha \bm{v})_{i} \beta \bm{b}$ and $\bm{y}= (\bm{a}\overline{\alpha} \bm{v})_{j} \overline{\beta} \bm{b}$.
For any $\bm{z}\in \mathcal{S}(\bm{x}_i, \bm{y}_j)$, we consider its $(m+\ell+1)$-th entry $z_{m+\ell+1}$.
If $z_{m+\ell+1}=\beta$, since $\bm{z}\in \mathcal{S}(\bm{y}_j)$, we can conclude that $\bm{z}= (\bm{a} \overline{\alpha} \bm{v})_j \beta \bm{b}\in \mathcal{D}(\bm{a} \overline{\alpha} \bm{v} \beta \bm{b})$.
Similarly, if $z_{m+\ell+1}=\overline{\beta}$, since $\bm{z}\in \mathcal{S}(\bm{x}_i)$, we can conclude that $\bm{z}= (\bm{a}\alpha \bm{v})_{i} \overline{\beta} \bm{b}\in \mathcal{D}(\bm{a} \alpha \bm{v} \overline{\beta} \bm{b})$.
Then we can conclude that
\begin{align*}
\mathcal{S}(\bm{x}_i, \bm{y}_j)
\subseteq \mathcal{D}(\bm{a}\overline{\alpha} \bm{v}\beta \bm{b}) \cup \mathcal{D}(\bm{a} \alpha \bm{v} \overline{\beta} \bm{b})=D.
\end{align*}

\item If $i,j\geq m+2$, by considering the reversal of $\bm{x}$ and $\bm{y}$ and using the conclusion of the previous one, we can conclude that
\begin{align*}
\mathcal{S}(\bm{x}_i, \bm{y}_j)
\subseteq \mathcal{D}(\bm{a}\overline{\alpha} \bm{v}\beta \bm{b}) \cup \mathcal{D}(\bm{a} \alpha \bm{v} \overline{\beta} \bm{b})=D.
\end{align*}

\item If $i\le m+1$ and $j\ge m+\ell+2$, we have $\bm{x}= (\bm{a}\alpha)_{i}  \bm{v}\beta \bm{b}$ and $\bm{y}= \bm{a}\overline{\alpha} \bm{v} (\overline{\beta} \bm{b})_{j-m-\ell-1}$.
We can compute $d_H(\bm{x}_i,\bm{y}_{j})= d_H((\bm{a}\alpha)_i,\bm{a})+d_H(\bm{v}\beta,\overline{\alpha}\bm{v})+d_H(\bm{b},(\overline{\beta}\bm{b})_{j-m-\ell-1})$. 
Recall that $\overline{\alpha}\bm{v}\neq \bm{v}\beta$, we have $d_H(\bm{v}\beta,\overline{\alpha}\bm{v})\geq 1$.
\begin{itemize}
    \item If $d_H(\bm{v}\beta,\overline{\alpha}\bm{v})\geq 3$, we have $d_H(\bm{x}_i,\bm{y}_{j})\geq 3$. In this case, we get 
    \begin{align*}
        \bigcup_{i=1}^{m+1}\bigcup_{j= m+\ell+2}^{n} \mathcal{S}(\bm{x}_i, \bm{y}_j)=\emptyset.
    \end{align*}

    \item If $d_H(\bm{v}\beta,\overline{\alpha}\bm{v})=2$, it suffices to consider the scenario where $d_H(\bm{x}_i,\bm{y}_j)=2$, or equivalently $(\bm{a}\alpha)_i=\bm{a}$ and $\bm{b}=(\overline{\beta}\bm{b})_{j-m-\ell-1}$, as in the other cases, the intersection $\mathcal{S}(\bm{x}_i, \bm{y}_j)$ is empty. In this case, we have
    \begin{align*}
        \bigcup_{i=1}^{m+1}\bigcup_{j= m+\ell+2}^{n} \mathcal{S}(\bm{x}_i, \bm{y}_j)= \left\{\bm{a}\bm{u}\bm{b}:\bm{u}\in \mathcal{S}(\bm{v}\beta,\overline{\alpha}\bm{v}) \right\}.
    \end{align*}
    Then by Lemma \ref{lem:sub}, we get 
    \begin{align*}
        \left| \bigcup_{i=1}^{m+1}\bigcup_{j= m+\ell+2}^{n} \mathcal{S}(\bm{x}_i, \bm{y}_j)\right|= 2.
    \end{align*}

    \item If $d_H(\bm{v}\beta, \overline{\alpha}\bm{v})=1$, it suffices to consider the scenario where $d_H(\bm{x}_i,\bm{y}_j)\leq 2$, or equivalently $d_H((\bm{a}\alpha)_i,\bm{a}) +d_H(\bm{b},(\overline{\beta}\bm{b})_{j-m-\ell-1})\leq 1$, as in the other cases, the intersection $\mathcal{S}(\bm{x}_i, \bm{y}_j)$ is empty.    
    As a result, we need to further analyze the following three cases.
        \begin{itemize}
        \item If $d_H((\bm{a}\alpha)_i,\bm{a})=d_H(\bm{b},(\overline{\beta}\bm{b})_{j-m-\ell-1})=0$, then $(\bm{a}\alpha)_i=\bm{a}$ and $\bm{b}=(\overline{\beta}\bm{b})_{j-m-\ell-1}$. 
        We can compute $d_H(\bm{x}_i,\bm{y}_j)=d_H(\bm{v}\beta,\overline{\alpha}\bm{v})=1$. By Lemma \ref{lem:sub}, we get
        \[\mathcal{S}(\bm{x}_i, \bm{y}_j)=\{\bm{a}\overline{\alpha}\bm{v}\bm{b},\bm{a}\bm{v}\beta\bm{b}\}\subseteq \mathcal{D}(\bm{a}\overline{\alpha}\bm{v}\beta \bm{b})\subseteq D.
        \]
        \item If $d_H((\bm{a}\alpha)_i,\bm{a})=1$ and $d_H(\bm{b},(\overline{\beta}\bm{b})_{j-m-\ell-1})=0$, we have $d_H(\bm{x}_i,\bm{y}_j)=d_H((\bm{a}\alpha)_i,\bm{a})+d_H(\bm{v}\beta,\overline{\alpha}\bm{v})=2$. 
        Consider the sequence $\bm{a}\alpha$, since $d_H((\bm{a}\alpha)_i,\bm{a})=1$, it follows by Lemma \ref{lem:del_position} that $\bm{a}'\triangleq(\bm{a}\alpha)_i\neq \bm{a}$ is obtained from $\bm{a}\alpha$ by deleting a symbol of its last $\overline{\alpha}$-run.
        Then by Lemma \ref{lem:sub}, we get
        \[\mathcal{S}(\bm{x}_i, \bm{y}_j)=\{\bm{a}'\overline{\alpha}\bm{v}\bm{b},\bm{a}\bm{v}\beta\bm{b}\}.
        \]
        \item  If $d_H((\bm{a}\alpha)_i,\bm{a})=0$ and $d_H(\bm{b},(\overline{\beta}\bm{b})_{j-m-\ell-1})=1$, we have $d_H(\bm{x}_i,\bm{y}_j)=d_H(\bm{v}\beta,\overline{\alpha}\bm{v})+d_H((\overline{\beta}\bm{b})_{j-m-\ell-1},\bm{b})=2$. 
        Consider the sequence $\overline{\beta}\bm{b}$, since $d_H((\overline{\beta}\bm{b})_{j-m-\ell-1},\bm{b})=1$, it follows by Lemma \ref{lem:del_position} that  $\bm{b}'\triangleq(\overline{\beta}\bm{b})_{j-m-\ell-1}\neq \bm{b}$ is obtained from $\overline{\beta}\bm{b}$ by deleting a symbol of its first $\beta$-run.
        Then we get
        \[\mathcal{S}(\bm{x}_i, \bm{y}_j)=\{\bm{a}\bm{v}\beta\bm{b}',\bm{a}\overline{\alpha}\bm{v}\bm{b}\}.
        \]
        \end{itemize}     
        Observe that $\bm{av}\beta \bm{b}, \bm{a}\overline{\alpha} \bm{v} \bm{b}\in \mathcal{D}(\bm{a}\overline{\alpha} \bm{v} \beta \bm{b})$, we get
        \begin{align*}
        \left|\left( \bigcup_{i=1}^{m+1}\bigcup_{j= m+\ell+2}^{n}  \mathcal{S}(\bm{x}_i, \bm{y}_j) \right)\setminus D \right|\le 2.
        \end{align*}
    \end{itemize}
In all cases, we can conclude that
\begin{align*}
\left|\left( \bigcup_{i=1}^{m+1}\bigcup_{j= m+\ell+2}^{n}  \mathcal{S}(\bm{x}_i, \bm{y}_j) \right)\setminus D \right|\le 2.
\end{align*}

\item If $i\ge m+\ell+2$ and $j\le m+1$, by considering the reversal of $\bm{x}$ and $\bm{y}$ and using the conclusion of the previous one, we can conclude that
\begin{align*}
\left| \left(\bigcup_{i=m+\ell+2}^{n}\bigcup_{j= 1}^{m+1} \mathcal{S}(\bm{x}_i, \bm{y}_j) \right) \setminus D\right|\le 2.
\end{align*}
\end{itemize}
Therefore, we can conclude that $|B|= |\mathcal{B}(\bm{x},\bm{y})\setminus D|\le 4$, which completes the proof.
\end{IEEEproof}
    
Now, combining Equation (\ref{eq:(0,2)_B}) with Claims \ref{cla:(0,2)_D} and \ref{cla:(0,2)_B}, we have $|\mathcal{B}(\bm{x},\bm{y})|=|D|+|B|\leq \min\{2n+4, r(\bm{x})+r(\bm{y})+8\}$.
This completes the proof.

\subsection{Proof of Lemma \ref{lem:(1,0)}}\label{subsec:(1,0)}  

The proof follows a strategy similar to that of Lemma \ref{lem:(2,2)}, we include it here for completeness. 
We first characterize the structures of $\bm{x}$ and $\bm{y}$ when $|\mathcal{D}(\bm{x},\bm{y})|=1$ and $|\mathcal{S}(\bm{x},\bm{y})|=0$.
Since $|\mathcal{S}(\bm{x},\bm{y})|=0$, by Lemma \ref{lem:sub}, we have $d_H(\bm{x},\bm{y})\geq 3$. 
Moreover, since $|\mathcal{D}(\bm{x},\bm{y})|=1$, by Lemma \ref{lem:del}, we can express $\{\bm{x}, \bm{y}\}= \big\{\bm{a}\overline{\alpha}\alpha  \bm{c}\beta \bm{b}, \bm{a}\alpha \bm{c}\beta  \overline{\beta} \bm{b}\big\}$ with $\overline{\alpha}\alpha  \bm{c}\neq \bm{c}\beta  \overline{\beta}$ and $\alpha\bm{c}\neq \bm{c}\beta$ for some $\alpha, \beta\in \Sigma$ and $\bm{c}\in \Sigma^{\ast}$.
In this case, by Lemmas \ref{lem:sub} and \ref{lem:del'}, we can derive 
\begin{gather*}
    \mathcal{S}(\bm{x},\bm{y})= \emptyset,\quad
    \mathcal{D}(\bm{x},\bm{y})= \{\bm{a}\alpha\bm{c}\beta \bm{b}\}.
\end{gather*}
This implies that
\begin{equation}\label{eq:(1,0)}
\begin{gathered}
  D= \emptyset,\quad
  S= \mathcal{S}(\bm{a}\alpha\bm{c}\beta \bm{b}).
\end{gathered}
\end{equation}
Then by Equation (\ref{eq:B}), we can calculate 
\begin{equation}\label{eq:(1,0)_B}
    |\mathcal{B}(\bm{x},\bm{y})|= |B|+|S|,
\end{equation}
where $B \triangleq \mathcal{B}(\bm{x}, \bm{y}) \setminus S$.
In what follows, we will calculate the sizes of \(S\) and \(B\) respectively, which will then allow us to determine the size of \(\mathcal{B}(\bm{x},\bm{y})\) based on Equation (\ref{eq:(1,0)_B}).

\begin{claim}\label{cla:(1,0)_S}
    Let $S$ be defined in (\ref{eq:(1,0)}), we have $|S|=n$.
\end{claim}

\begin{IEEEproof}
    The conclusion follows by Lemma \ref{lem:size} directly.
\end{IEEEproof}

\begin{claim}\label{cla:(1,0)_B}
    Let $S$ be defined in Equation (\ref{eq:(1,0)}), we have $|B|= |\mathcal{B}(\bm{x}, \bm{y}) \setminus S|\leq 20$.
\end{claim}

\begin{IEEEproof}
    Recall that $\{\bm{x}, \bm{y}\}= \big\{\bm{a}\overline{\alpha}\alpha  \bm{c}\beta \bm{b}, \bm{a}\alpha \bm{c}\beta  \overline{\beta} \bm{b}\big\}$ with $\overline{\alpha}\alpha  \bm{c}\neq \bm{c}\beta  \overline{\beta}$ and $\alpha\bm{c}\neq \bm{c}\beta$ for some $\alpha, \beta\in \Sigma$ and $\bm{c}\in \Sigma^{\ast}$. 
    Without loss of generality, we assume $\bm{x}=\bm{a}\overline{\alpha}\alpha  \bm{c}\beta \bm{b}$ and $\bm{y}=\bm{a}\alpha \bm{c}\beta  \overline{\beta} \bm{b}$.
    Let $\bm{x}_i\triangleq \bm{x}_{[n]\setminus\{i\}}$ and $\bm{y}_i\triangleq \bm{y}_{[n]\setminus\{i\}}$ for $i\in [n]$, we have  
    \begin{align*}
        \mathcal{B}(\bm{x},\bm{y})
        &= \bigcup_{i=1}^{n} \bigcup_{j=1}^{n} \mathcal{S}(\bm{x}_i, \bm{y}_j).
    \end{align*}
    We first determine the intersection $\mathcal{S}(\bm{x}_i, \bm{y}_j)$. Let $\ell\triangleq |\bm{c}|$.
    \begin{itemize}
        \item If $i=m+1$ or $j=m+\ell+3$, we have
        \begin{align*}
            \mathcal{S}(\bm{x}_i, \bm{y}_j)\subseteq \mathcal{S}(\bm{a}\alpha \bm{c}\beta  \bm{b})=S.
        \end{align*}

        \item If $i=m+\ell+3$, we have $\bm{x}_i=\bm{a}\overline{\alpha} \alpha \bm{c}\bm{b}$. We further consider the choice for $j$. 
        \begin{itemize}
            \item If $j\geq m+\ell+4$, we have $\bm{y}_j= \bm{a}\alpha\bm{c}\beta\overline{\beta} \bm{b}_{j-m-\ell-3}$. Since we have discussed $j=m+\ell+3$ in the first case, we assume that $\bm{y}_j\neq \bm{y}_{m+\ell+3}$, i.e., $\overline{\beta}\bm{b}_{j-m-\ell-3}\neq \bm{b}$.
            Recall that $\alpha\bm{c}\neq\bm{c}\beta$, we can compute $d_H(\bm{x}_i,\bm{y}_j)= d_H(\bm{a}\overline{\alpha},\bm{a}\alpha)+ d_H(\alpha\bm{c},\bm{c}\beta)+d_H(\bm{b},\overline{\beta}\bm{b}_{j-m-\ell-3})\geq 3$. 
            In this case, we have
            \begin{align*}
                \mathcal{S}(\bm{x}_i, \bm{y}_j)=\emptyset.
            \end{align*}

            \item If $m+2\leq j \leq m+\ell+2$, we have $\bm{y}_j=\bm{a}\alpha (\bm{c}\beta)_{j-m-1} \overline{\beta} \bm{b}$.
            We can compute $d_H(\bm{x}_i,\bm{y}_j)= d_H(\bm{a}\overline{\alpha},\bm{a}\alpha)+d_H(\alpha\bm{c}, (\bm{c}\beta)_{j-m-1}\overline{\beta})$.
            It then suffices to consider the scenario where $d_H(\alpha\bm{c}, (\bm{c}\beta)_{j-m-1}\overline{\beta})\leq 1$, as in the other cases, the intersection $\mathcal{S}(\bm{x}_i, \bm{y}_j)$ is empty.
            Now, by Lemma \ref{lem:del_Hamming}, we can conclude that there are at most three choices for $(\bm{c}\beta)_{j-m-1}\overline{\beta}$, which, in the worst case, we denote these sequences as $\alpha\bm{c}, \bm{v}^{(1)}, \bm{v}^{(2)}$.
            In this case, by Lemma \ref{lem:sub}, we have
            \begin{align*}
                \bigcup_{j=m+2}^{m+\ell+2} \mathcal{S}(\bm{x}_{m+\ell+3},\bm{y}_j) =\{\bm{a} \overline{\alpha} \alpha\bm{c}\bm{b},~\bm{a} \overline{\alpha} \bm{v}^{(1)}\bm{b},~\bm{a} \overline{\alpha} \bm{v}^{(2)}\bm{b},~\bm{a} \alpha \alpha \bm{c} \bm{b}\}.
            \end{align*}

            \item If $j \leq m+1$, we have $\bm{y}_j=(\bm{a}\alpha)_j \bm{c}\beta\overline{\beta} \bm{b}$.
            We can compute $d_H(\bm{x}_i,\bm{y}_j)= d_H(\bm{a},(\bm{a}\alpha)_j)+d_H(\overline{\alpha}\alpha\bm{c}, \bm{c}\beta\overline{\beta})$.
            Recall that $\overline{\alpha}\alpha\bm{c}\neq \bm{c}\beta\overline{\beta}$.
            \begin{itemize}
                \item If $d_H(\overline{\alpha}\alpha\bm{c}, \bm{c}\beta\overline{\beta})\geq 3$, then the intersection $\mathcal{S}(\bm{x}_i, \bm{y}_j)$ is empty.

                \item If $d_H(\overline{\alpha}\alpha\bm{c}, \bm{c}\beta\overline{\beta})=2$, it suffices to consider the scenario where $d_H(\bm{a},(\bm{a}\alpha)_j)=0$, or equivalently $(\bm{a}\alpha)_j=\bm{a}$, as in the other cases, the intersection $\mathcal{S}(\bm{x}_i, \bm{y}_j)$ is empty. In this case, we have
                \begin{align*}
                    \bigcup_{j=1}^{m+1} \mathcal{S}(\bm{x}_{m+\ell+3}, \bm{y}_j)= \left\{\bm{a}\bm{u}\bm{b}:\bm{u}\in \mathcal{S}(\overline{\alpha}\alpha\bm{c}, \bm{c}\beta\overline{\beta}) \right\}.
                \end{align*}
                Then by Lemma \ref{lem:sub}, we get 
                \begin{align*}
                    \left| \bigcup_{j=1}^{m+1} \mathcal{S}(\bm{x}_{m+\ell+3}, \bm{y}_j)\right|= 2.
                \end{align*}   

                \item If $d_H(\overline{\alpha}\alpha\bm{c}, \bm{c}\beta\overline{\beta})=1$, it suffices to consider the scenario where $d_H(\bm{a},(\bm{a}\alpha)_j)\leq 1$, as in the other cases, the intersection $\mathcal{S}(\bm{x}_i, \bm{y}_j)$ is empty.
                By Lemma \ref{lem:del_position}, we can conclude that there are at most two choices for $(\bm{a}\alpha)_{j}$, which, in the worst case, we can denote these sequences as $\bm{a}, \bm{a}'\neq \bm{a}$.
                In this case, by Lemma \ref{lem:sub}, we get
                \begin{align*}
                \bigcup_{j=1}^{m+1} \mathcal{S}(\bm{x}_{m+\ell+3},\bm{y}_j) =\{\bm{a} \bm{c}\beta\overline{\beta} \bm{b},~\bm{a}\overline{\alpha}\alpha\bm{c}\bm{b},~\bm{a}'\overline{\alpha}\alpha\bm{c}\bm{b}\}.
                \end{align*}
                Note that $\bm{a}\overline{\alpha}\alpha\bm{c}\bm{b}\in \bigcup_{j=m+2}^{m+\ell+2} \mathcal{S}(\bm{x}_{m+\ell+3},\bm{y}_j)$, we get 
                \begin{align*}
                \left|\left(\bigcup_{j=1}^{m+1} \mathcal{S}(\bm{x}_{m+\ell+3},\bm{y}_j)\right)\setminus \left(\bigcup_{j=m+2}^{m+\ell+2} \mathcal{S}(\bm{x}_{m+\ell+3},\bm{y}_j)\right)\right|\leq 2.
                \end{align*}
            \end{itemize}
        In all cases, we can conclude that   
        \begin{align*}
           \left|\left(\bigcup_{j=1}^{m+1} \mathcal{S}(\bm{x}_{m+\ell+3},\bm{y}_j)\right)\setminus \left(\bigcup_{j=m+2}^{m+\ell+2} \mathcal{S}(\bm{x}_{m+\ell+3},\bm{y}_j)\right)\right|\leq 2.
        \end{align*}
    \end{itemize}
    Consequently, we can conclude that
    \begin{align*}
        \left|\bigcup_{j=1}^{n} \mathcal{S}(\bm{x}_{m+\ell+3},\bm{y}_j)\right|\leq 6.
    \end{align*}

    \item If $i\geq m+\ell+4$, we have $\bm{x}_i= \bm{a} \overline{\alpha} \alpha \bm{c}\beta \bm{b}_{i-m-\ell-3}$.
    Since we have discussed $i=m+\ell+3$ in the previous case, we assume that $\bm{x}_i\neq \bm{x}_{m+\ell+3}$, i.e., $\beta\bm{b}_{i-m-\ell-3}\neq \bm{b}$.
    We further consider the choice for $j$.
    \begin{itemize}
        \item If $j\geq m+\ell+4$, we have $\bm{y}_j= \bm{a} \alpha \bm{c}\beta \overline{\beta} \bm{b}_{j-m-\ell-3}$. We can compute $d_H(\bm{x}_i,\bm{y}_j)\geq d_H(\bm{a}\overline{\alpha}\alpha\bm{c}\beta, \bm{a}\alpha\bm{c}\beta\overline{\beta})= d_H(\bm{x},\bm{y})\geq 3$.
        In this case, we have
        \begin{align*}
            \mathcal{S}(\bm{x}_i, \bm{y}_j)=\emptyset.
        \end{align*}
        
        \item If $j\leq m+1$, we have $\bm{y}_j= (\bm{a} \alpha)_j \bm{c}\beta \overline{\beta} \bm{b}$.
        Recall that $\overline{\alpha}\alpha\bm{c}\neq \bm{c}\beta\overline{\beta}$ and $\beta \bm{b}_{i-m-\ell-3}\neq \bm{b}$.
        We can compute $d_H(\bm{x}_i,\bm{y}_j)= d_H(\bm{a},(\bm{a} \alpha)_j)+ d_H(\overline{\alpha}\alpha\bm{c}, \bm{c}\beta\overline{\beta})+d_H(\beta \bm{b}_{i-m-\ell-3},\bm{b})\geq 2$.
        It then suffices to consider the scenario where $d_H(\bm{x}_i,\bm{y}_j)=2$, or equivalently, $\bm{a}=(\bm{a} \alpha)_j$ and $d_H(\overline{\alpha}\alpha\bm{c}, \bm{c}\beta\overline{\beta})=d_H(\beta \bm{b}_{i-m-\ell-3},\bm{b})=1$, as in the other cases, the intersection $\mathcal{S}(\bm{x}_i, \bm{y}_j)$ is empty.
        By Lemma \ref{lem:del_position}, we can conclude that there is at most one choice for $\beta \bm{b}_{i-m-\ell-3}$, which, in the worst case, we denote this sequence as $\bm{b}'$. 
        In this case, by Lemma \ref{lem:sub}, we get
        \begin{align*}
            \bigcup_{i=m+\ell+4}^{n} \bigcup_{j=1}^{m+1} \mathcal{S}(\bm{x}_i, \bm{y}_j)=\{\bm{a}\overline{\alpha}\alpha\bm{c}\bm{b},~ \bm{a}\bm{c}\beta\overline{\beta}\bm{b}'\}.
        \end{align*}

        \item If $m+2\leq j\leq m+\ell+2$, we have $\bm{y}_j= \bm{a}\alpha  (\bm{c}\beta)_{j-m-1} \overline{\beta} \bm{b}$. 
        We can compute $d_H(\bm{x}_i,\bm{y}_j)= d_H(\bm{a}\overline{\alpha},\bm{a}\alpha)+ d_H(\alpha\bm{c}, (\bm{c}\beta)_{j-m-1}\overline{\beta})+d_H(\beta\bm{b}_{i-m-\ell-3},\bm{b})\geq 2$. 
        It then suffices to consider the scenario where $d_H(\bm{x}_i,\bm{y}_j)=2$, or equivalently $(\bm{c}\beta)_{j-m-1}\overline{\beta}=\alpha\bm{c}$ and $d_H(\beta\bm{b}_{i-m-\ell-3},\bm{b})=1$, as in the other cases, the intersection $\mathcal{S}(\bm{x}_i, \bm{y}_j)$ is empty.
        By Lemma \ref{lem:del_position}, we can conclude that there is at most one choice for $\beta \bm{b}_{i-m-\ell-3}$, which in the worst case, we denote this sequence as $\bm{b}'$.
        In this case, by Lemma \ref{lem:sub}, we have 
        \begin{align*}
            \bigcup_{i=m+\ell+4}^{n} \bigcup_{j=m+2}^{m+\ell+2} \mathcal{S}(\bm{x}_i, \bm{y}_j)= \{\bm{a}\overline{\alpha} \alpha \bm{c}\bm{b}, \bm{a}\alpha \alpha\bm{c}\bm{b}' \}.
        \end{align*}

    \end{itemize}
    Note that $\bm{a}\overline{\alpha}\alpha\bm{c}\bm{b}\in \bigcup_{j=m+2}^{m+\ell+2} \mathcal{S}(\bm{x}_{m+\ell+3},\bm{y}_j)$, we get 
    \begin{align*}
        \left|\left( \bigcup_{i=m+\ell+4}^{n} \bigcup_{j=1}^{n} \mathcal{S}(\bm{x}_i,\bm{y}_j)\right)\setminus \left(\bigcup_{j=m+2}^{m+\ell+2} \mathcal{S}(\bm{x}_{m+\ell+3},\bm{y}_j)\right)\right|\leq 2.
    \end{align*}

    \item If $m+2\leq i\leq m+\ell+2$, we have $\bm{x}_i= \bm{a}\overline{\alpha}(\alpha \bm{c})_{i-m-1} \beta \bm{b}$. Since we have discussed $i=m+\ell+3$ in the second case, we assume that $\bm{x}_i\neq \bm{x}_{m+\ell+3}$, i.e., $(\alpha \bm{c})_{i-m-1} \beta\neq \alpha \bm{c}$.
    We further consider the choice for $j$.
    \begin{itemize}
        \item If $j\geq m+\ell+4$, we have $\bm{y}_j= \bm{a}\alpha\bm{c}\beta\overline{\beta} \bm{b}_{j-m-\ell-3}$. Since we have discussed $j=m+\ell+3$ in the first case, we assume that $\bm{y}_j\neq \bm{y}_{m+\ell+3}$, i.e., $\overline{\beta}\bm{b}_{j-m-\ell-3}\neq \bm{b}$.
        We can compute $d_H(\bm{x}_i,\bm{y}_j)= d_H(\bm{a}\overline{\alpha},\bm{a}\alpha)+ d_H((\alpha \bm{c})_{i-m-1} \beta, \bm{c}\beta)+d_H(\bm{b},\overline{\beta}\bm{b}_{j-m-\ell-3})\geq 2$. 
        It then suffices to consider the scenario where $(\alpha \bm{c})_{i-m-1} \beta= \bm{c}\beta$ and $d_H(\bm{b},\overline{\beta}\bm{b}_{j-m-\ell-3})=1$, as in the other cases, the intersection $\mathcal{S}(\bm{x}_i, \bm{y}_j)$ is empty. 
        In this case, by Lemma \ref{lem:del_position}, we can conclude that there is at most one choice for $\overline{\beta}\bm{b}_{j-m-\ell-3}$, which, in the worst case, we denote this sequence as $\bm{b}''$.
        By Lemma \ref{lem:sub}, we have
        \begin{align*}
            \bigcup_{i=m+2}^{m+\ell+2} \bigcup_{j=m+\ell+4}^n \mathcal{S}(\bm{x}_{i},\bm{y}_j) =\{\bm{a} \overline{\alpha} \bm{c} \beta\bm{b}'',~\bm{a} \alpha \bm{c} \beta\bm{b}\}.
        \end{align*}
        Note that $\bm{a} \alpha \bm{c} \beta\bm{b}\in \mathcal{S}(\bm{a} \alpha \bm{c} \beta\bm{b})=S$, we get 
        \begin{align*}
            \left(\bigcup_{i=m+2}^{m+\ell+2} \bigcup_{j=m+\ell+4}^n \mathcal{S}(\bm{x}_{i},\bm{y}_j)\right) \setminus S =\{\bm{a} \overline{\alpha} \bm{c} \beta\bm{b}''\}.
        \end{align*}

        \item If $m+2\leq j \leq m+\ell+2$, we have $\bm{y}_j=\bm{a}\alpha (\bm{c}\beta)_{j-m-1} \overline{\beta} \bm{b}$.
        We can compute $d_H(\bm{x}_i,\bm{y}_j)= d_H(\bm{a}\overline{\alpha},\bm{a}\alpha)+d_H((\alpha\bm{c})_{i-m-1}, (\bm{c}\beta)_{j-m-1})+d_H(\beta\bm{b},\overline{\beta}\bm{b})\geq 2$.
        It then suffices to consider the scenario where $(\alpha\bm{c})_{i-m-1}=(\bm{c}\beta)_{j-m-1}$, as in the other cases, the intersection $\mathcal{S}(\bm{x}_i, \bm{y}_j)$ is empty.
        Recall that $\alpha\bm{c}\neq \bm{c}\beta$, by Lemma \ref{lem:del}, we can conclude that there are at most two choices for $(\alpha\bm{c})_{i-m-1}$, which, in the worst case, we denote these sequences as $\bm{u}^{(1)}, \bm{u}^{(2)}$.
        In this case, by Lemma \ref{lem:sub}, we have
        \begin{align*}
            \bigcup_{i=m+2}^{m+\ell+2} \bigcup_{j=m+2}^{m+\ell+2} \mathcal{S}(\bm{x}_{i},\bm{y}_j) =\{\bm{a} \overline{\alpha} \bm{u}^{(1)} \overline{\beta}\bm{b},~\bm{a} \alpha \bm{u}^{(1)} \beta\bm{b},~\bm{a} \overline{\alpha} \bm{u}^{(2)} \overline{\beta}\bm{b},~\bm{a} \alpha \bm{u}^{(2)} \beta\bm{b}\}.
        \end{align*}

        \item If $j \leq m+1$, we have $\bm{y}_j=(\bm{a}\alpha)_j \bm{c}\beta\overline{\beta} \bm{b}$.
        We can compute $d_H(\bm{x}_i,\bm{y}_j)= d_H(\bm{a},(\bm{a}\alpha)_j)+d_H(\overline{\alpha}(\alpha\bm{c})_{i-m-1}, \bm{c}\beta)+d_H(\beta\bm{b},\overline{\beta}\bm{b})\geq 1$.
        It then suffices to consider the scenario where $d_H(\bm{a},(\bm{a}\alpha)_j)+d_H(\overline{\alpha}(\alpha\bm{c})_{i-m-1}, \bm{c}\beta)\leq 1$, as in the other cases, the intersection $\mathcal{S}(\bm{x}_{i},\bm{y}_j)$ is empty.
        \begin{itemize}
            \item If $d_H(\bm{a},(\bm{a}\alpha)_j)\geq 2$, then the intersection $\mathcal{S}(\bm{x}_i, \bm{y}_j)$ is empty.

            \item If $d_H(\bm{a},(\bm{a}\alpha)_j)=1$, we need $d_H(\overline{\alpha}(\alpha\bm{c})_{i-m-1}, \bm{c}\beta)=0$, i.e., $\overline{\alpha}(\alpha\bm{c})_{i-m-1}=\bm{c}\beta$.
            In this case, by Lemma \ref{lem:del_position}, we can conclude that there is at most one choice for $(\bm{a}\alpha)_j$, which, in the worst case, we assume this sequence as $\bm{a}''$.
            By Lemma \ref{lem:sub}, we get
            \begin{align*}
            \mathcal{S}(\bm{x}_{i},\bm{y}_j) =\{\bm{a} \bm{c}\beta \overline{\beta}\bm{b},~\bm{a}'' \bm{c}\beta\beta \bm{b}\}.
            \end{align*}

            \item If $d_H(\bm{a},(\bm{a}\alpha)_j)=0$, i.e., $(\bm{a}\alpha)_j=\bm{a}$,  we need $d_H(\overline{\alpha}(\alpha\bm{c})_{i-m-1}, \bm{c}\beta)\leq 1$.
            By Lemma \ref{lem:del_position}, we can conclude that there are at most three choices for $\overline{\alpha}(\alpha\bm{c})_{i-m-1}$, which, in the worst case, we can denote these sequences as $\bm{c}\beta,\bm{w}^{(1)},\bm{w}^{(2)}$.
            In this case, by Lemma \ref{lem:sub}, we get
            \begin{align*}
            \mathcal{S}(\bm{x}_{i},\bm{y}_j) =\{\bm{a} \bm{c}\beta\beta\bm{b},~\bm{a}\bm{c}\beta\overline{\beta}\bm{b},~\bm{a}\bm{w}^{(1)}\overline{\beta}\bm{b},~\bm{a}\bm{w}^{(2)}\overline{\beta}\bm{b}\}.
            \end{align*}
        \end{itemize}
        In this case, we get
        \begin{align*}
            \left|\bigcup_{i=m+2}^{m+\ell+2} \bigcup_{j=1}^{m+1}  \mathcal{S}(\bm{x}_{i},\bm{y}_j)\right|\leq 5.
        \end{align*}
    \end{itemize}
    Consequently, we can conclude that
    \begin{align*}
        \left|\bigcup_{i=m+2}^{m+\ell+2} \bigcup_{j=1}^{n} \mathcal{S}(\bm{x}_{i},\bm{y}_j)\right|\leq 10.
    \end{align*}

    \item If $i\leq m$, we have $\bm{x}_i= \bm{a}_i \overline{\alpha} \alpha \bm{c}\beta \bm{b}$.
    Since we have discussed $i=m+1$ in the first case, we assume that $\bm{x}_i\neq \bm{x}_{m+1}$, i.e., $\bm{a}_i\overline{\alpha}\neq \bm{a}$.
    We further consider the choice for $j$.
    \begin{itemize}
        \item If $j\leq m$, we have $\bm{y}_j= \bm{a}_j \alpha \bm{c}\beta \overline{\beta} \bm{b}$.
        We can compute $d_H(\bm{x}_i,\bm{y}_j)\geq d_H(\overline{\alpha}\alpha\bm{c}\beta \bm{b}, \alpha\bm{c}\beta\overline{\beta} \bm{b})= d_H(\bm{x},\bm{y})\geq 3$.
        In this case, we have
        \begin{align*}
            \mathcal{S}(\bm{x}_i, \bm{y}_j)=\emptyset.
        \end{align*}

        \item If $m+1\leq j\leq m+\ell+2$, we have $\bm{y}_j= \bm{a} (\alpha \bm{c}\beta)_{j-m} \overline{\beta} \bm{b}$. 
        We can compute $d_H(\bm{x}_i,\bm{y}_j)= d_H(\bm{a}_i\overline{\alpha},\bm{a})+ d_H(\alpha\bm{c}, (\alpha\bm{c}\beta)_{j-m})+d_H(\beta\bm{b},\overline{\beta}\bm{b})\geq 2$. 
        It then suffices to consider the scenario where $d_H(\bm{x}_i,\bm{y}_j)=2$, or equivalently $d_H(\bm{a}_i\overline{\alpha},\bm{a})=1$ and $\alpha\bm{c}= (\alpha\bm{c}\beta)_{j-m}$, as in the other cases, the intersection $\mathcal{S}(\bm{x}_i, \bm{y}_j)$ is empty.
        In this case, by Lemma \ref{lem:sub}, we have 
        \begin{align*}
            \mathcal{S}(\bm{x}_i, \bm{y}_j)= \{\bm{a}_i\overline{\alpha} \alpha \bm{c}\overline{\beta}\bm{b}, \bm{a}\alpha \bm{c}\beta \bm{b} \}.
        \end{align*}
        We now consider the sequence $\bm{a}\overline{\alpha}$.
        Since $d_H(\bm{a}_i\overline{\alpha},\bm{a})=1$, it follows by Lemma \ref{lem:del_position} that there is exactly one choice for $\bm{a}_i \overline{\alpha}$, namely, $\bm{a}'\triangleq \bm{a}_i \overline{\alpha}\neq \bm{a}$ is obtained from $\bm{a}\overline{\alpha}$ by deleting a symbol of its last $\alpha$-run.
        It can be easily checked that $\bm{a}\alpha \bm{c}\beta \bm{b}\in \mathcal{S}(\bm{a}\alpha \bm{c}\beta \bm{b})=S$ and $\bm{a}' \alpha \bm{c}\overline{\beta}\bm{b}\notin \mathcal{S}(\bm{a}\alpha \bm{c}\beta \bm{b})=S$.
        Consequently, we obtain 
        \begin{align*}
            \left(\bigcup_{i=1}^m \bigcup_{j=m+1}^{m+\ell+2} \mathcal{S}(\bm{x}_i,\bm{y}_j)\right) \setminus S\subseteq \{\bm{a}' \alpha \bm{c}\overline{\beta}\bm{b}\}.
        \end{align*}

        \item If $j\geq m+\ell+4$, we have $\bm{y}_j= \bm{a} \alpha \bm{c}\beta \overline{\beta} \bm{b}_{j-m-\ell-3}$.
        Since we have discussed $j=m+\ell+3$ in the first case, we assume that $\bm{y}_j\neq \bm{y}_{m+\ell+3}$, i.e., $\overline{\beta}\bm{b}_{j-m-\ell-3}\neq \bm{b}$.
        We can compute $d_H(\bm{x}_i,\bm{y}_j)= d_H(\bm{a}_i\overline{\alpha},\bm{a})+ d_H(\bm{b},\overline{\beta}\bm{b}_{j-m-\ell-3})\geq 2$. 
        It then suffices to consider the scenario where $d_H(\bm{x}_i,\bm{y}_j)=2$, or equivalently $d_H(\bm{a}_i\overline{\alpha},\bm{a})= d_H(\bm{b},\overline{\beta}\bm{b}_{j-m-\ell-3})=1$, as in the other cases, the intersection $\mathcal{S}(\bm{x}_i, \bm{y}_j)$ is empty.
        In this case, by Lemma \ref{lem:sub}, we have 
        \begin{align*}
            \mathcal{S}(\bm{x}_i, \bm{y}_j)= \{\bm{a}_i\overline{\alpha} \alpha \bm{c}\beta \overline{\beta}\bm{b}_{j-m-\ell-3}, \bm{a}\alpha \bm{c}\beta \bm{b} \}.
        \end{align*}
        We now consider the sequences $\bm{a}\overline{\alpha}$ and $\overline{\beta}\bm{b}$.
        Since $d_H(\bm{a}_i\overline{\alpha},\bm{a})= d_H(\bm{b},\overline{\beta}\bm{b}_{j-m-\ell-3})=1$, it follows by Lemma \ref{lem:del_position} that there is exactly one choice for $\bm{a}_i \overline{\alpha}$ and $\overline{\beta}\bm{b}_{j-m-\ell-3}$, namely, $\bm{a}'\triangleq \bm{a}_i \overline{\alpha}\neq \bm{a}$ is obtained from $\bm{a}\overline{\alpha}$ by deleting a symbol of its last $\alpha$-run and $\bm{b}'\triangleq \overline{\beta}\bm{b}_{j-m-\ell-3}\neq \bm{b}$ is obtained from $\overline{\beta}\bm{b}$ by deleting a symbol of its first $\beta$-run.
        It can be easily checked that $\bm{a}\alpha \bm{c}\beta \bm{b}\in \mathcal{S}(\bm{a}\alpha \bm{c}\beta \bm{b})=S$ and $\bm{a}' \alpha \bm{c}\beta\bm{b}'\notin \mathcal{S}(\bm{a}\alpha \bm{c}\beta \bm{b})=S$.
        We obtain 
        \begin{align*}
            \left(\bigcup_{i=1}^m \bigcup_{j=m+\ell+4}^{n} \mathcal{S}(\bm{x}_i,\bm{y}_j)\right) \setminus S\subseteq \{\bm{a}' \alpha \bm{c}\beta\bm{b}'\}.
        \end{align*}
    \end{itemize}
    Consequently, we can conclude that 
    \begin{align*}
        \left| \left(\bigcup_{i=1}^m \bigcup_{j=1}^{n} \mathcal{S}(\bm{x}_i,\bm{y}_j)\right) \setminus S \right|\leq 2.
    \end{align*}
    \end{itemize}
    Therefore, we can calculate $|B|=|\mathcal{B}(\bm{x},\bm{y})\setminus S|\leq 20$, which completes the proof.
\end{IEEEproof}

Now, combining Equation (\ref{eq:(1,0)_B}) with Claims \ref{cla:(1,0)_S} and \ref{cla:(1,0)_B}, we have $|\mathcal{B}(\bm{x},\bm{y})|=|S|+|B|\leq n+20$.
This completes the proof.

\subsection{Proof of Lemma \ref{lem:(0,0)}}\label{IV-F}
The proof follows a strategy similar to that of Lemma \ref{lem:(2,2)}, we include it here for completeness. 
We first characterize the structures of $\bm{x}$ and $\bm{y}$ when $|\mathcal{D}(\bm{x},\bm{y})|=0$ and $|\mathcal{S}(\bm{x},\bm{y})|=0$.
Since $|\mathcal{S}(\bm{x},\bm{y})|=0$, by Lemma \ref{lem:sub}, we have $d_H(\bm{x},\bm{y})\geq 3$. 
Then we can write $\bm{x}=\bm{a}\alpha\bm{u} \beta\bm{b}$ and $\bm{y}=\bm{a}\overline{\alpha}\bm{v} \overline{\beta}\bm{b}$ with $\bm{u}\neq \bm{v}$ for some $\alpha\,\beta \in \Sigma$ and $\bm{u},\bm{u}\in \Sigma^{\ast}$.
Moreover, since $|\mathcal{D}(\bm{x},\bm{y})|=0$, by Lemma \ref{lem:del}, we can conclude that $\bm{u} \beta\neq \overline{\alpha}\bm{v}$ and $\alpha\bm{u}\neq \bm{v} \overline{\beta}$. 
In this case, we have
\begin{align*}
  \mathcal{D}(\bm{x},\bm{y})=\emptyset, \quad   \mathcal{S}(\bm{x},\bm{y})=\emptyset.
\end{align*}
This implies that
\begin{align*}
  S=\emptyset, \quad D=\emptyset.  
\end{align*}
Therefore, we shall calculate $\mathcal{B}(\bm{x},\bm{y})$ directly. 

Let $\bm{x}_i\triangleq \bm{x}_{[n]\setminus\{i\}}$ and $\bm{y}_i\triangleq \bm{y}_{[n]\setminus\{i\}}$ for $i\in [n]$, we have  
\begin{align*}
    \mathcal{B}(\bm{x},\bm{y})
    &= \bigcup_{i=1}^{n} \bigcup_{j=1}^{n} \mathcal{S}(\bm{x}_i, \bm{y}_j).
\end{align*}
We first determine the intersection $\mathcal{S}(\bm{x}_i, \bm{y}_j)$.
Let $m\triangleq |\bm{a}|$ and $\ell\triangleq |\bm{u}|=|\bm{v}|$.
\begin{itemize}
    \item If $i=m+1$, we have $\bm{x}_i=\bm{a}\bm{u} \beta\bm{b}$. We further consider the choice for $j$.
    \begin{itemize}
        \item If $m+1\leq j\leq m+\ell+1$, we have $\bm{y}_j=\bm{a}(\overline{\alpha}\bm{v})_{j-m}\overline{\beta}\bm{b}$. We can compute $d_H(\bm{x}_i,\bm{y}_j)=d_H(\bm{u},(\overline{\alpha}\bm{v})_{j-m})+d_H(\beta\bm{b},\overline{\beta}\bm{b})\ge 1$. It suffices to consider the scenario where $d_H(\bm{x}_i,\bm{y}_j)\le 2$, or equivalently $d_H(\bm{u},(\overline{\alpha}\bm{v})_{j-m})\le 1$, as in the other cases, the intersection $\mathcal{S}(\bm{x}_i, \bm{y}_j)$ is empty. 
        By Lemma \ref{lem:del_position}, we can conclude that there are at most three choices for $(\overline{\alpha}\bm{v})_{j-m}$, which, in the worst case, we can denote these sequences as $\bm{u},\bm{w}^{(1)},\bm{w}^{(2)}$.
        By Lemma \ref{lem:sub}, we get
        \begin{align*}
        \bigcup_{j=m+1}^{m+\ell+1}  \mathcal{S}(\bm{x}_{m+1},\bm{y}_j) =\{\bm{a} \bm{u}\overline{\beta}\bm{b},~\bm{a}\bm{u}\beta\bm{b},~\bm{a}\bm{w}^{(1)}\beta\bm{b},~\bm{a}\bm{w}^{(2)}\beta\bm{b}\}.
        \end{align*}
        In this case, for any $\bm{z}=\bm{x}(i,\hat{i})=\bm{y}(j,\hat{j})$, we have $i,j\leq m+\ell+1$ and $m+\ell+2\in \{\hat{i},\hat{j}\}$.
        This implies that the sequences $\bm{a} \bm{u}\overline{\beta}\bm{b},\bm{a}\bm{u}\beta\bm{b},\bm{a}\bm{w}^{(1)}\beta\bm{b},\bm{a}\bm{w}^{(2)}\beta\bm{b}$ are good.
        
        \item If $j\le m$, we have $\bm{y}_j=\bm{a}_j\overline{\alpha}\bm{v} \overline{\beta}\bm{b}$. Since we have discussed $j=m+1$ in the previous case, we assume that $\bm{y}_j\neq \bm{y}_{m+1}$, i.e., $\bm{a}_j\overline{\alpha}\neq \bm{a}$. Recall that $\bm{u}\neq \bm{v}$, we can compute $d_H(\bm{x}_i,\bm{y}_j)=d_H(\bm{a},\bm{a}_j\overline{\alpha})+d_H(\bm{u} ,\bm{v})+d_H(\beta \bm{b},\overline{\beta}\bm{b})\ge 3$. Then we get
        \begin{align*}
            \mathcal{S}(\bm{x}_{i},\bm{y}_j)= \emptyset.
        \end{align*}
        
        \item If $j\ge m+\ell+2$, we have $\bm{y}_j=\bm{a}\overline{\alpha}\bm{v} (\overline{\beta}\bm{b})_{j-m-\ell-1}$. We can compute $d_H(\bm{x}_i,\bm{y}_j)=d_H(\bm{u} \beta, \overline{\alpha}\bm{v})+d_H(\bm{b}, (\overline{\beta}\bm{b})_{j-m-\ell-1})$. Recall that $\bm{u} \beta\neq \overline{\alpha}\bm{v}$.
        \begin{itemize}
             \item If $d_H(\bm{u} \beta, \overline{\alpha}\bm{v})\ge 3$, then the intersection $\mathcal{S}(\bm{x}_i, \bm{y}_j)$ is empty. 
           
            \item If $d_H(\bm{u} \beta, \overline{\alpha}\bm{v})=2$, it suffices to consider the scenario where $d_H(\bm{b},(\overline{\beta}\bm{b})_{j-m-\ell-1})=0$, or equivalently $(\overline{\beta}\bm{b})_{j-m-\ell-1}=\bm{b}$, as in the other cases, the intersection $\mathcal{S}(\bm{x}_i, \bm{y}_j)$ is empty. In this case, we have
            \begin{align*}
                \bigcup_{j=m+\ell+2}^{n} \mathcal{S}(\bm{x}_{m+1}, \bm{y}_j)= \left\{\bm{a}\bm{w}\bm{b}:\bm{w}\in \mathcal{S}(\bm{u}\beta, \overline{\alpha}\bm{v}) \right\}.
            \end{align*}
            Then by Lemma \ref{lem:sub}, we get 
            \begin{align*}
                \left| \bigcup_{j=m+\ell+2}^{n} \mathcal{S}(\bm{x}_{m+1}, \bm{y}_j)\right|= 2.
            \end{align*}   
    
            \item If $d_H(\bm{u} \beta, \overline{\alpha}\bm{v})=1$, it suffices to consider the scenario where $d_H(\bm{b},(\overline{\beta}\bm{b})_{j-m-\ell-1})\leq 1$, as in the other cases, the intersection $\mathcal{S}(\bm{x}_i, \bm{y}_j)$ is empty. By Lemma \ref{lem:del_position}, we can conclude that there are at most two choices for $(\overline{\beta}\bm{b})_{j-m-\ell-1}$, which, in the worst case, we can denote these sequences as $\bm{b},\bm{b}'\neq \bm{b}$. In this case, by Lemma \ref{lem:sub}, we get
            \begin{align*}
                    \bigcup_{j=m+\ell+2}^{n}  \mathcal{S}(\bm{x}_{m+1},\bm{y}_j) =\{\bm{a} \overline{\alpha}\bm{v}\bm{b},~\bm{a}\bm{u}\beta\bm{b},~\bm{a}\bm{u}\beta\bm{b}'\}.
            \end{align*}
Note that $\bm{a}\bm{u}\beta\bm{b}\in \bigcup_{j=m+1}^{m+\ell+1}  \mathcal{S}(\bm{x}_{m+1},\bm{y}_j)$, in this case, we get
    \begin{align*}
        \left|\left(\bigcup_{j=m+\ell+2}^{n}  \mathcal{S}(\bm{x}_{m+1},\bm{y}_j)\right)\setminus \left(\bigcup_{j=m+1}^{m+\ell+1}  \mathcal{S}(\bm{x}_{m+1},\bm{y}_j)\right)\right|\le 2.
        \end{align*}
        \end{itemize}     
        In all cases, we get
        \begin{align*}
        \left|\left(\bigcup_{j=m+\ell+2}^{n}  \mathcal{S}(\bm{x}_{m+1},\bm{y}_j)\right)\setminus \left(\bigcup_{j=m+1}^{m+\ell+1}  \mathcal{S}(\bm{x}_{m+1},\bm{y}_j)\right)\right|\le 2.
        \end{align*}
    \end{itemize}
    Consequently, we obtain
    \begin{align*}
        \left|\bigcup_{j=1}^{n}  \mathcal{S}(\bm{x}_{m+1},\bm{y}_j)\right|\le 6. 
        \end{align*} 
    Moreover, there are at least four good sequences, which are  $\bm{a} \bm{u}\overline{\beta}\bm{b},\bm{a}\bm{u}\beta\bm{b},\bm{a}\bm{w}^{(1)}\beta\bm{b},\bm{a}\bm{w}^{(2)}\beta\bm{b}$.

    \item If $i=m+\ell+2$, by considering the reversal of $\bm{x}$ and $\bm{y}$ and using the conclusions of case $i=m+1$, in the worst case, we can obtain
    \begin{align*}
        \left|\bigcup_{j=1}^{n}  \mathcal{S}(\bm{x}_{m+\ell+2},\bm{y}_j)\right|\le 6. 
    \end{align*} 
    In detail, we have
      \begin{align*}
        \bigcup_{j=m+2}^{m+\ell+2}  \mathcal{S}(\bm{x}_{m+\ell+2},\bm{y}_j) =\{\bm{a} \overline{\alpha}\bm{u}\bm{b},~ \bm{a}\alpha\bm{u}\bm{b},~ \bm{a}\alpha\bm{w}^{(3)}\bm{b},~ \bm{a}\alpha\bm{w}^{(4)}\bm{b}\},
        \end{align*}
 and
   \begin{align*}
        \left| \bigcup_{j=1}^{m+1} \bigcup_{j=m+\ell+3}^{n} \mathcal{S}(\bm{x}_{m+\ell+2}, \bm{y}_j)\right|\le 2,
     \end{align*} 
where $\bm{w}^{(3)}, \bm{w}^{(4)}\in \mathcal{D}(\bm{v}\overline{\beta})\cap \mathcal{S}(\bm{u})$. Moreover, 
     $\bm{a} \overline{\alpha}\bm{u}\bm{b}, \bm{a}\alpha\bm{u}\bm{b}, \bm{a}\alpha\bm{w}^{(3)}\bm{b}, \bm{a}\alpha\bm{w}^{(4)}\bm{b}$ are good sequences.
    

    \item If $i\le m$, we have $\bm{x}_i=\bm{a}_i\alpha\bm{u} \beta\bm{b}$. Since we have discussed $i=m+1$ in the first case, we assume that $\bm{x}_i\neq \bm{x}_{m+1}$, i.e., $\bm{a}_i\alpha\neq \bm{a}$. We further consider the choice for $j$.
    \begin{itemize}
        \item If $j\le m$, we have $\bm{y}_j=\bm{a}_j\overline{\alpha}\bm{v} \overline{\beta}\bm{b}$. We can compute $d_H(\bm{x}_i,\bm{y}_j)\geq d_H(\alpha\bm{u} \beta,\overline{\alpha}\bm{v} \overline{\beta})=d_H(\bm{x},\bm{y})\geq 3$. This implies that
        \begin{align*}
            \mathcal{S}(\bm{x}_i, \bm{y}_j)=\emptyset.
        \end{align*}
    
        \item If $m+1\leq j\leq m+\ell+1$, we have $\bm{y}_j=\bm{a}(\overline{\alpha}\bm{v})_{j-m}\overline{\beta}\bm{b}$. Recall that $\bm{a}_i\alpha\neq \bm{a}$, we can compute $d_H(\bm{x}_i,\bm{y}_j)=d_H(\bm{a}_i\alpha,\bm{a})+d_H(\bm{u},(\overline{\alpha}\bm{v})_{j-m})+d_H(\beta\bm{b},\overline{\beta}\bm{b})\ge 2$. It suffices to consider the scenario where $d_H(\bm{x}_i,\bm{y}_j)=2$, or equivalently $(\overline{\alpha}\bm{v})_{j-m}=\bm{u}$ and $d_H(\bm{a}_i\alpha,\bm{a})=1$, as in the other cases, the intersection $\mathcal{S}(\bm{x}_i, \bm{y}_j)$ is empty. By Lemma \ref{lem:del_position}, we can conclude that $\bm{a}''\triangleq \bm{a}_i\alpha\neq \bm{a}$ is obtained from $\bm{a}\alpha$ by deleting its last $\overline{\alpha}$-run. By Lemma \ref{lem:sub}, we get
        \begin{align*}
        \bigcup_{i=1}^{m}\bigcup_{j=m+1}^{m+\ell+1}  \mathcal{S}(\bm{x}_{i},\bm{y}_j) =\{\bm{a}''\bm{u}\overline{\beta}\bm{b},~\bm{a}\bm{u}\beta\bm{b}\}.
        \end{align*}
        In this case, for any $\bm{z}=\bm{x}(i,\hat{i})=\bm{y}(j,\hat{j})$, we have $i,j\leq m+\ell+1$ and $m+\ell+2\in \{\hat{i},\hat{j}\}$.
        This implies that the sequences $\bm{a}''\bm{u}\overline{\beta}\bm{b},\bm{a}\bm{u}\beta\bm{b}$ are good.

        \item If $j\ge m+\ell+2$, we have $\bm{y}_j=\bm{a}\overline{\alpha}\bm{v} (\overline{\beta}\bm{b})_{j-m-\ell-1}$. Recall that $\bm{a}_i \alpha\neq \bm{a}$ and $\bm{u}\beta\neq \overline{\alpha}\bm{v}$, we can compute $d_H(\bm{x}_i,\bm{y}_j)=d_H(\bm{a}_i\alpha,\bm{a})+d_H(\bm{u}\beta,\overline{\alpha}\bm{v})+d_H(\bm{b},(\overline{\beta}\bm{b})_{j-m-\ell-1})\ge 2$. It suffices to consider the scenario where $d_H(\bm{x}_i,\bm{y}_j)=2$, or equivalently $d_H(\bm{a}_i\alpha,\bm{a})=d_H(\bm{u}\beta,\overline{\alpha}\bm{v})=1$ and $(\overline{\beta}\bm{b})_{j-m-\ell-1}=\bm{b}$, as in the other cases, the intersection $\mathcal{S}(\bm{x}_i, \bm{y}_j)$ is empty. By Lemma \ref{lem:del_position}, we can conclude that $\bm{a}''\triangleq \bm{a}_i\alpha\neq \bm{a}$ is obtained from $\bm{a}\alpha$ by deleting its last $\overline{\alpha}$-run. By Lemma \ref{lem:sub}, we get
        \begin{align*}
        \bigcup_{i=1}^{m}\bigcup_{j=m+\ell+2}^{n}  \mathcal{S}(\bm{x}_{i},\bm{y}_j) =\{\bm{a}''\overline{\alpha}\bm{v}\bm{b},~\bm{a}\bm{u}\beta\bm{b}\}.
        \end{align*}
        Note that $\bm{a}\bm{u}\beta\bm{b}\in \bigcup_{j=m+1}^{m+\ell+1}  \mathcal{S}(\bm{x}_{m+1},\bm{y}_j)$, we get
        \begin{align*}
        \left(\bigcup_{i=1}^{m}\bigcup_{j=1}^{n}  \mathcal{S}(\bm{x}_{i},\bm{y}_j) \right)\setminus \left(\bigcup_{j=m+1}^{m+\ell+1}  \mathcal{S}(\bm{x}_{m+1},\bm{y}_j) \right)=\{\bm{a}''\bm{u}\overline{\beta}\bm{b},~\bm{a}''\overline{\alpha}\bm{v}\bm{b}\}.
        \end{align*}
    \end{itemize}

    \item If $i\ge m+\ell+3$, by considering the reversal of $\bm{x}$ and $\bm{y}$ and using the conclusion of case $i\le m$, in the worst case, we can obtain
    \begin{align*}
        \left(\bigcup_{i=m+\ell+3}^{n}\bigcup_{j=1}^{n}  \mathcal{S}(\bm{x}_{i},\bm{y}_j) \right)\setminus \left(\bigcup_{j=m+2}^{m+\ell+2}  \mathcal{S}(\bm{x}_{m+\ell+2},\bm{y}_j) \right)= \{\bm{a}\overline{\alpha}\bm{u}\bm{b}'',~\bm{a}\bm{v}\overline{\beta}\bm{b}''\},
    \end{align*}
    where $\bm{b}''\neq \bm{b}$ is obtained from $\beta\bm{b}$ by deleting its first $\overline{\beta}$-run. Moreover, the sequence $\bm{a}\overline{\alpha}\bm{u}\bm{b}''$ is good.

    \item If $m+2\leq i\leq m+\ell+1$, we have $\bm{x}_i=\bm{a}\alpha\bm{u}_{i-m-1} \beta\bm{b}$. We further consider the choice for $j$.
    \begin{itemize}
        \item If $j\leq m+1$, we have $\bm{y}_j=(\bm{a}\overline{\alpha})_{j}\bm{v} \overline{\beta}\bm{b}$. We can compute $d_H(\bm{x}_i,\bm{y}_j)=d_H(\bm{a},(\bm{a}\overline{\alpha})_{j})+d_H(\alpha\bm{u}_{i-m-1},\bm{v})+d_H(\beta\bm{b},\overline{\beta}\bm{b})\ge 1$. It suffices to consider the scenario where $d_H(\bm{x}_i,\bm{y}_j)\le 2$, or equivalently, $d_H(\bm{a},(\bm{a}\overline{\alpha})_{j})+d_H(\alpha\bm{u}_{i-m-1},\bm{v})\le 1$, as in the other cases, the intersection $\mathcal{S}(\bm{x}_i, \bm{y}_j)$ is empty.
        \begin{itemize}
            \item If $d_H(\bm{a},(\bm{a}\overline{\alpha})_{j})=0$, then $d_H(\alpha\bm{u}_{i-m-1},\bm{v})\le 1$. By Lemma \ref{lem:del_position}, we can conclude that there are at most three choices for $\alpha\bm{u}_{i-m-1}$, which, in the worst case, we can denote these sequences as $\bm{v},\tilde{\bm{w}}^{(1)},\tilde{\bm{w}}^{(2)}$.
            By Lemma \ref{lem:sub}, we get
            \begin{align*}
            \mathcal{S}(\bm{x}_{i},\bm{y}_j) =\{\bm{a}\bm{v}\beta\bm{b},~\bm{a} \bm{v}\overline{\beta}\bm{b},~\bm{a}\tilde{\bm{w}}^{(1)}\overline{\beta}\bm{b},~\bm{a}\tilde{\bm{w}}^{(2)}\overline{\beta}\bm{b}\}.
            \end{align*}
    
            \item If $d_H(\bm{a},(\bm{a}\overline{\alpha})_{j})=1$, then $\alpha\bm{u}_{i-m-1}=\bm{v}$. By Lemma \ref{lem:del_position}, we can conclude that $\bm{a}'\triangleq (\bm{a}\overline{\alpha})_{j}\neq \bm{a}$ is obtained from $\bm{a}\overline{\alpha}$ by deleting its last $\alpha$-run. By Lemma \ref{lem:sub}, we get
            \begin{align*}
            \mathcal{S}(\bm{x}_{i},\bm{y}_j) =\{\bm{a}\bm{v}\overline{\beta}\bm{b},~\bm{a}'\bm{v}\beta\bm{b}\}.
            \end{align*}
        \end{itemize}
        Consequently, we obtain
        \begin{align*}
        \bigcup_{i=m+2}^{m+\ell+1} \bigcup_{j=1}^{m+1}  \mathcal{S}(\bm{x}_{i},\bm{y}_j)= \{\bm{a}\bm{v}\beta\bm{b},~\bm{a} \bm{v}\overline{\beta}\bm{b},~\bm{a}\tilde{\bm{w}}^{(1)}\overline{\beta}\bm{b},~\bm{a}\tilde{\bm{w}}^{(2)}\overline{\beta}\bm{b},~\bm{a}'\bm{v}\beta\bm{b} \}.
        \end{align*}
        In this case, for any $\bm{z}=\bm{x}(i,\hat{i})=\bm{y}(j,\hat{j})$, we have $i,j\leq m+\ell+1$ and $m+\ell+2\in \{\hat{i},\hat{j}\}$.
        This implies that the sequences $\bm{a} \bm{v}\overline{\beta}\bm{b}, \bm{a}\bm{v}\beta\bm{b}, \bm{a}\tilde{\bm{w}}^{(1)}\overline{\beta}\bm{b}, \bm{a}\tilde{\bm{w}}^{(2)}\overline{\beta}\bm{b}, \bm{a}'\bm{v}\beta\bm{b}$ are good.
    
        \item If $j\geq m+\ell+2$, by considering the reversal of $\bm{x}$ and $\bm{y}$ and using the conclusion of the previous case $j\le m+1$, we can obtain
        \begin{align*}
        \bigcup_{i=m+2}^{m+\ell+1} \bigcup_{j=m+\ell+2}^{n}  \mathcal{S}(\bm{x}_{i},\bm{y}_j)= \{\bm{a}\alpha\bm{v}\bm{b},~ \bm{a} \overline{\alpha}\bm{v}\bm{b},~ \bm{a}\overline{\alpha}\tilde{\bm{w}}^{(3)}\bm{b},~ \bm{a}\overline{\alpha}\tilde{\bm{w}}^{(4)}\bm{b},~ \bm{a}\alpha\bm{v}\bm{b}'\},
        \end{align*}
        where $\tilde{\bm{w}}^{(3)},\tilde{\bm{w}}^{(4)}\in \mathcal{D}(\bm{u}\beta)\cap \mathcal{S}(\bm{v})$ and $\bm{b}'\neq \bm{b}$ is obtained from $\overline{\beta}\bm{b}$ by deleting its first $\beta$-run.
        Moreover, the sequences $\bm{a} \overline{\alpha}\bm{v}\bm{b},  \bm{a}\alpha\bm{v}\bm{b}, \bm{a}\overline{\alpha}\tilde{\bm{w}}^{(3)}\bm{b}, \bm{a}\overline{\alpha}\tilde{\bm{w}}^{(4)}\bm{b}, \bm{a}\alpha\bm{v}\bm{b}'$ are good.
        
        \item If $m+2\leq j\leq m+\ell+1$, we have $\bm{y}_j=\bm{a}\overline{\alpha}\bm{v}_{j-m-1}\overline{\beta}\bm{b}$. We can compute $d_H(\bm{x}_i,\bm{y}_j)=d_H(\bm{a}\alpha,\bm{a}\overline{\alpha})+d_H(\bm{u}_{i-m-1},\bm{v}_{j-m-1})+d_H(\beta\bm{b},\overline{\beta}\bm{b})\ge 2$. It suffices to consider the scenario where $d_H(\bm{x}_i,\bm{y}_j)= 2$, or equivalently, $\bm{u}_{i-m-1}=\bm{v}_{j-m-1}$, as in the other cases, the intersection $\mathcal{S}(\bm{x}_i, \bm{y}_j)$ is empty. Recall that $\bm{u}\neq \bm{v}$, by Lemma \ref{lem:del}, we can conclude that there are at most two choices for $\bm{u}_{i-m-1}$, namely, $\bm{u}^{(1)}$ and $\bm{u}^{(2)}$. By Lemma \ref{lem:sub}, we get
        \begin{align*}
        \bigcup_{i=m+2}^{m+\ell+1} \bigcup_{j=m+2}^{m+\ell+1}  \mathcal{S}(\bm{x}_{i},\bm{y}_j) =\{\bm{a}\alpha\bm{u}^{(1)}\overline{\beta}\bm{b},~\bm{a}\overline{\alpha}\bm{u}^{(1)}\beta\bm{b},~\bm{a}\alpha\bm{u}^{(2)}\overline{\beta}\bm{b},~\bm{a}\overline{\alpha}\bm{u}^{(2)}\beta\bm{b}\}.
        \end{align*}
        In this case, for any $\bm{z}=\bm{x}(i,\hat{i})=\bm{y}(j,\hat{j})$, we have $i,j\leq m+\ell+1$ and $m+\ell+2\in \{\hat{i},\hat{j}\}$.
        This implies that the sequences $\bm{a}\alpha\bm{u}^{(1)}\overline{\beta}\bm{b}, ~\bm{a}\overline{\alpha}\bm{u}^{(1)}\beta\bm{b}, \bm{a}\alpha\bm{u}^{(2)}\overline{\beta}\bm{b}, \bm{a}\overline{\alpha}\bm{u}^{(2)}\beta\bm{b}$ are good.
    \end{itemize}
    Consequently, we can conclude that
    \begin{align*}
    \bigcup_{i=m+2}^{m+\ell+1} \bigcup_{j=1}^{n}  \mathcal{S}(\bm{x}_{i},\bm{y}_j)
    &= \big\{\bm{a}\bm{v}\beta\bm{b},~ \bm{a} \bm{v}\overline{\beta}\bm{b},~ \bm{a}\tilde{\bm{w}}^{(1)}\overline{\beta}\bm{b},~ \bm{a}\tilde{\bm{w}}^{(2)}\overline{\beta}\bm{b},~ \bm{a}'\bm{v}\beta\bm{b},~ \bm{a}\alpha\bm{v}\bm{b},~ \bm{a} \overline{\alpha}\bm{v}\bm{b},~ \bm{a}\overline{\alpha}\tilde{\bm{w}}^{(3)}\bm{b},\\
    &\quad \quad \quad \quad \bm{a}\overline{\alpha}\tilde{\bm{w}}^{(4)}\bm{b},~\bm{a}\bm{v}\beta\bm{b}',~ \bm{a}\alpha\bm{u}^{(1)}\overline{\beta}\bm{b},~\bm{a}\overline{\alpha}\bm{u}^{(1)}\beta\bm{b},~\bm{a}\alpha\bm{u}^{(2)}\overline{\beta}\bm{b},~\bm{a}\overline{\alpha}\bm{u}^{(2)}\beta\bm{b}\big\}.
    \end{align*}
\end{itemize}  

Combining the five cases above, we establish the upper bound
\begin{align*}
   | \mathcal{B}(\bm{x},\bm{y})|\le 6+6+2+2+14=30.
\end{align*}
Furthermore, there exist $24$ good sequences in the subsequent set listing:
\begin{align*}
\Big\{&\bm{a} \bm{u}\overline{\beta}\bm{b},~\bm{a}\bm{u}\beta\bm{b},~\bm{a}\bm{w}^{(1)}\beta\bm{b},~\bm{a}\bm{w}^{(2)}\beta\bm{b},~\bm{a} \overline{\alpha}\bm{u}\bm{b},~\bm{a}\alpha\bm{u}\bm{b},~\bm{a}\alpha\bm{w}^{(3)}\bm{b},~\bm{a}\alpha\bm{w}^{(4)}\bm{b},~\bm{a}''\bm{u}\overline{\beta}\bm{b},~\bm{a}\overline{\alpha}\bm{u}\bm{b}'',~
\bm{a}\bm{v}\beta\bm{b},~ \bm{a} \bm{v}\overline{\beta}\bm{b},~ \\
&\bm{a}\tilde{\bm{w}}^{(1)}\overline{\beta}\bm{b},~ \bm{a}\tilde{\bm{w}}^{(2)}\overline{\beta}\bm{b},~ \bm{a}'\bm{v}\beta\bm{b},~  \bm{a}\alpha\bm{v}\bm{b},~ \bm{a} \overline{\alpha}\bm{v}\bm{b},~\bm{a}\overline{\alpha}\tilde{\bm{w}}^{(3)}\bm{b},~\bm{a}\overline{\alpha}\tilde{\bm{w}}^{(4)}\bm{b},~\bm{a}\bm{v}\beta\bm{b}'~
\bm{a}\alpha\bm{u}^{(1)}\overline{\beta}\bm{b},~\bm{a}\overline{\alpha}\bm{u}^{(1)}\beta\bm{b},~\bm{a}\alpha\bm{u}^{(2)}\overline{\beta}\bm{b},~\bm{a}\overline{\alpha}\bm{u}^{(2)}\beta\bm{b}\Big\}.
\end{align*}
This completes the proof.

\section{conclusion}\label{sec:concl}

In this paper, we study \((n, N; \mathcal{B})\)-reconstruction codes.  
Our results show that when \(N\) is set to \(4n - 8\), \(3n - 4\), \(2n + 9\), \(n + 21\), \(31\), and \(7\), the redundancy of binary \((n, N; \mathcal{B})\)-reconstruction codes can be \(0\), \(1\), \(2\), \(\log \log n + 3\), \(\log n + 1\), and \(3 \log n + 4\), respectively.  
In future work, we will explore the construction of \((n, N; \mathcal{B})\)-reconstruction codes for other values of \(N\), particularly when \(N\) is a small constant.  
One possible approach is to use Lemma \ref{lem:(0,0)} to characterize the structure of pairs of sequences whose single-deletion single-substitution balls intersect in at most \(N-1\) elements, and then impose suitable constraints to forbid such pairs.

\end{document}